\documentclass[aps,prb,reprint,superscriptaddress,longbibliography]{revtex4-1}
\pdfoutput=1

\usepackage{graphicx}
\usepackage{amsmath}
\usepackage{amssymb}
\usepackage[makeroom]{cancel}
\usepackage{color}

\newcommand{\dresden}{Max-Planck-Institut f\"ur Physik komplexer Systeme, D-01187 Dresden, Germany}
\newcommand{\aarhus}{On leave from Department of Physics and Astronomy, Aarhus University, DK-8000 Aarhus C, Denmark}

\begin{document}

\title{Chain and ladder models with two-body interactions and analytical ground states}

\begin{abstract}
We consider a family of spin-$1/2$ models with few-body, SU(2) invariant Hamiltonians and analytical ground states related to the 1D Haldane-Shastry wavefunction. The spins are placed on the surface of a cylinder, and the standard 1D Haldane-Shastry model is obtained by placing the spins with equal spacing in a circle around the cylinder. Here, we show that another interesting family of models with two-body exchange interactions is obtained if we instead place the spins along one or two lines parallel to the cylinder axis, giving rise to chain and ladder models, respectively. We can change the scale along the cylinder axis without changing the radius of the cylinder. This gives us a parameter that controls the ratio between the circumference of the cylinder and all other length scales in the system. We use Monte Carlo simulations and analytical investigations to study how this ratio affects the properties of the models. If the ratio is large, we find that the two legs of the ladder decouple into two chains that are in a critical phase with Haldane-Shastry-like properties. If the ratio is small, the wavefunction reduces to a product of singlets. In between, we find that the behavior of the correlations and the Renyi entropy depends on the distance considered. For small distances the behavior is critical, and for long distances the correlations decay exponentially and the entropy shows an area law behavior. The distance up to which there is critical behavior gets larger and larger as the ratio increases.
\end{abstract}

% % % % % % % % % % % % % % % % % % % % % % % % % % % % % % % % % %	
	
\author{Sourav Manna}
\affiliation{\dresden}
\author{Anne E. B. Nielsen}
\affiliation{\dresden}
\affiliation{\aarhus}
\maketitle

% % % % % % % % % % % % % % % % % % % % % % % % % % % % % % % % % %

\section{Introduction}\label{SEC:Introduction}

Models that can be solved partially or fully by using analytical tools play a crucial role to illuminate the physics of strongly correlated quantum many-body systems. They overcome, in particular cases, the bottleneck that the resources needed to do numerical computations generally grow exponentially with system size, they provide insight into mechanisms lying behind many-body phenomena, and they can be used to test numerical approximation schemes.

A number of different exactly solvable models have been found in 1D systems. These models can be grouped into three main categories.\cite{solvable_models,ha} The first one is the Heisenberg spin model\cite{Heisenberg_model} (and other related models in 1D\cite{Ising_1, Ising_2}) with its exact solution by Bethe's ansatz.\cite{bethe} The second member is the Tomonaga-Luttinger liquids,\cite{haldane1981,luttinger,tomonaga} solved by bosonization techniques. This model reveals the non Fermi-liquid properties of 1D fermionic systems. Finally, the third family are models related to the Calogero-Sutherland model\cite{calogero_sutherland} with long range interactions. The Calogero-Sutherland model is defined in the continuum, and a lattice spin version of the model was found by Haldane and Shastry.\cite{HSmodel_H,HSmodel_S} In addition, tensor networks provide an efficient tool to find models with known ground states and short range interactions.\cite{AKLT,fannes,fernandez}

Important work has also been done in the context of exactly solvable ladder models (see e.g.\ \onlinecite{ladder_1,ladder_6,ladder_4, ladder_5,ladder_7,ladder_8,ladder_2,ladder_3}). An exactly solvable spin ladder with biquadratic interactions has been obtained via Bethe's ansatz in \onlinecite{ladder_1}. In \onlinecite{ladder_5}, a spin ladder model with interactions between spins on neighboring rungs, and in \onlinecite{ladder_3}, behavior of the two leg frustrated quantum spin $\frac{1}{2}$ ladder containing Heisenberg intra rung and Ising inter rung interactions has been studied. A three leg spin ladder with isotropic Heisenberg interactions and additional many-body terms in the context of magnetization is discussed in \onlinecite{ladder_4}, and recently entanglement entropy has been investigated for an exactly solvable two leg spin ladder which contains three body interactions.\cite{ladder_2}

\begin{figure}
\includegraphics[width=0.8\columnwidth]{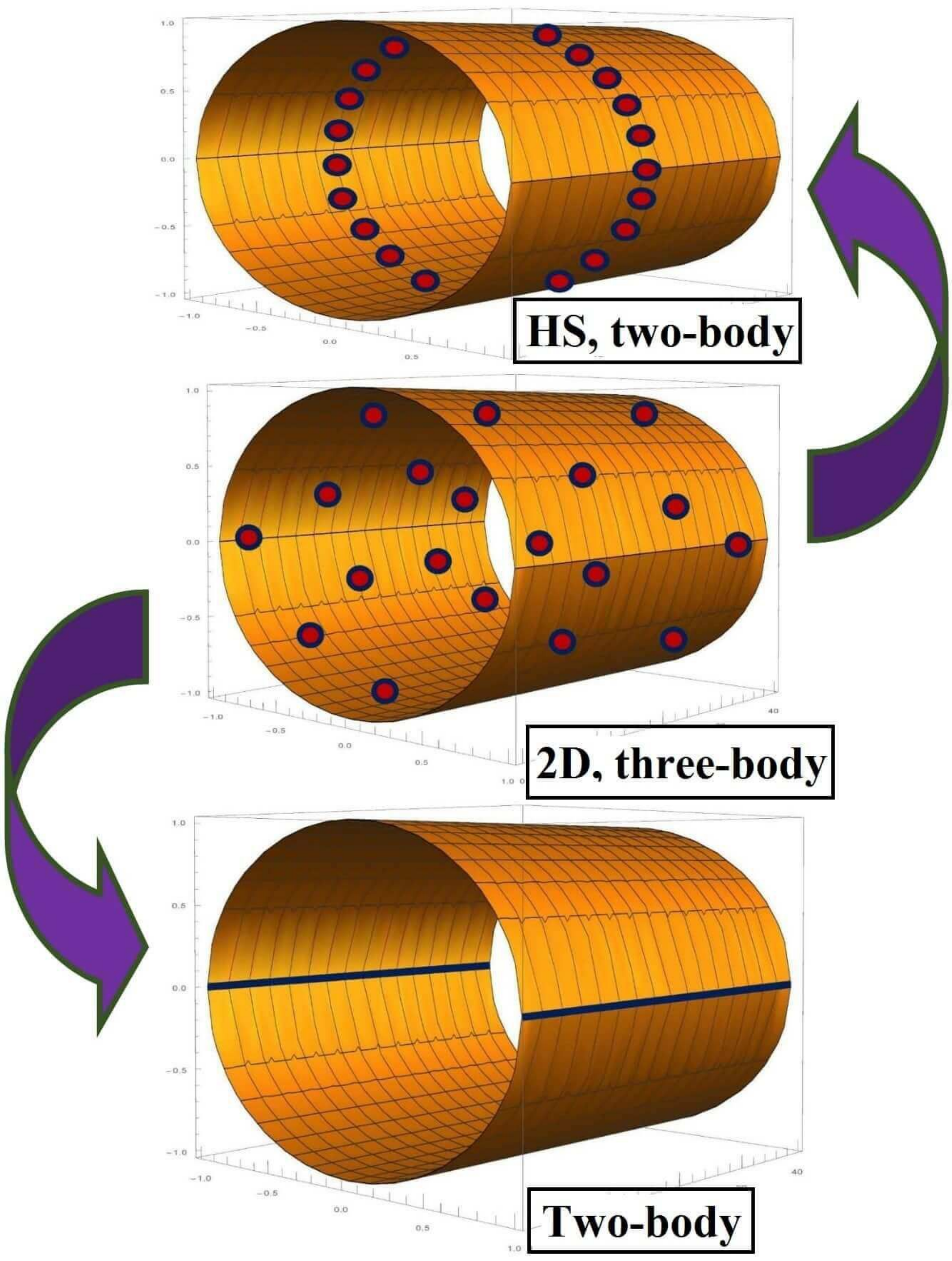}
\caption{We consider a model, in which the lattice points (spins) are placed on the surface of a cylinder. The middle cylinder depicts the 2D generalization of the 1D HS model, which can be defined for spins on an arbitrary lattice. In general, it has two- and three-body interactions. The 1D HS model shown on the upper most cylinder is a special case with only two-body interactions. In this article, we show that if the positions of the spins are restricted to the blue lines shown on the lower most cylinder, we also get a model with only two-body interactions.}\label{FIG-1}
\end{figure}

In the present paper, we construct chain and ladder models with two-body interactions and analytical ground states that are related to the 1D Haldane-Shastry (HS) model \cite{stephan_Pollman}. In the original 1D HS model, $N$ equidistant spin $\frac{1}{2}$ particles are arranged on a 1D circle and interact antiferromagnetically through an exchange interaction. The interaction strength is inversely proportional to the square of the chord distance between two spins on the circle. The Hamiltonian of the 1D HS model
\begin{equation}\label{HamHS}
H_\mathrm{HS}=\sum_{i\neq j}^{N}\left[\frac{N}{\pi} \sin\left(\frac{i-j}{N}\pi\right)\right]^{-2}\mathbf{S}_i\cdot \mathbf{S}_j
\end{equation}
is exactly solvable up to all of its ground and excited states. An interesting feature of this model is that it contains elementary excitations named spinons, which are spin $\frac{1}{2}$ particles obeying semion statistics. The possibility of having a 1D hyperbolic version of the HS model with infinitely many spins has been investigated by Inozemtsev.\cite{Inozemtsev}

A generalization of the 1D HS model, which is valid for arbitrary lattices on a cylinder surface, has been found recently\cite{Anne_PRL} and is illustrated in Fig.\ \ref{FIG-1}. The model has two- and three-body interactions, and the ground state, but not the excited states, is known analytically. In 2D, the ground state is closely related to the Kalmeyer-Laughlin state\cite{KL_state} which is the spin version of the bosonic Laughlin state at half filling. It is known that if one restricts the positions of the spins to a circle around the cylinder, one gets a two-body model. If, in addition, the spins are uniformly distributed on the circle, as visualized on the upper most cylinder in Fig.\ \ref{FIG-1}, the model reduces to the 1D HS model.\cite{jstat} A different choice of the spin positions gives a 1D HS model with open boundary conditions.\cite{openHS}

Here, we show that there is also a different way to obtain a family of two-body models, and we investigate the properties of some members of this family. More specifically, the two-body models are obtained, when the spin positions are restricted to be on the two blue lines depicted on the lower most cylinder in Fig.\ \ref{FIG-1}. In particular, this allows us to construct a family of 1D models and of ladder models with only two-body interactions and analytical ground states. It is interesting to ask, whether the properties of these models are similar to those of the original 1D HS model or not. Our investigations show that the properties of the models depend on how large the circumference of the cylinder is compared to the other length scales in the system. If this ratio is large, the models have properties close to those of the 1D HS model, and if the ratio is small, the wavefunction reduces to a product of singlets. In between, we find an interesting behavior, where the correlations and entropy display critical properties over short distances and exponential decay of correlations and area law entropy for large distances. As the parameter controlling the ratio varies, the length scale separating the two behaviors changes.

The paper is organized as follows. In Sec.\ \ref{SEC:2Dmodel}, we briefly recall the 2D HS model for spins on an arbitrary lattice on the cylinder. In Sec.\ \ref{SEC:1Dmodel}, we discuss the 1D HS model on the circle. In Sec.\ \ref{SEC:twobody}, we show that the 2D HS model reduces to a two-body model for particular choices of the lattice. Special cases include spin chain models, which we analyze in Sec.\ \ref{SEC:chain}, and ladder models, which we analyze in Sec.\ \ref{SEC:ladder}. Section \ref{SEC:conclusion} concludes the paper.

\section{The 2D HS Model}\label{SEC:2Dmodel}

We first briefly recall the 2D HS model\cite{Anne_PRL} on the cylinder. The position of the $j$th spin on the cylinder is specified by the complex number $W_j$. $\mathrm{Re}(W_j)$ is the position in the direction along the cylinder axis, and $\mathrm{Im}(W_j)$ is the position in the perpendicular direction around the cylinder. We take the circumference of the cylinder to be $2\pi$, and therefore $\mathrm{Im}(W_j)$ is periodic with period $2\pi$. We also define a corresponding set of points $z_j$ in the complex plane through the mapping $z_j = e^{W_j}$. We shall assume throughout that all spins are at different positions, i.e., $z_j\neq z_k$ whenever $j\neq k$.

The local Hilbert space on site number $j$ is spanned by the states $|s_j\rangle$ with $s_j\in\{-1,1\}$. In the following, we choose the number of spins $N$ to be even and consider the many-body state
\begin{equation}
|\psi\rangle=\sum_{s_1,s_2,\ldots,s_N}\psi_{s_1,s_2,\ldots,s_N}(z_1,z_2,\ldots,z_N) |s_1,s_2,\ldots,s_N\rangle
\end{equation}
with
\begin{multline}\label{wf}
\psi_{s_1,s_2,\ldots,s_N}(z_1,z_2,\ldots,z_N)=\\
\delta_\mathbf{s}\prod_{p=1}^N\chi_{p,s_p}\prod_{j<k}^N(z_j-z_k)^{\frac{1}{2} (s_js_k-1)}.
\end{multline}
Here, $\delta_\mathbf{s}=1$ for $\sum_{j=1}^N s_j=0$ and $\delta_\mathbf{s}=0$ otherwise, and the phase factors are $\chi_{p,s_p}=\exp[i\pi(p-1)(s_p+1)/2]$, since this ensures that \eqref{wf} is a spin singlet.\cite{Anne_PRL} The state \eqref{wf} is invariant under relabelling of the indices,\cite{torus} and we can hence choose the numbering of the spins after convenience.

We define a set of positive semi-definite and Hermitian operators
\begin{multline} \label{Ham}
H_i=\frac{1}{2}\sum_{j(\neq i)}|w_{ij}|^2
-\frac{2i}{3}\sum_{j\neq k(\neq i)}\bar{w}_{ij}w_{ik}
\, \mathbf{S}_i\cdot(\mathbf{S}_j\times\mathbf{S}_k)\\
+\frac{2}{3}\sum_{j(\neq i)}|w_{ij}|^2 \, \mathbf{S}_i\cdot \mathbf{S}_j+\frac{2}{3}\sum_{j\neq k(\neq i)}\bar{w}_{ij}w_{ik} \, \mathbf{S}_j\cdot \mathbf{S}_k
\end{multline}
acting on the $N$ spins. Here, $w_{ij}=g(z_i)/(z_i-z_j)+h(z_i)$, $g$ and $h$ are arbitrary functions of $z_i$, $\bar{w}_{ij}$ is the complex conjugate of $w_{ij}$, and $\mathbf{S}_i=(S_i^x,S_i^y,S_i^z)$ is the spin operator acting on the spin positioned at $z_i$ ($|s_i\rangle$ are the eigenstates of $S_i^z$ with eigenvalues $s_i/2$). We use the notation $\sum_{p\neq q}$ as the sum over $p$ and $q$ and $\sum_{p(\neq q)}$ as the sum over $p$ only. Likewise, $\sum_{p\neq q (\neq r)}$ means the sum over $p$ and $q$ with $p\neq q$, $p\neq r$, and $q\neq r$.

It can be shown\cite{Anne_PRL} that all the $H_i$, and also $\sum_i\mathbf{S}_i$, annihilate the state \eqref{wf}. Any linear combination of the $H_i$ and $\sum_{i,j}\mathbf{S}_i\cdot \mathbf{S}_j$ with nonnegative coefficients is hence a parent Hamiltonian for \eqref{wf}. In this paper, we will use
\begin{equation}\label{hamiltonian}
H=\frac{1}{4}\sum_i H_i
\end{equation}
as our Hamiltonian, unless specified otherwise.

\section{The 1D HS model}\label{SEC:1Dmodel}

The standard 1D HS model \eqref{HamHS} is obtained as a special case of the 2D HS model by choosing the Hamiltonian as\cite{jstat}
\begin{equation}
H_\textrm{HS} = \frac{\pi^2}{2N^2}\sum_i H_i + \frac{\pi^2(N+1)}{3N^2}\sum_{i,j}\mathbf{S}_i\cdot \mathbf{S}_j-\frac{\pi^2(N^2+5)}{12N}
\end{equation}
and putting $z_j=\exp(2\pi i j/N)$ and $w_{ij}=2z_i/(z_i-z_j)-1$. Note that the three-body term in $H_i$ vanishes in this case, since $\bar{w}_{ij}=-w_{ij}$. The ground state is again given by \eqref{wf} with $z_j=\exp(2\pi i j/N)$. The standard 1D HS model is a critical model belonging to the SU(2)$_1$ Wess-Zumino-Witten universality class.\cite{FSSbook} For later comparison, we will now discuss a few important properties of this model.

We first consider the spin-spin interaction strength $b_{ij}^\textrm{HS}$, which is defined such that $H_\textrm{HS}=\sum_{i\neq j}b_{ij}^\textrm{HS} \mathbf{S}_i\cdot \mathbf{S}_j+C_\textrm{HS}$, where $C_\textrm{HS}$ is a constant. Hence
\begin{equation}
b^\textrm{HS}_{ij}=\left[\frac{N}{\pi} \sin\left(\frac{i-j}{N}\pi\right)\right]^{-2}=d_{ij}^{-2}.
\end{equation}
Here, $d_{ij}$ is the chord distance between spins $i$ and $j$, when the spins are put on a circle with circumference $N$. The spin-spin interaction hence decays as the inverse of the square of the chord distance between the spins. For spins that are nearby each other ($|i-j|\ll N$), the chord distance is approximately the same as the distance along the circle, and the expression simplifies to
\begin{equation}\label{bapp}
b^\textrm{HS}_{ij}\approx(i-j)^{-2}, \qquad (|i-j|\ll N).
\end{equation}

We next consider the spin-spin correlation function
\begin{equation}\label{corr_def}
\langle S_j^z S_k^z \rangle = \frac{\sum_{s_1,\ldots,s_N}s_js_k| \psi_{s_1,\ldots,s_N}(z_1,\ldots,z_N) |^2}{4\sum_{s_1,\ldots,s_N}| \psi_{s_1,\ldots,s_N}(z_1,\ldots,z_N) |^2},
\end{equation}
which is the expectation value of a product of two spin operators $S_j^z$ acting on different lattice sites. In the standard 1D HS model, the spin-spin correlation function can be computed analytically.\cite{jstat} The analytical expression for the correlation function simplifies to
\begin{equation}\label{1D HS corr}
\langle S_{j+k}^z S_j^z \rangle\approx \frac{\pi(-1)^k}{8N\sin(\pi k/N)}-\frac{1}{4N^2\sin^2(\pi k/N)}
\end{equation}
in the limit $k\gg 1$ and $N \gg 1$ with $k/N$ fixed. It follows that the correlation function shows critical behavior with the power law decay $(-1)^k/(8k)$ for $1\ll k \ll N$. This is consistent with Haldane's conjecture.\cite{1DHS}

Finally, we consider the Renyi entropy of order two, which is defined as follows. We divide the system into two parts $A$ and $B$. In our case, $A$ is the first $x$ spins, and $B$ is the remaining $N-x$ spins. The Renyi entanglement entropy gives the entanglement of one part with the other. Now, construct the density matrix $\rho=|\psi\rangle\langle\psi|$ for the whole system and evaluate the reduced density matrix of part $A$ as $\rho_A = \mathrm{Tr}_B(\rho)$. Here, $\mathrm{Tr}_B(\rho)$ is the trace of $\rho$ over the spins in part $B$. The Renyi entanglement entropy of order two is then given by
\begin{equation}
S_x = -\ln[\mathrm{Tr}(\rho_A^2)].
\end{equation}
The reason for considering this entanglement entropy is that it can be computed efficiently using a metropolis Monte Carlo algorithm and the replica trick.\cite{ciracsierra,kallin}

The leading order behavior of the Renyi entropy of order $\alpha$ in a 1D critical system is generally given by\cite{RE,Ref_1,Ref_2,Ref_3,Ref_4, RE_Our}
\begin{equation}\label{Scritical}
S_x^{(\alpha)}\approx\frac{c}{6\eta}\left(1+\frac{1}{\alpha}\right) \ln\left[\eta N\sin(\pi x/N)/\pi\right] + \mathrm{constant},
\end{equation}
where $\alpha$ is the order of the Renyi entropy, $c$ is the central charge of the underlying conformal field theory, and $\eta=1$ ($\eta=2$) for periodic (open) boundary conditions. The expected leading order behavior for the standard 1D HS model is hence
\begin{equation}\label{1D HS RE}
S_x \approx \frac{c}{4}\ln\left[\frac{N}{\pi}\sin\left(\frac{\pi x}{N}\right)\right] + \mathrm{constant},
\end{equation}
which agrees with numerics\cite{ciracsierra} for $c=1$. The numerical results for the HS model also show an oscillation with period 2, which is present in the subleading terms.

\section{Two-body chain and ladder models}\label{SEC:twobody}

We now demonstrate that the Hamiltonian \eqref{hamiltonian} also reduces to a two-body Hamiltonian in other particular cases. Specifically, if we take all $w_{ij}$ to be real, the three-body terms in \eqref{Ham} vanish, and the Hamiltonian simplifies to
\begin{equation}
H=\frac{1}{4}\sum_{i}H_i = \sum_{i\neq j}b_{ij} \, \mathbf{S}_i\cdot\mathbf{S}_j + C,
\end{equation}
where $C$ is a constant and
\begin{equation}\label{b_ij}
b_{ij}=\frac{1}{6}w_{ij}^2+
\frac{1}{6}\sum_{k(\neq i\neq j)}w_{ki}w_{kj} \qquad (i\neq j)
\end{equation}
expresses the strength of the interaction between the spins at positions $i$ and $j$ (note that $b_{ij}=b_{ji}$).

Since $w_{ij}=g(z_i)/(z_i-z_j)+h(z_i)$, where $g$ and $h$ are arbitrary functions of $z_i$, we can achieve that $w_{ij}$ are real by choosing all $z_i$ real and taking $g$ and $h$ to be real functions. Requiring $z_i$ to be real corresponds to restricting $\mathrm{Im}(W_i)$ to be an integer times $\pi$. In other words, all the lattice points should be on the blue lines on the lower most cylinder in Fig.\ \ref{FIG-1}.

Lattice points on the blue lines can be expressed in the form
\begin{equation}\label{zj}
z_j=\sigma_j e^{\Lambda f(j)},
\end{equation}
where $\sigma_j\in\{-1,+1\}$, $\Lambda$ is a positive number, and $f(j)\in\mathbb{R}$ is a real valued function of $j$. If we take all $\sigma_j$ to be $+1$, we get a 1D chain model, and if we take some $\sigma_j$ to be $+1$ and some to be $-1$, we get a ladder model. Note that these models have open boundary conditions by construction. In the following, we shall refer to the spins with positive (negative) $\sigma_j$ as the spins on the front (back) of the cylinder. The circumference of the cylinder is fixed to $2\pi$, and changing $\Lambda$ corresponds to a scale transformation in the direction parallel to the cylinder axis. If $\Lambda$ is very small (large), the circumference of the cylinder will be large (small) compared to the other length scales in the system.

\subsection{Symmetries}

The wavefunction \eqref{wf} can be written as a conformal block times a normalization constant, and it is therefore invariant under all global conformal transformations, i.e.\ transformations of the type
\begin{equation}
z_j\to \frac{az_j+b}{cz_j+d},
\end{equation}
where $a$, $b$, $c$, and $d$ are complex numbers fulfilling $ad-bc=1$. If we do the same transformation on the Hamiltonian, we still have the same expression for the Hamiltonian, but $g$ and $h$ are, in general, modified. The Hamiltonian is hence not invariant under the full set of conformal transformations (unless $g=h=0$), but for particular choices of $g$ and $h$, the Hamiltonian is invariant under a smaller set of transformations.\cite{glasser}

A particularly natural choice of Hamiltonian for the models we are looking at is to take $w_{ij} = 2z_i/(z_i-z_j)-1= (z_i+z_j)/(z_i-z_j)$. In that case, the Hamiltonian is invariant under the transformations $z_j\to az_j$ and $z_j\to z_j^{-1}$, where $a$ is a constant number. These two transformations correspond, respectively, to displacing the lattice points along the blue lines in Fig.\ \ref{FIG-1} (plus a rotation around the cylinder axis if $a$ is complex) and to inverting the directions of the blue lines.

\subsection{Spin-spin correlations and Renyi entropy}

As part of our investigations of the properties of the models, we shall below compute the spin-spin correlation function and the Renyi entropy of order two for particular cases. It was found in \onlinecite{jstat} that if the spins are put on a circle around the cylinder and $w_{ij}=(z_i+z_j)/(z_i-z_j)$, then the spin-spin correlations fulfil the following set of linear
equations
\begin{equation}\label{cor}
w_{ij}\langle S^z_iS^z_j\rangle + \sum_{k(\neq i \neq j)}w_{ik} \langle S^z_kS^z_j\rangle +\frac{1}{4} w_{ij} = 0.
\end{equation}
Following the same steps as in \onlinecite{jstat}, we find that \eqref{cor} also applies whenever all the $w_{ij}=g(z_i)/(z_i-z_j)+h(z_i)$ are real (as long as $w_{ij}$ are real, we can choose $g(z_i)$ and $h(z_i)$ after convenience). This allows us to easily compute the spin-spin correlations for quite large systems. We compute the Renyi entropy using Monte Carlo simulations.

\subsection{Small $\Lambda$ limit: Decoupling of the legs} \label{SEC:decoupling}

When $\Lambda$ is sufficiently small, the circumference of the cylinder is large compared to all other relevant length scales in the system, and it would be natural if the two legs decouple in that limit. Specifically, we shall assume that
\begin{equation}\label{limit}
\Lambda |f(j)-f(k)|\ll 1
\end{equation}
for all $j$ and $k$. In this section, we shall label the $N_+$ spins with $\sigma_j>0$ from $1$ to $N_+$ and the $N_-$ spins with $\sigma_j<0$ from $N_++1$ to $N=N_++N_-$.

Let us first look at the Hamiltonian for $w_{ij}=(z_i+z_j)/(z_i-z_j)$. Using \eqref{limit}, we get
\begin{align}
w_{ij}&=\frac{\sigma_i e^{\Lambda f(i)}+\sigma_j e^{\Lambda f(j)}}{\sigma_i e^{\Lambda f(i)}-\sigma_j e^{\Lambda f(j)}}\nonumber\\
&\approx \left\{\begin{array}{ll}
2/\{\Lambda[f(i)-f(j)]\} &\textrm{for } \sigma_i=\sigma_j\\
\Lambda[f(i)-f(j)]/2  &\textrm{for } \sigma_i=-\sigma_j
\end{array}\right..
\end{align}
In other words, $w_{ij}$ is large if the $i$th and the $j$th spin sit on the same leg and small if they sit on different legs. Inserting this into \eqref{b_ij}, we observe that the spin-spin interaction strength between spins sitting on the same leg is larger by a factor of $\Lambda^{-2}$ compared to the spin-spin interaction strength between spins sitting on different legs. When \eqref{limit} applies, we can hence neglect the interactions between spins on different legs.

In appendix \ref{appB}, we show that if there is an even number of spins on both of the legs, then the wavefunction \eqref{wf} reduces to
\begin{multline}\label{wfeven}
\psi_{s_1,\ldots,s_N}(z_1,\ldots,z_N)\approx \\
\textrm{constant}\times\psi_{s_1,\ldots,s_{N_+}}(z_1,\ldots,z_{N_+})\\
\times \psi_{s_{N_++1},\ldots,s_N}(z_{N_++1},\ldots,z_N),
\end{multline}
when \eqref{limit} applies. In other words, the ladder model reduces to two copies of the chain model with $N_+$ and $N_-$ spins, respectively. If there is an odd number of spins on each of the legs, the wavefunction is a sum of two terms
\begin{multline}\label{wfodd}
\psi_{s_1,\ldots,s_N}(z_1,\ldots,z_N)\approx \textrm{constant}\\
\times[\psi^{(1)}_{s_1,\ldots,s_{N_+}}(z_1,\ldots,z_{N_+})
\psi^{(-1)}_{s_{N_++1},\ldots,s_N}(z_{N_++1},\ldots,z_N)\\
-\psi^{(-1)}_{s_1,\ldots,s_{N_+}}(z_1,\ldots,z_{N_+})
\psi^{(1)}_{s_{N_++1},\ldots,s_N}(z_{N_++1},\ldots,z_N)].
\end{multline}
Here, $\psi^{(1)}$ ($\psi^{(-1)}$) is defined as in \eqref{wf}, except that we now take $\delta_\mathbf{s}$ to be one if the sum of the spin variables $s_j$ is $1$ ($-1$). Note, however, that since we found above that the Hamiltonian with $w_{ij}=(z_i+z_j)/(z_i-z_j)$ does not couple the two legs, each of these terms are individually zero energy eigenstates of the Hamiltonian. The ground state is hence degenerate in that case.

\subsection{Large $\Lambda$ limit: Product of singlets} \label{SEC:singlet}

In appendices \ref{appC} and \ref{appD}, we show that the wavefunction \eqref{wf} reduces to a product of $N/2$ singlets in the limit of sufficiently large $\Lambda$ for almost all choices of the lattice coordinates \eqref{zj}. Without loss of generality, we will here label the lattice sites such that $f(j+1)\geq f(j)$ for all $j\in\{1,2,\ldots,N-1\}$. Stated more precisely, we find that
\begin{multline}
\psi_{s_1,s_2,\ldots,s_N}(z_1,z_2,\ldots,z_N)\propto\\
\psi_s(s_1,s_2)\otimes \psi_s(s_3,s_4) \otimes \cdots \otimes \psi_s(s_{N-1},s_N)
\end{multline}
when
\begin{equation}\label{condition}
\exp\{\Lambda [f(2j+1)-f(2j)]\}\gg 1
\end{equation}
for all $j\in\{1,2,\ldots,N/2-1\}$. Here,
\begin{equation}
\psi_s(s_{2j-1},s_{2j})=(|+1,-1\rangle-|-1,+1\rangle)/\sqrt{2}
\end{equation}
is the singlet wavefunction of the spins $s_{2j-1}$ and $s_{2j}$. Note that \eqref{condition} can only be fulfilled provided $f(2j+1)>f(2j)$ for all $j\in\{1,2,\ldots,N/2-1\}$. The wavefunction hence reduces to a product of singlets for sufficiently large $\Lambda$ unless there is a $j$ for which lattice site number $2j+1$ and lattice site number $2j$ are placed on opposite sides of the cylinder. The same result applies also in the case, where the $\sigma_k$ in \eqref{zj} are general phase factors. Finally, we comment that the pattern of singlets in the state is fixed, because the model has open boundary conditions per construction. It is hence always the first spin that forms a singlet with the second, the third spin that forms a singlet with the fourth, and so on.

\section{Uniform 1D spin chain}\label{SEC:chain}

\begin{figure}
\includegraphics[width=0.8\columnwidth]{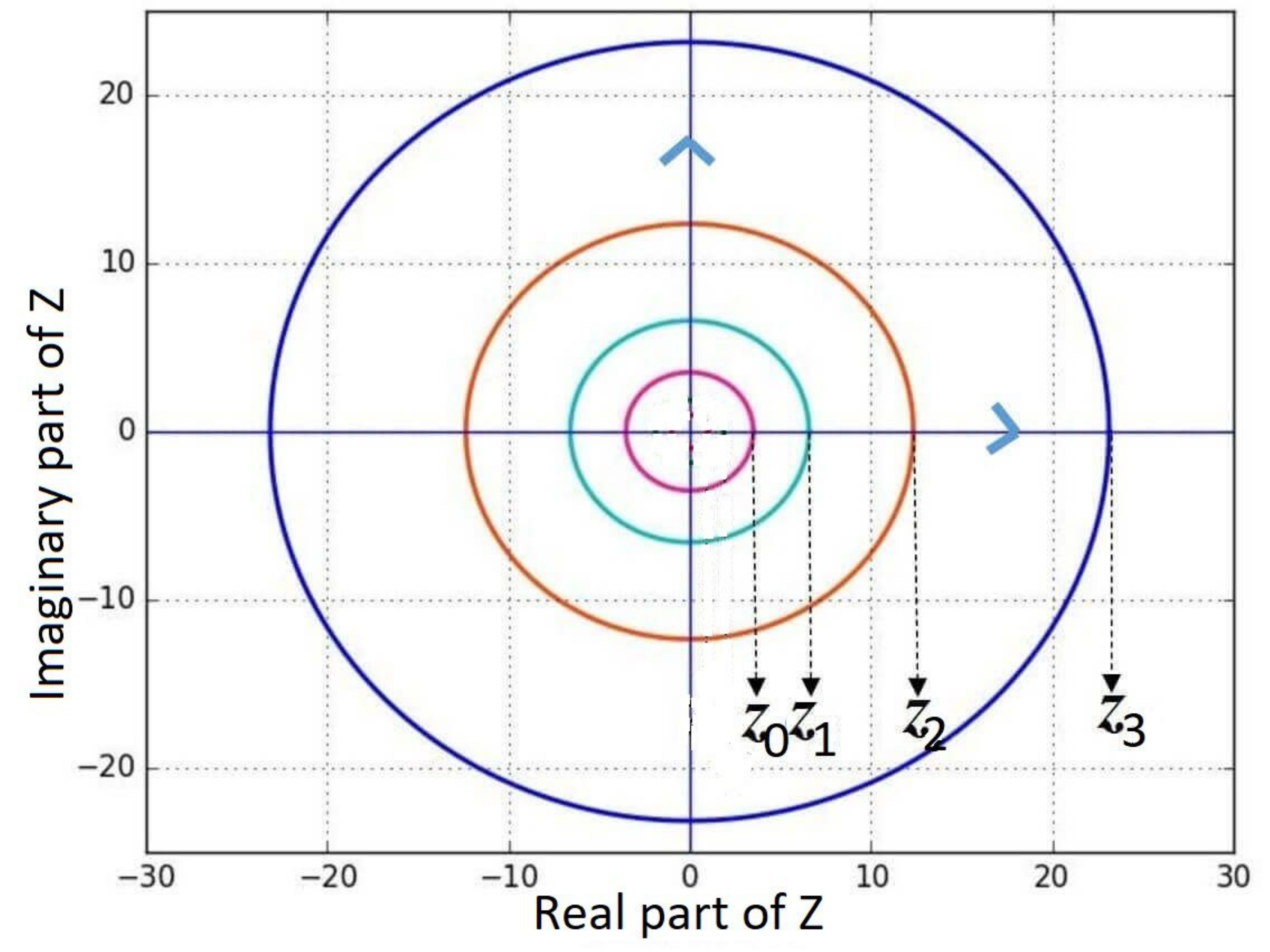}\vfill
\includegraphics[width=0.8\columnwidth]{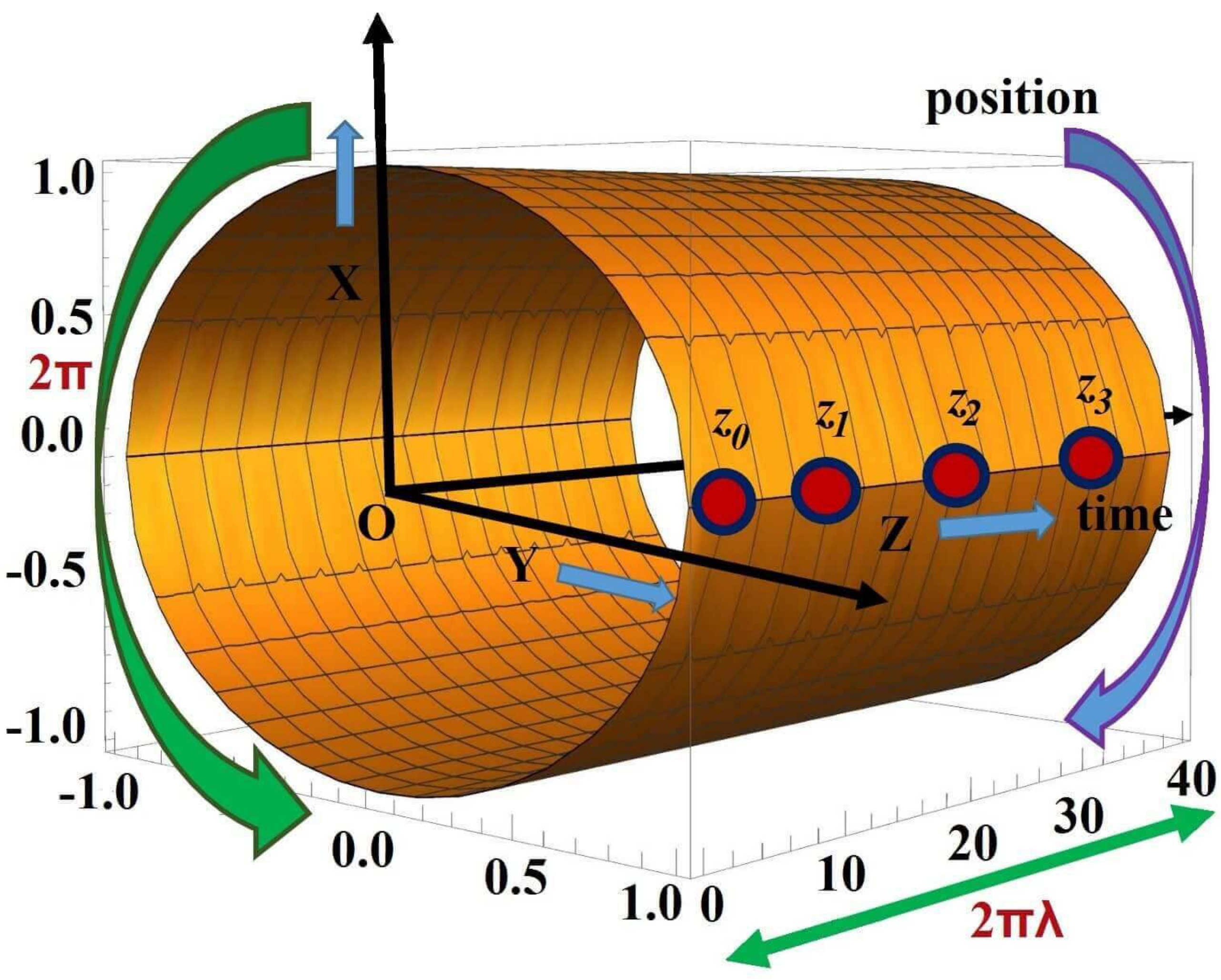}\vfill
\caption{Mapping of the spin positions from the complex plane (upper plot) to the cylinder surface (lower plot) for the uniform 1D chain. The radii of the consecutive circles in the plane are $z_j = \exp(2\pi \lambda j/N)$.}\label{fig-4}
\end{figure}

In this section, we study the 1D model obtained by choosing $z_j=\exp({2\pi\lambda j/N})$ in more detail. Here, $N$ is the number of sites in the chain, which must be even, and we shall take $j\in\{0,1,\ldots,N-1\}$. The positive number $\lambda$ controls the ratio between the total length of the chain, which is $2\pi\lambda$, and the circumference of the cylinder, which is $2\pi$. Note that in this case a possible choice of $\omega_{jk}$ is
\begin{equation}\label{wchain}
w_{jk}=\frac{z_j+z_k}{z_j-z_k}=\frac{1}{\tanh[\pi\lambda(j-k)/N]}.
\end{equation}
Figure \ref{fig-4} shows the lattice both in the complex plane and on the cylinder. In the following, we first study the physics of the ground state by computing the spin-spin correlations and the Renyi entropy. We then investigate the spin-spin interaction strengths in the Hamiltonian. Finally, we briefly discuss possibilities to construct models with an odd number of spins.

\subsection{Spin-spin correlations}

\begin{figure*}
\includegraphics[width=0.325\textwidth]{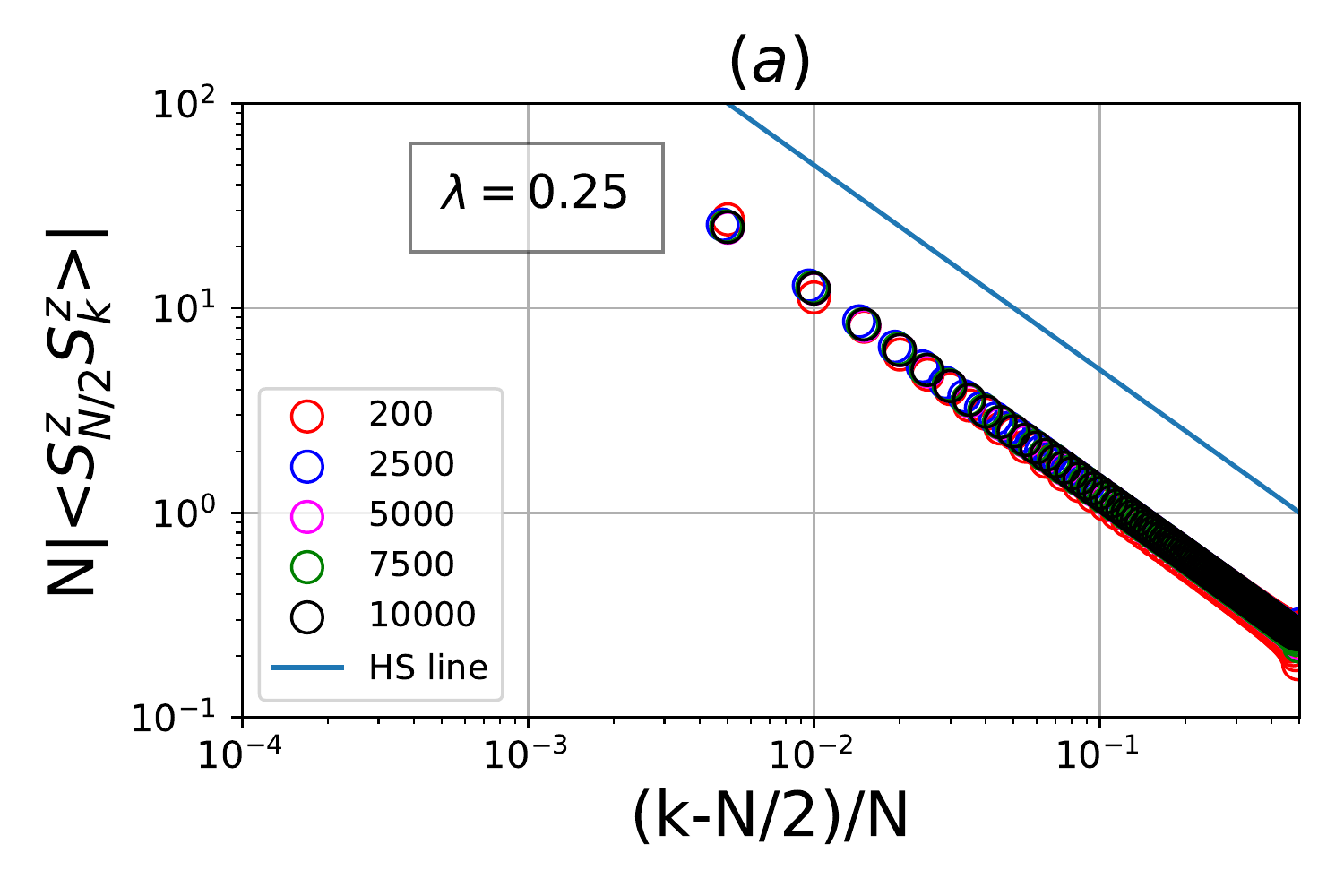}\hfill
\includegraphics[width=0.325\textwidth]{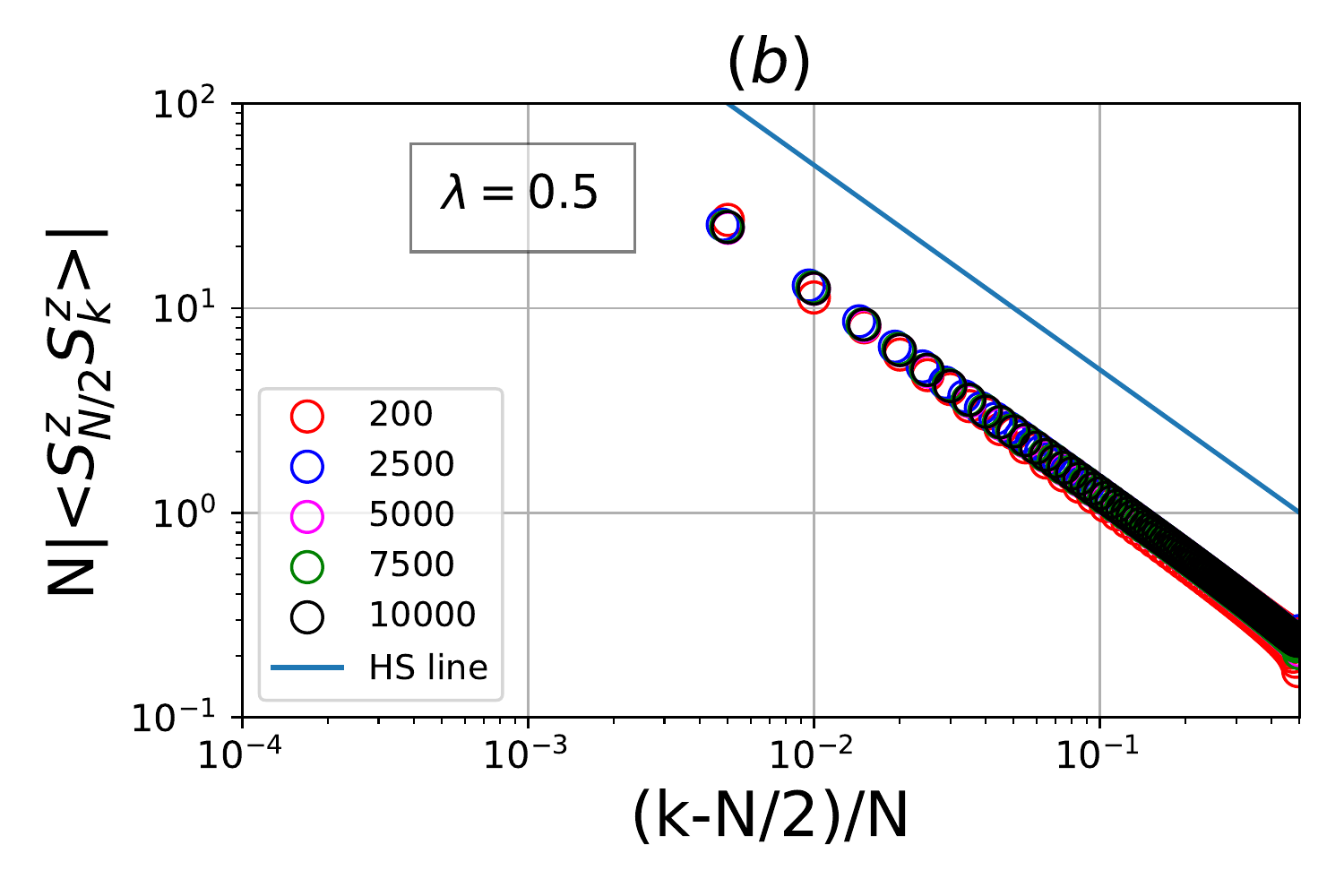}\hfill
\includegraphics[width=0.325\textwidth]{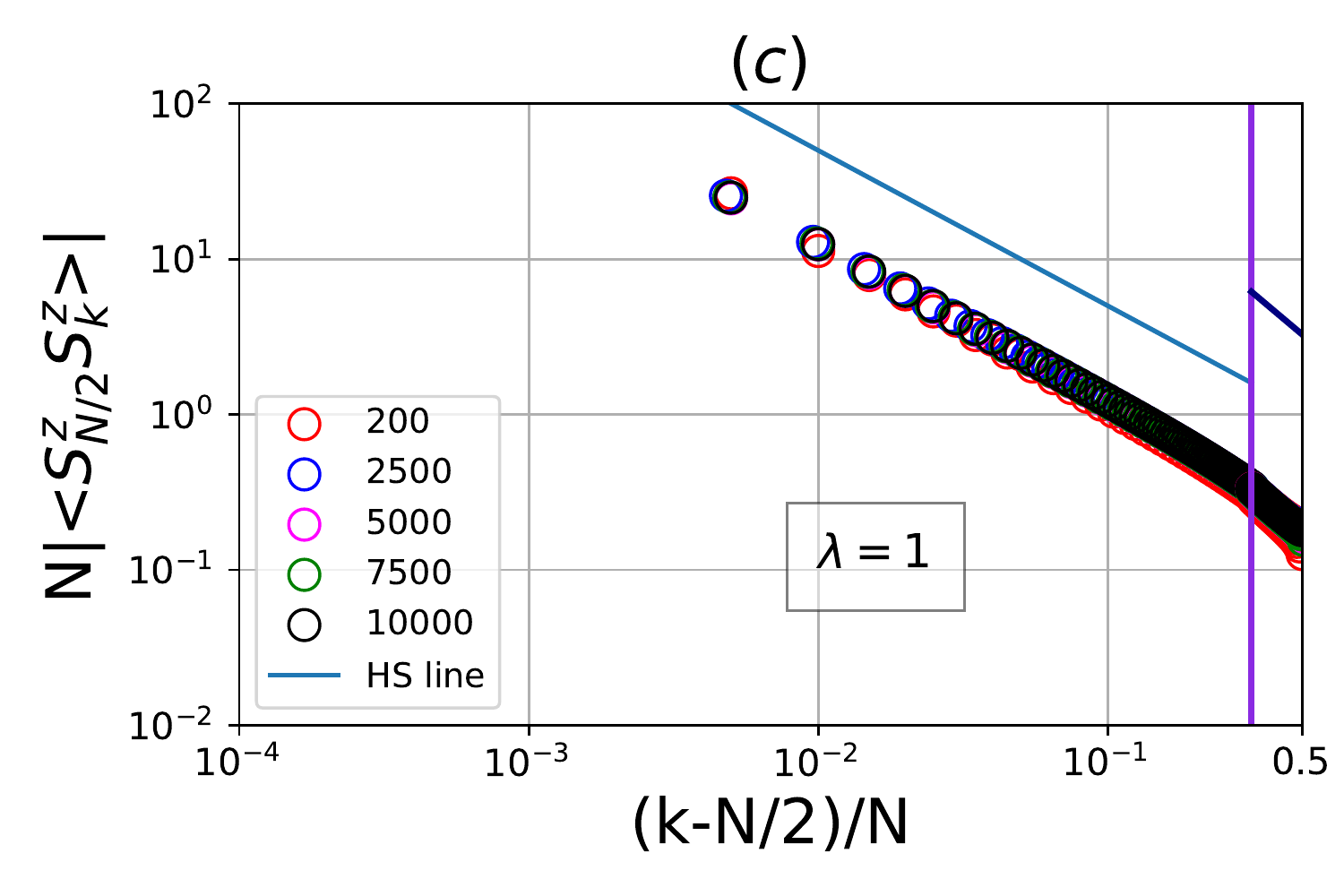}
\includegraphics[width=0.325\textwidth]{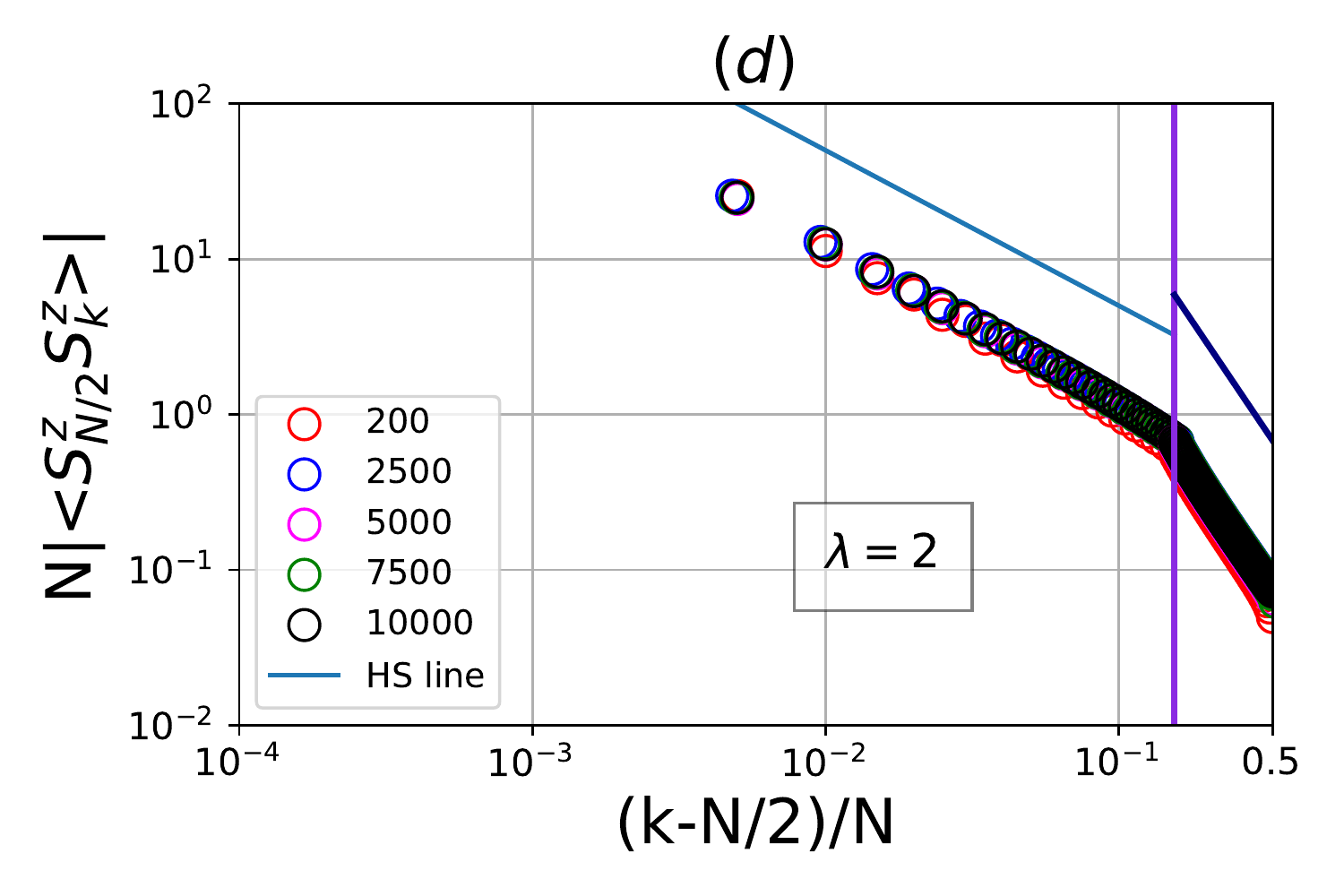}\hfill
\includegraphics[width=0.325\textwidth]{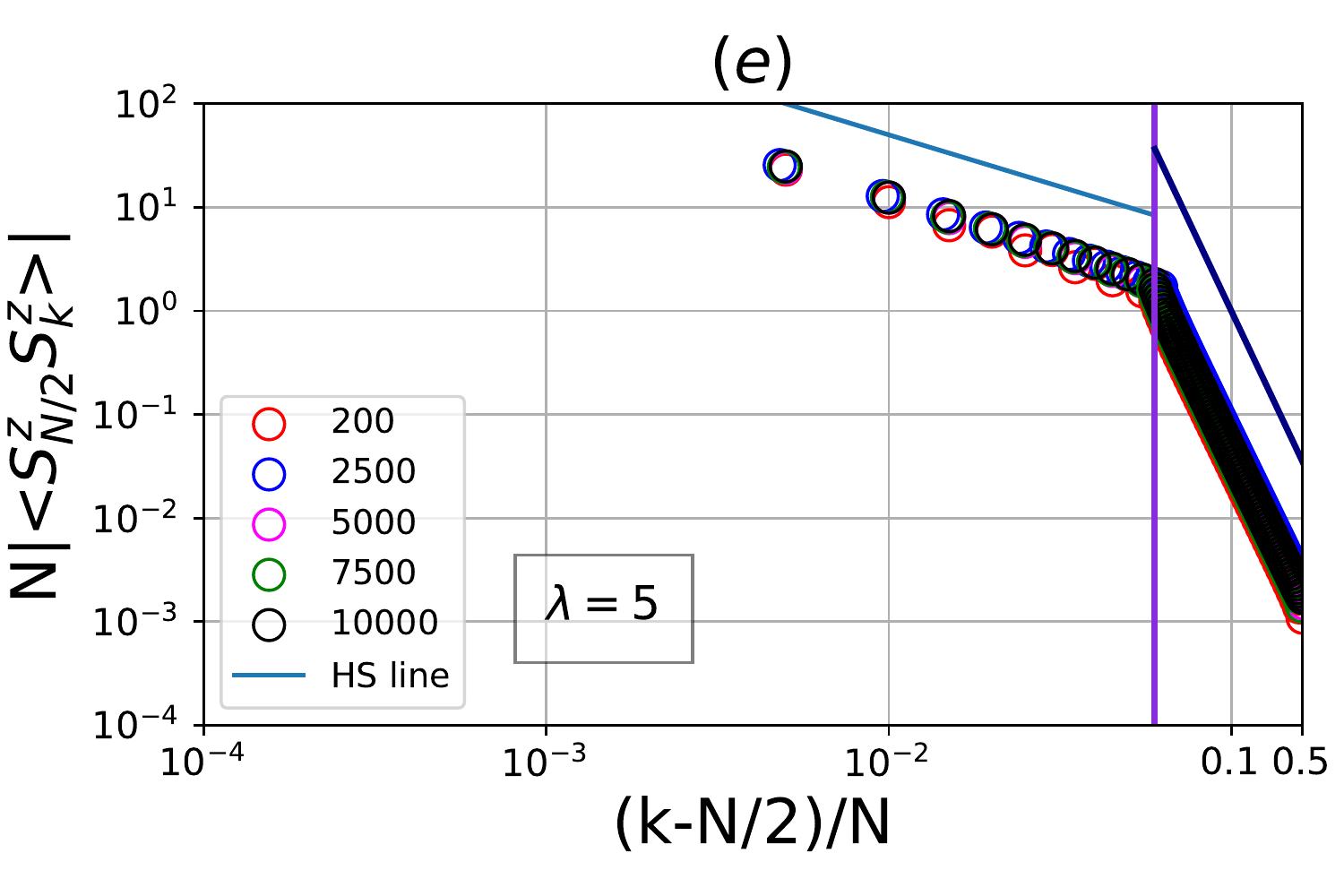}\hfill
\includegraphics[width=0.325\textwidth]{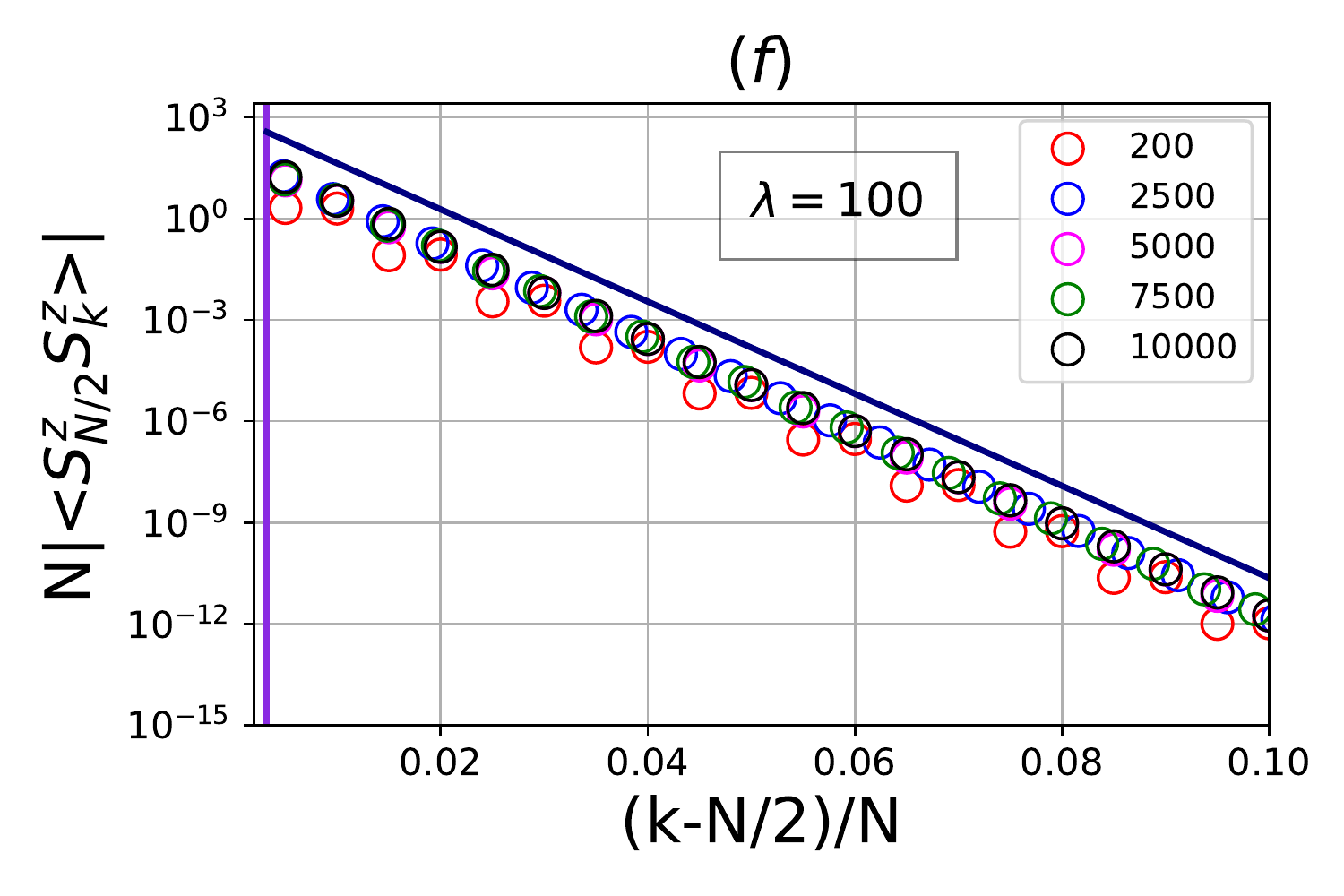}
\caption{Absolute value of the spin-spin correlation $\langle S^z_j S^z_k \rangle$ for the uniform 1D chain as a function of $(k-j)/N$ for $j=N/2$ (bulk spin) and $k\in\{N/2+1,N/2+2,\ldots,N-1\}$ (for clarity we plot only some of these $k$ values). The different plots are for different values of $\lambda$, and there are $N=200$ (red), $N=2500$ (blue), $N=5000$ (magenta), $N=7500$ (green), or $N=10000$ (black) spins in the chain. Note that in (c-f) the $x$-axis is in log scale to the left of the vertical line and in linear scale to the right of the vertical line. For $\lambda=0.25$, the correlations are seen to follow a power law, and for $\lambda=100$, the correlations decay exponentially. For intermediate values of $\lambda$, the correlations decay as a power law for short distances and exponentially for large distances, and the transition is seen to occur approximately at the vertical line, which is positioned at $(k-N/2)/N=1/(\pi\lambda)$. In the standard 1D HS model the correlations decay as the inverse of the distance, and in the region, where the $x$-axis is in log scale, we plot a straight line with slope $-1$ for comparison. The straight line plotted in the region to the right of the vertical line is proportional to $\exp(-\pi\lambda (k-N/2)/N)$.} \label{g-5-to-6_middle}
\end{figure*}

The spin-spin correlations are, in general, an important tool to extract information about the physics of a system. The typical situation is that the ground state is either critical with correlations that decay as a power law or noncritical with correlations that decay exponentially. We now take a look at the spin-spin correlations \eqref{corr_def} for the uniform 1D model by solving \eqref{cor} numerically. We find that $\langle S^z_j S^z_k \rangle$ is positive for $|j-k|$ even and negative for $|j-k|$ odd. To simplify the plots, we hence only consider the absolute value of the correlations in the following.

\begin{figure*}
\includegraphics[width=0.325\textwidth]{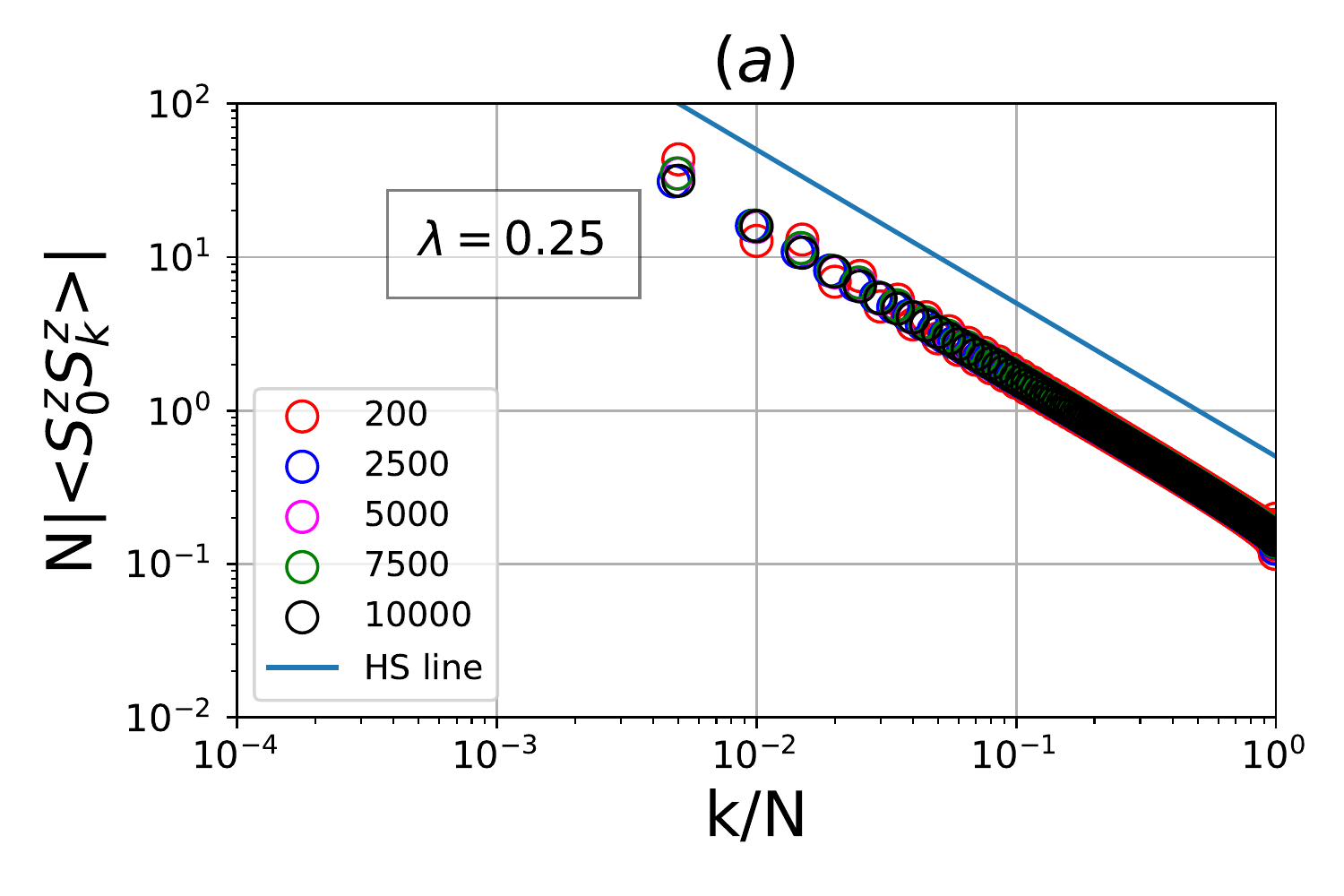}\hfill
\includegraphics[width=0.325\textwidth]{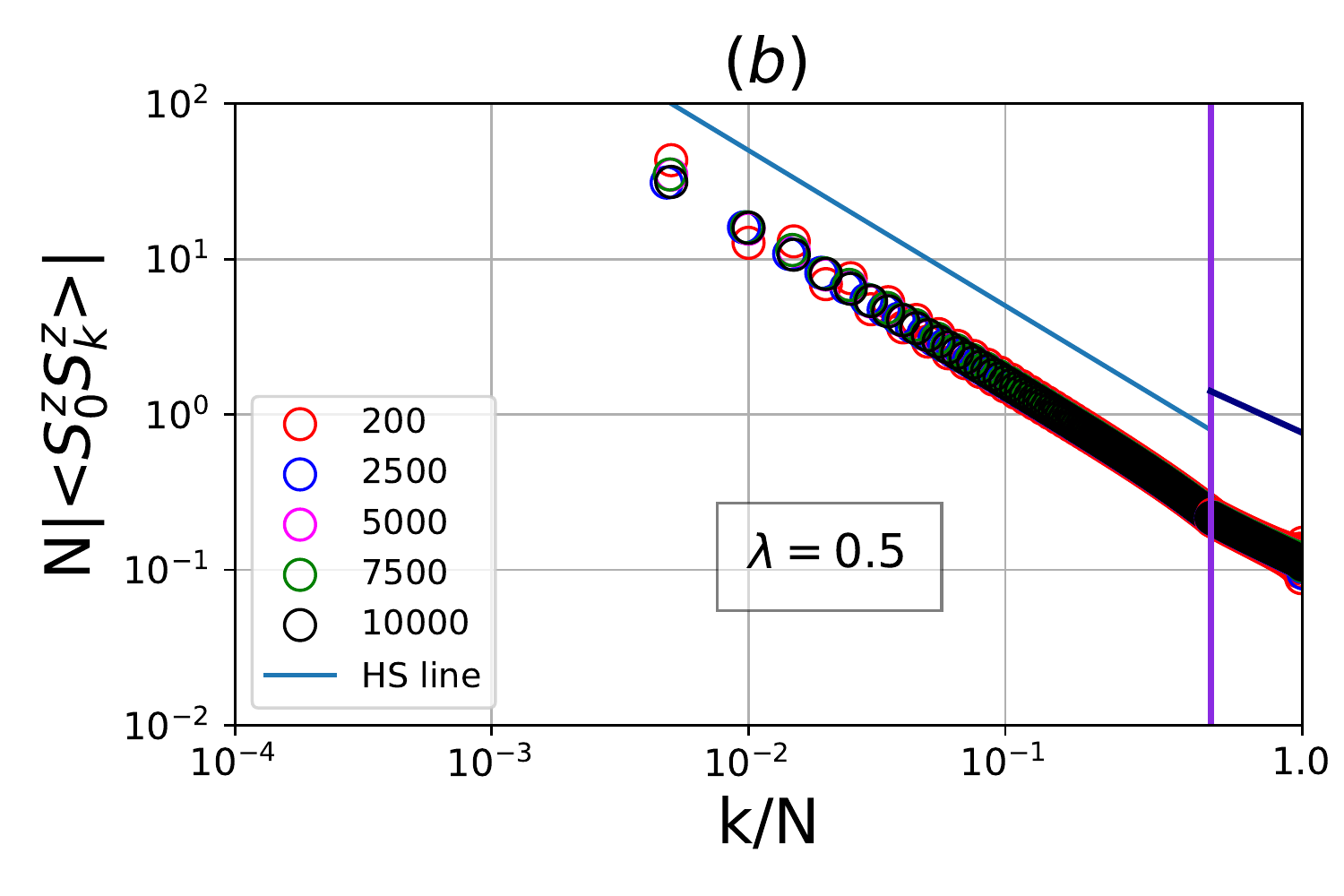}\hfill
\includegraphics[width=0.325\textwidth]{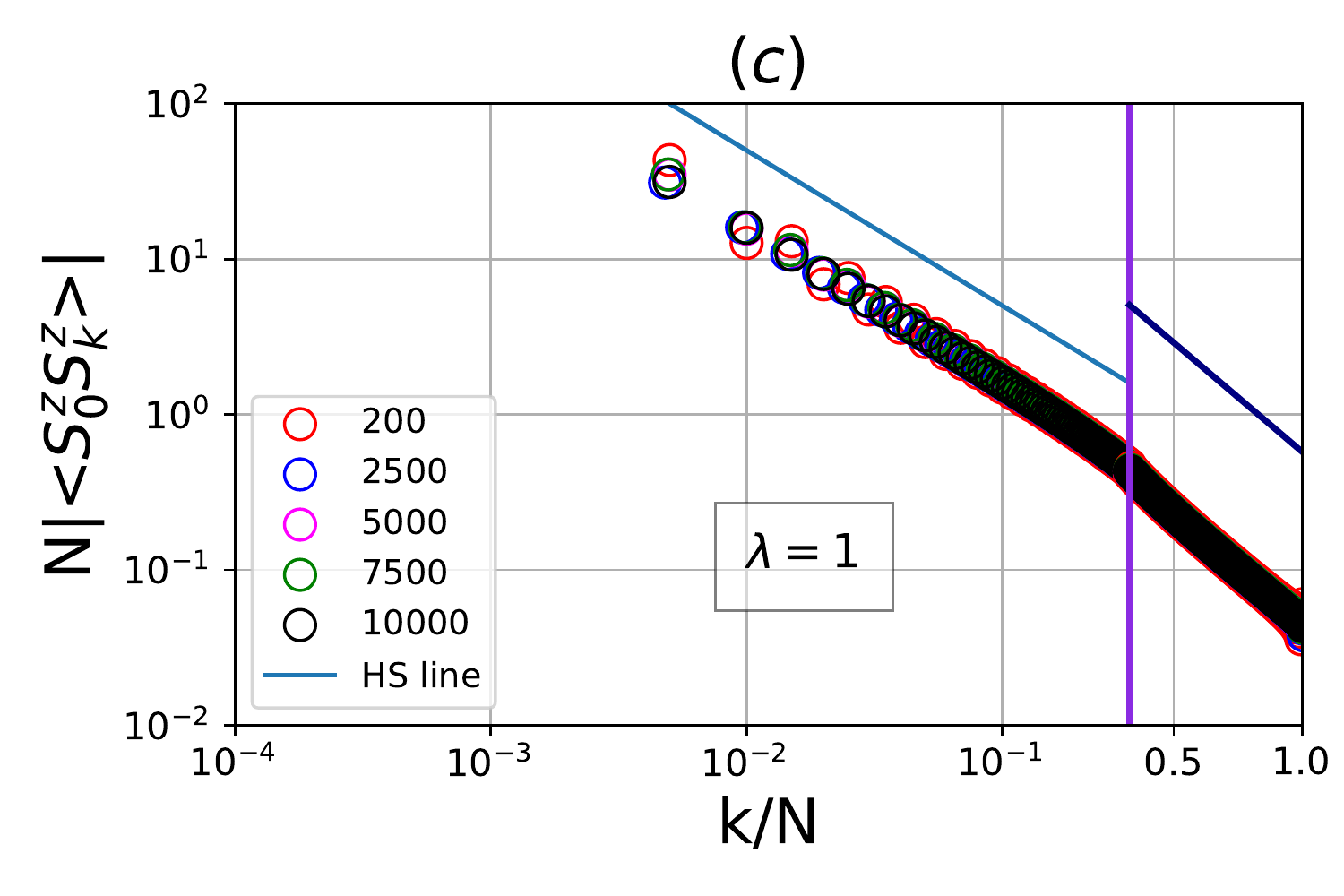}
\includegraphics[width=0.325\textwidth]{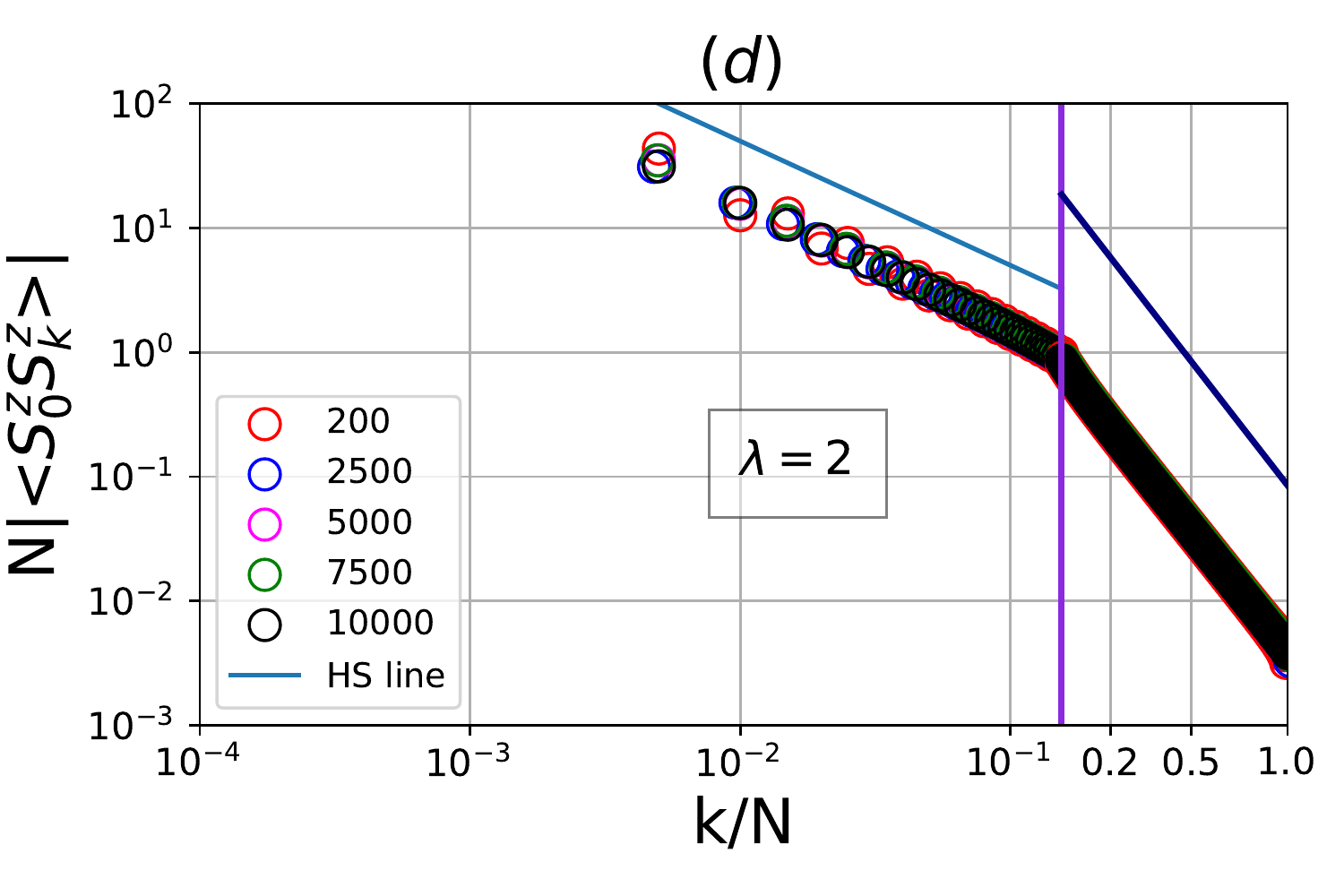}\hfill
\includegraphics[width=0.325\textwidth]{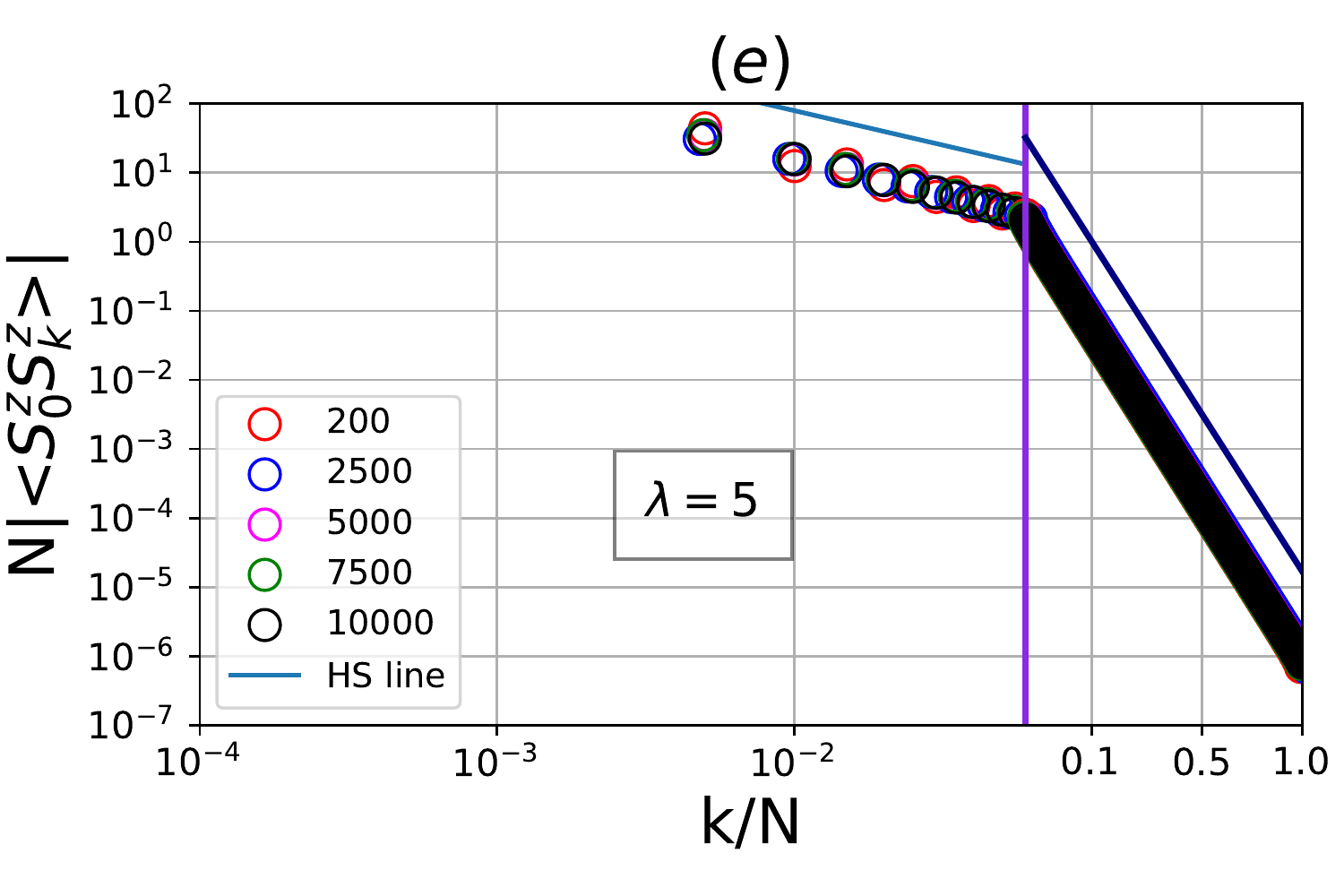}\hfill
\includegraphics[width=0.325\textwidth]{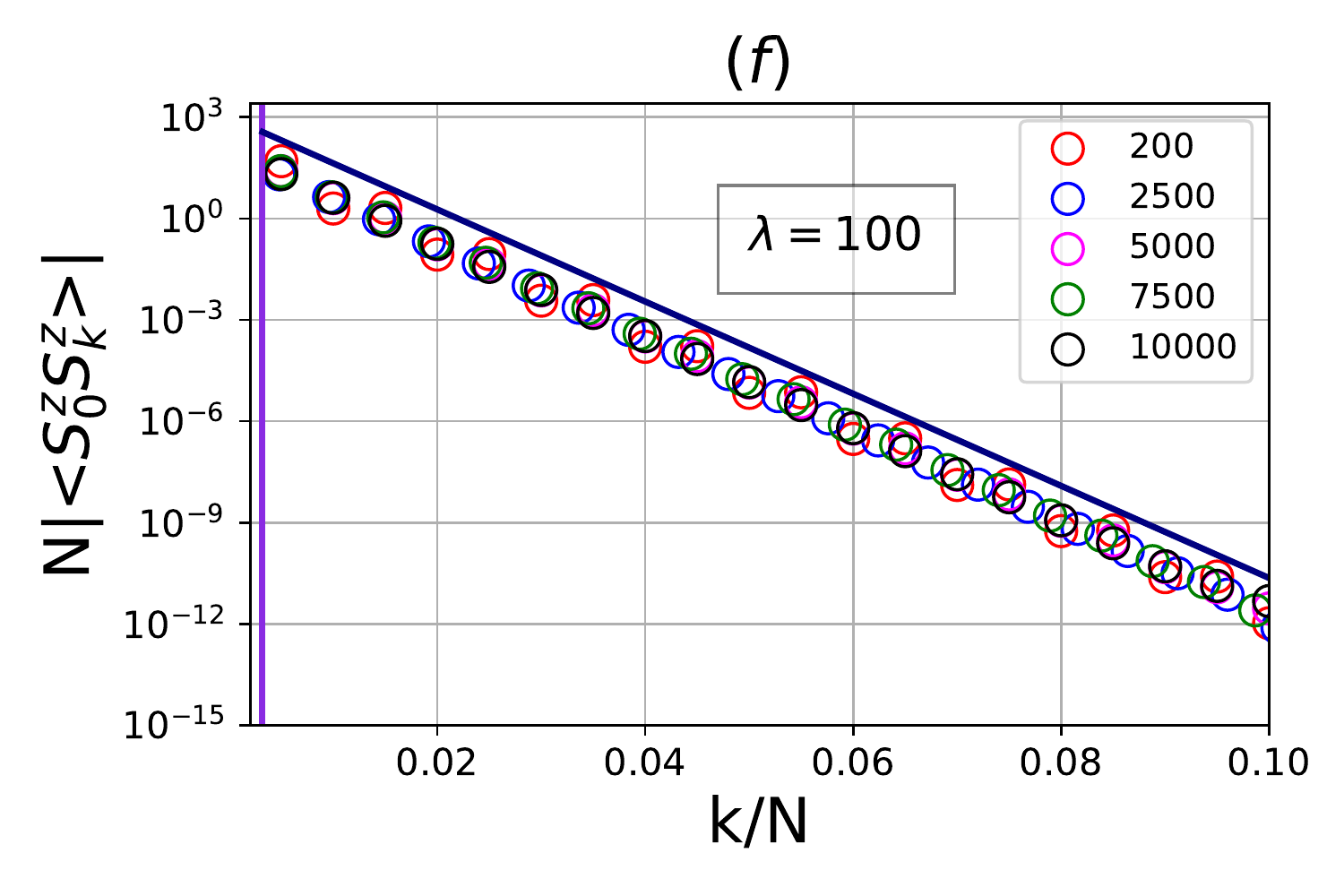}
\caption{Absolute value of the spin-spin correlation $\langle S^z_j S^z_k \rangle$ for the uniform 1D chain as a function of $(k-j)/N$ for $j=0$ (edge spin) and $k\in\{1,2,\ldots,N-1\}$ (for clarity we plot only some of these $k$ values). The different plots are for different values of $\lambda$, and there are $N=200$ (red), $N=2500$ (blue), $N=5000$ (magenta), $N=7500$ (green), or $N=10000$ (black) spins in the chain. Note that in (c-f) the $x$-axis is in log scale to the left of the vertical line and in linear scale to the right of the vertical line. For $\lambda=0.25$, the correlations are seen to follow a power law, and for $\lambda=100$, the correlations decay exponentially. For intermediate values of $\lambda$, the correlations decay as a power law for short distances and exponentially for large distances, and the transition is seen to occur approximately at the vertical line, which is positioned at $k/N=1/(\pi\lambda)$. In the standard 1D HS model the correlations decay as the inverse of the distance, and in the region, where the $x$-axis is in log scale, we plot a straight line with slope $-1$ for comparison. The straight line plotted in the region to the right of the vertical line is proportional to $\exp(-\pi\lambda k/N)$.} \label{g-5-to-6_edge}
\end{figure*}

Figures \ref{g-5-to-6_middle} and \ref{g-5-to-6_edge} show the spin-spin correlations $\langle S^z_j S^z_k \rangle$ for a spin in the bulk of the chain and for a spin on the edge, respectively. For the bulk spin, we fix $j=N/2$ and plot the correlations as a function of $k$ for $k> j$, and for the edge spin, we fix $j=0$ and plot the correlations as a function of $k$. These figures show several interesting features, as we now discuss.

For $\lambda=0.25$, we observe that the correlations decay as the power law $|\langle S^z_j S^z_k \rangle| \propto |j-k|^{-1}$. Here, $|j-k|$ is proportional to the distance between the spins. This is the same behavior as for the standard 1D HS model, where the correlations also decay as the inverse of the distance between the spins when $|j-k|$ is large compared to 1 and small compared to $N$ (see the discussion below Eq.\ \eqref{1D HS corr}). In the opposite limit of large $\lambda$, we observe that the correlations decay exponentially. In this limit, the model is hence qualitatively different from the standard 1D HS model. This is expected, since we found in Sec.\ \ref{SEC:singlet} that the state reduces to a product of singlets in the large $\lambda$ limit.

Given the qualitatively different behavior for small and large $\lambda$, the natural next question is how the transition from one behavior to the other occurs. The figures show that the transition happens gradually in the sense that for intermediate $\lambda$, the correlations decay as a power law for short distances and exponentially for large distances. As $\lambda$ increases, the range of distances for which there is exponential decay increases. A look at Eq.\ \eqref{wchain} suggests that the point
\begin{equation}\label{transition}
|j-k|/N=1/(\pi\lambda)
\end{equation}
plays a special role, and from the figures we observe that the transition from power law to exponential decay indeed occurs around this point. The power law decay at short distances again follows the behavior
\begin{equation}\label{corshort}
|\langle S^z_j S^z_k \rangle| \propto |j-k|^{-1},
\end{equation}
and at long distances the exponential decay is described by
\begin{equation}\label{corlong}
|\langle S^z_j S^z_k \rangle| \propto \frac{1}{N}\exp\left(\frac{\pi\lambda|j-k|}{N}\right).
\end{equation}
The curves in the figures are practically independent of the number of spins $N$, when $N$ is large enough, and this shows that the proportionality constants in \eqref{corshort} and \eqref{corlong} are independent of $N$. The independence of $N$ is also interesting because it shows that the possibility to have power law decay at short distances and exponential decay at long distances remains in the thermodynamic limit.

It is relevant to note that in the above discussion, short and long distances refer to $|j-k|/N$ taking a value close enough to zero and close enough to unity, respectively. The distances in question are hence measured relative to the length of the chain and do not refer to how many spins there are between the two considered spins. When $|j-k|/N$ is kept fixed, the number of spins between the considered spins grows linearly with $N$, when $N$ increases. We could instead consider the correlations between spins that are $|j-k|$ spins apart with $|j-k|$ of order unity. Since the transition from power law to exponential decay occurs around $|j-k|=N/(\pi\lambda)$, we are always on the left hand side of the transition, when $N$ is large enough. In other words, if we take the thermodynamic limit $N\to\infty$ with fixed $|j-k|$, the correlations decay as the inverse of the distance as in the HS model, independent of $\lambda$.

\begin{figure}
\includegraphics[width=0.8\columnwidth]{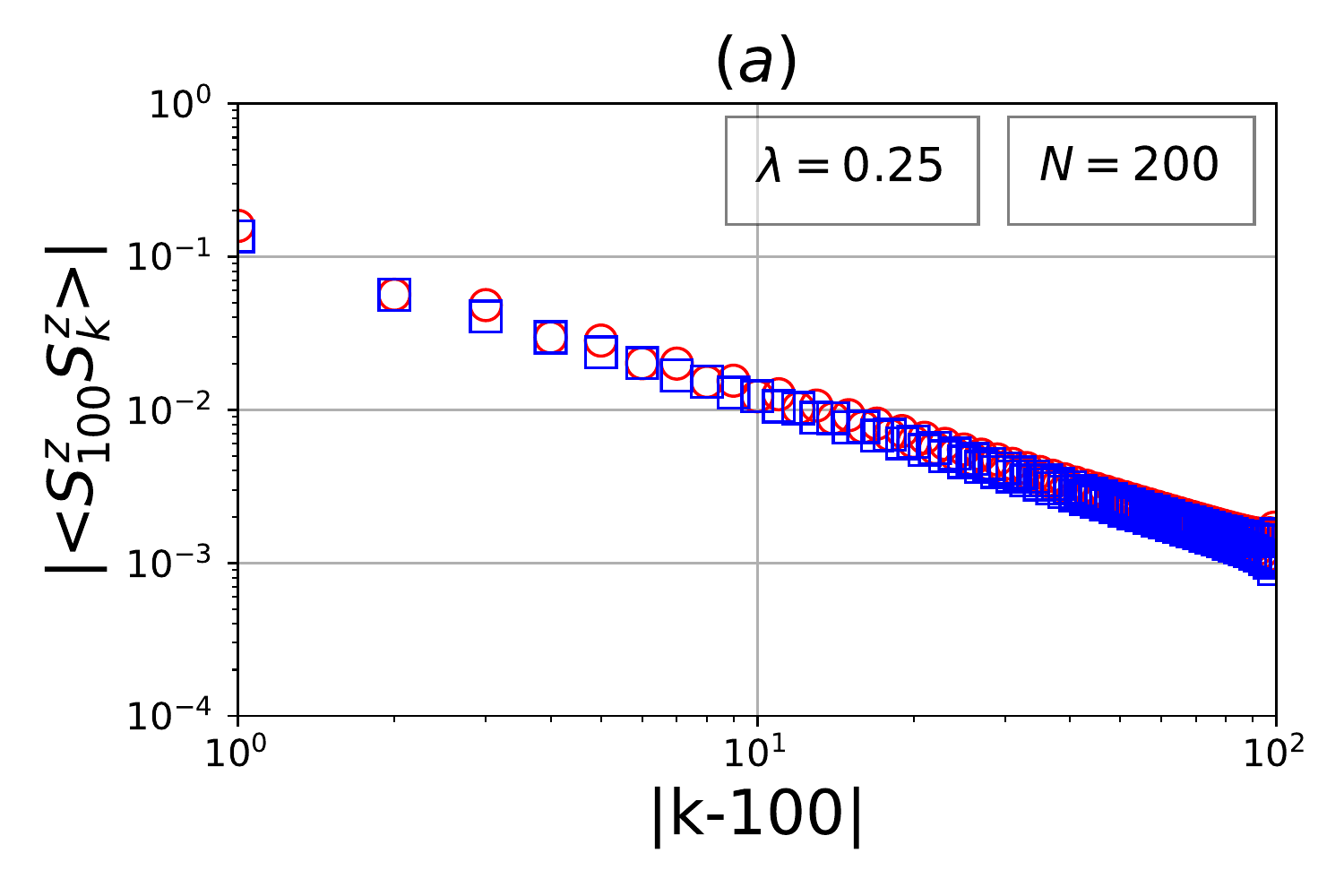}
\includegraphics[width=0.8\columnwidth]{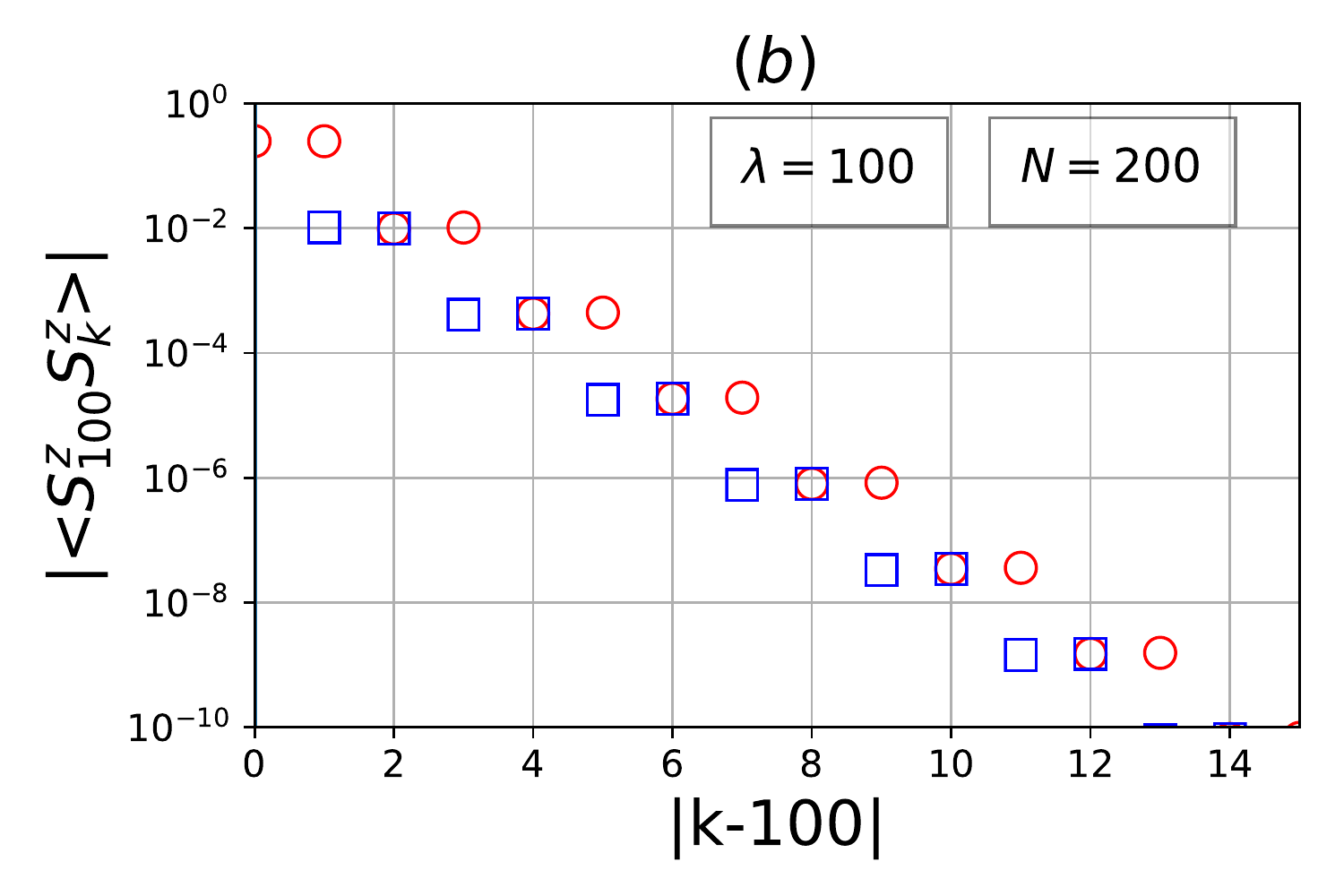}
\caption{Absolute value of the spin-spin correlation $\langle S^z_j S^z_k \rangle$ for the uniform 1D chain as a function of $|k-j|$ for $j=100$ (bulk spin) and $k\in\{0,1,\ldots,199\}$. We have plotted both halves of the spin chain. The red circles are for $k-j\geq 0$, and the blue squares are for $k-j<0$. Note that the spin with $k=100$ is more strongly correlated with the spin with $k=101$ than with the spin with $k=99$.} \label{g-5-to-6}
\end{figure}

We have only plotted the correlations for $k-j > 0$ in Fig.\ \ref{g-5-to-6_middle} for clarity. The conclusions regarding power law and exponential decay are the same for $k-j<0$. It is interesting to note, however, that there is not a perfect symmetry between the left and the right hand side of the chain, simply because the number of spins in the chain is even. This means that on one side of the bulk spin there is an odd number of spins, and on the other side of the bulk spin there is an even number of spins. We find that the bulk spin is generally more strongly correlated with the first neighbor sitting on the side with an odd number of spins than with the first neighbor sitting on the side with an even number of spins. This effect is particularly strong for large $\lambda$, where the bulk spin forms a singlet with the nearest neighbor sitting on the side, where there is an odd number of spins. The effect is illustrated in Fig.\ \ref{g-5-to-6} for both small and large $\lambda$. We note that this effect does not occur in the standard HS model, since this model is defined on a circle, where there is symmetry between the left and the right hand side.

\begin{figure}
\includegraphics[width=0.5\textwidth]{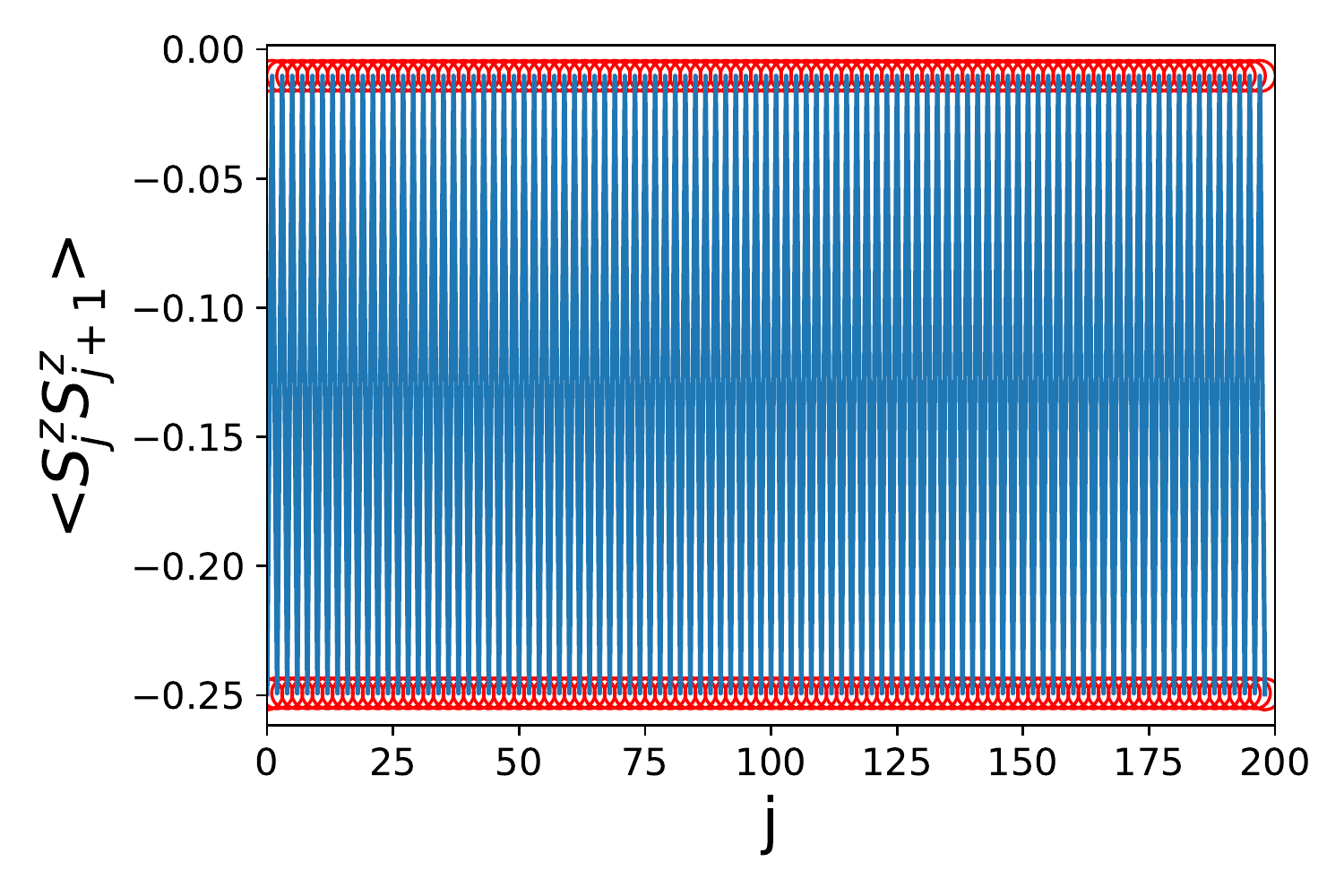}\hfill
\caption{Variation of $\langle S^z_jS^z_{j+1}\rangle$ as a function of $j$ for the 1D spin chain with $\lambda = 100$ and $N = 200$. It is seen that for $j$ even, spin number $j$ is almost perfectly anticorrelated with spin number $j+1$ and almost not correlated with spin number $j-1$.} \label{dimer}
\end{figure}

We saw in Sec.\ \ref{SEC:singlet} that the chain is perfectly dimerized into a product of singlets in the limit $\lambda\to\infty$. To investigate the behavior for large but finite $\lambda$, we plot numerical results for the dimer order parameter in Fig.\ \ref{dimer} for $\lambda=100$ and $N=200$. Since our model is $SU(2)$ invariant, we have $\langle S^x_jS^x_{j+1}\rangle = \langle S^y_jS^y_{j+1}\rangle = \langle S^z_jS^z_{j+1}\rangle$, and it is sufficient to focus on $\langle S^z_jS^z_{j+1}\rangle$ only. The figure shows that $\langle S^z_jS^z_{j+1}\rangle$ oscillates as a function of $j$. For $j$ even, $\langle S^z_jS^z_{j+1}\rangle$ is close to $-0.25$, and for $j$ odd, $\langle S^z_jS^z_{j+1}\rangle$ is almost zero. This is the expected behavior for a chain that is close to a product of singlets.

Finally, we note that the Hamiltonian is nonlocal, and we cannot conclude from the behavior of the correlation functions, whether there is an energy gap or not to the first excited state in the thermodynamic limit.

\subsection{Renyi Entropy of order two}

The Renyi entropy is another general tool to extract important information about the behavior of a spin system. As already noted in \eqref{Scritical}, the Renyi entropy grows logarithmically with subsystem size for critical systems. For noncritical systems, the entanglement entropy of the ground state typically follows an area law, which means that the Renyi entropy grows linearly with the boundary area of the selected region. In 1D, the boundary area is independent of subsystem size, and the Renyi entropy is hence constant.

\begin{figure}
\includegraphics[trim=28mm 16mm 18mm 15mm, clip,width=0.7\columnwidth]{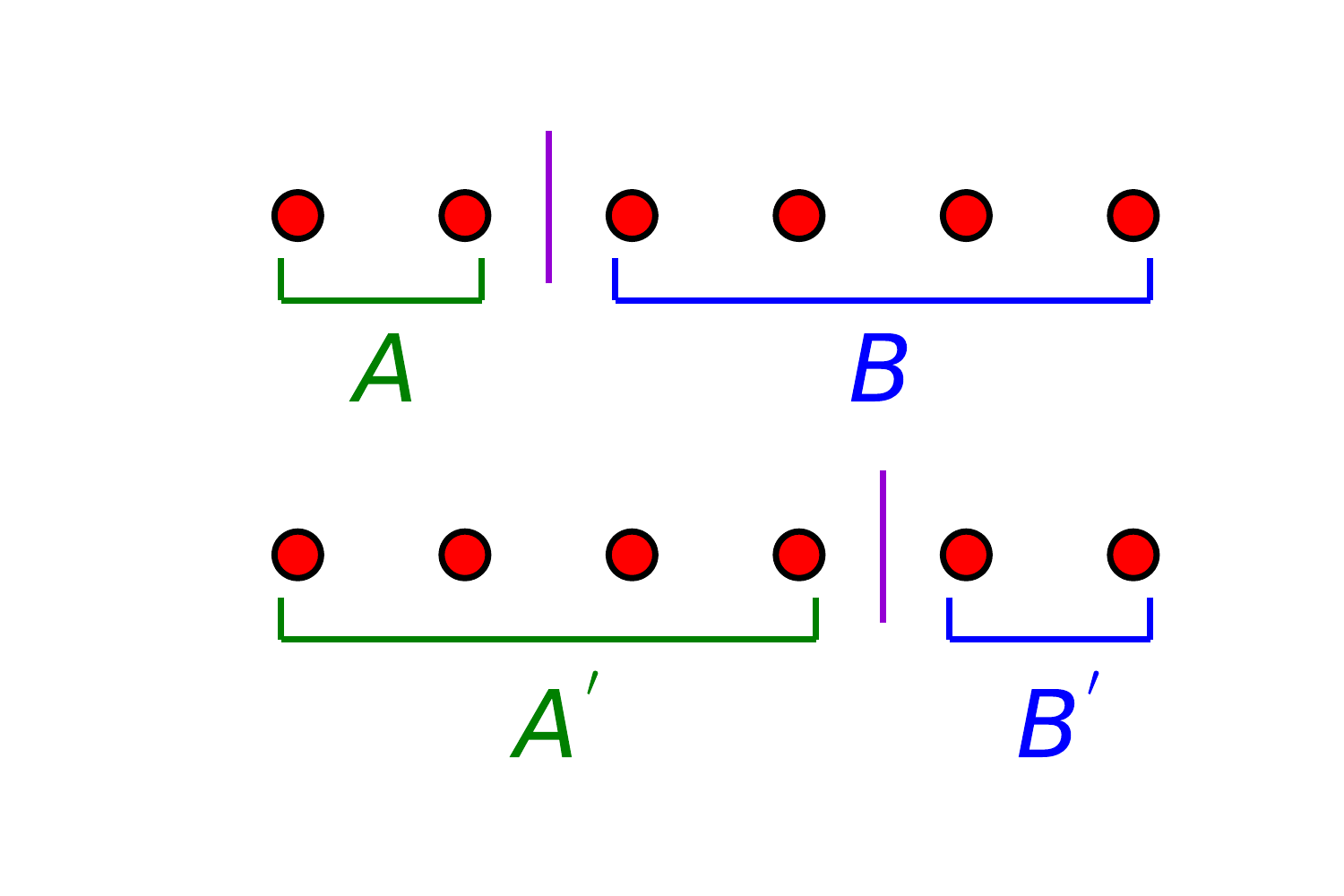}
\caption{Relation between Renyi entropies for different cuts of a spin chain in a pure quantum state. The upper part of the figure shows a spin chain partitioned into two regions $A$ and $B$, and the lower part of the figure shows the same spin chain partitioned into two different regions $A'$ and $B'$. We choose the regions such that $A$ and $B'$ contain $x$ spins each, while $A'$ and $B$ contain $N-x$ spins each. It is always the case that $S_A=S_B$ and $S_{A'}=S_{B'}$, but it is not necessarily the case that $S_{A}$ and $S_{A'}$ are the same. When the state of the chain has inversion symmetry, however, it is ensured that $S_A=S_{B'}$, and hence that $S_A=S_{A'}$.} \label{RE_demo}
\end{figure}

\begin{figure*}
\includegraphics[width=0.325\textwidth]{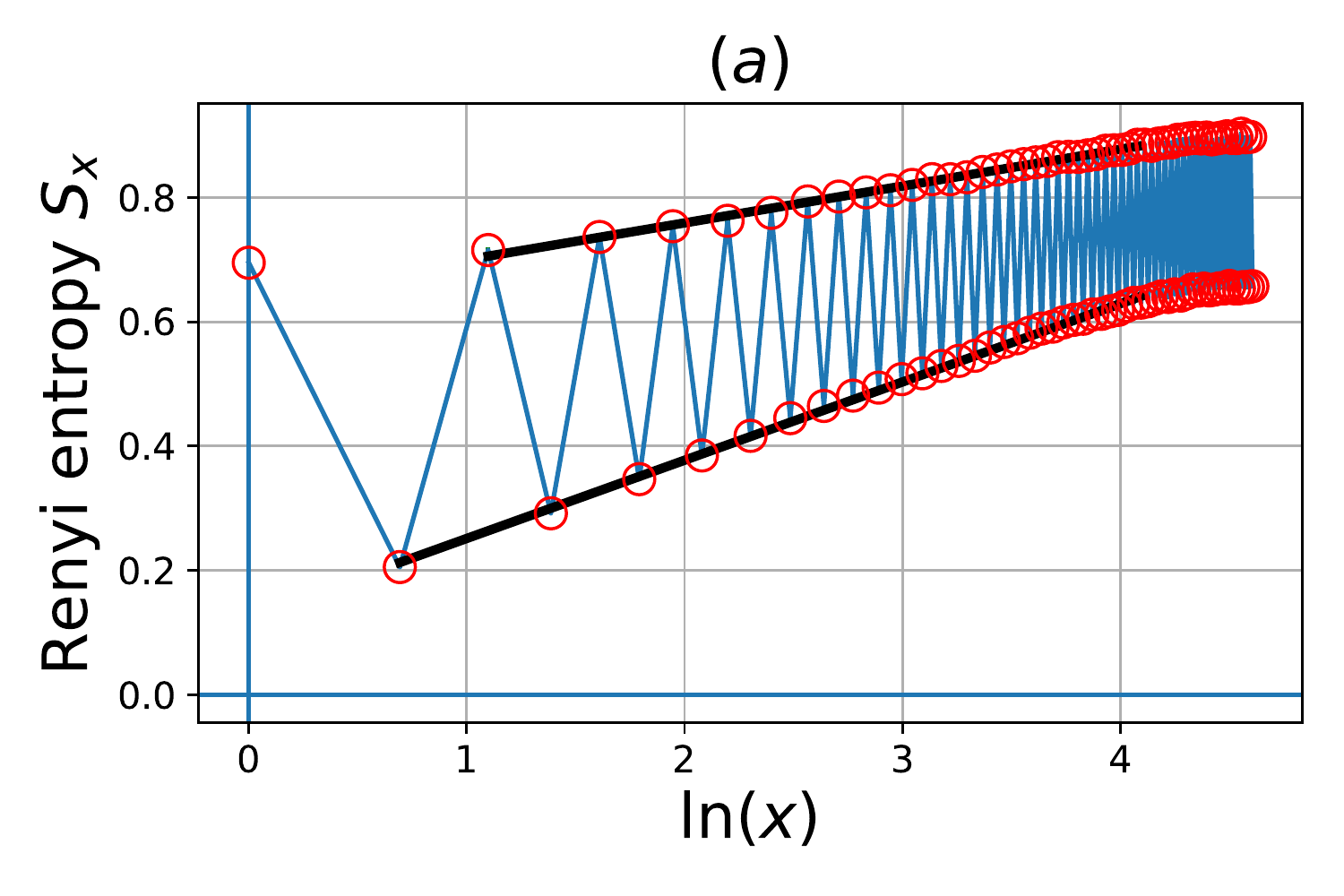}
\includegraphics[width=0.325\textwidth]{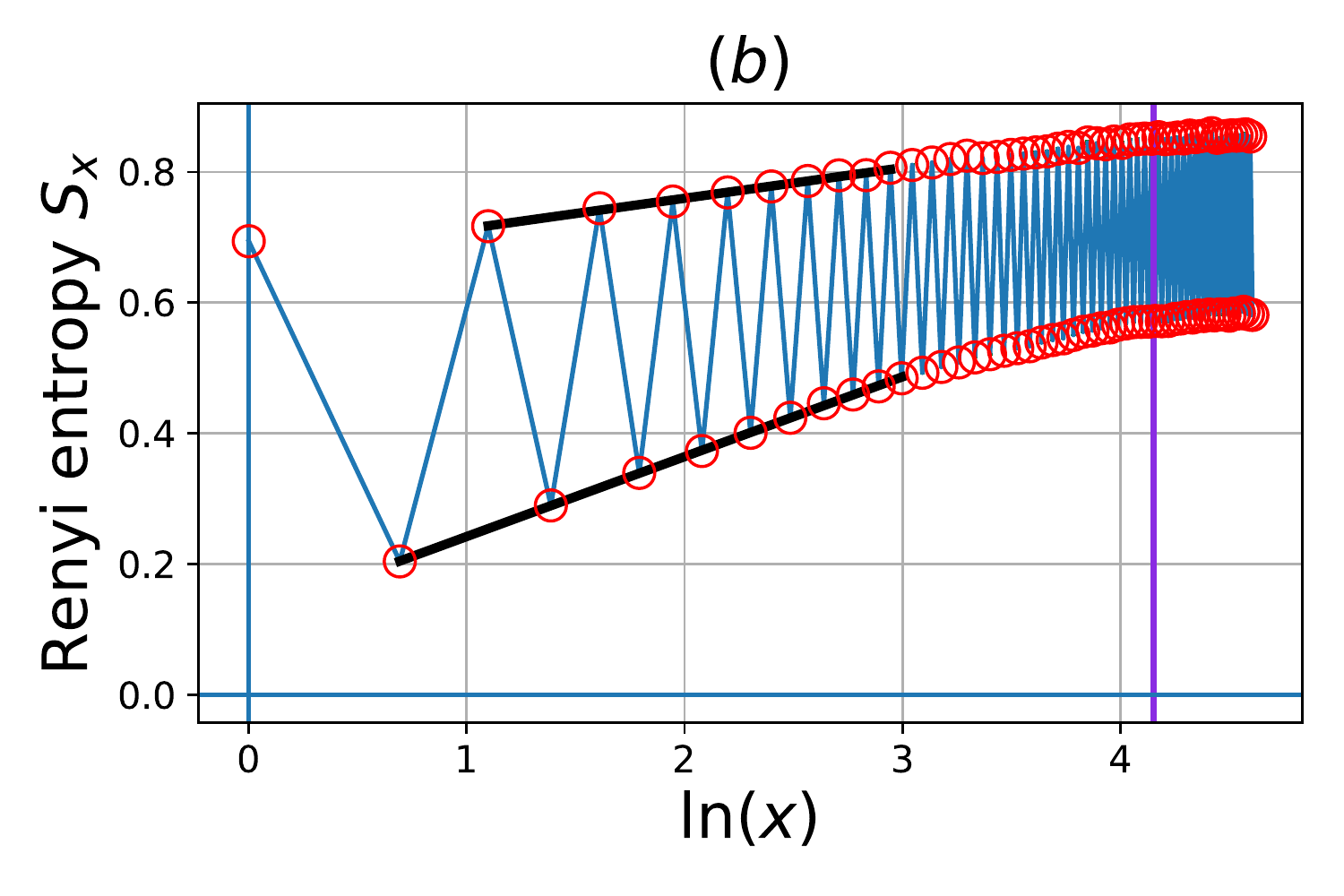}\\
\includegraphics[width=0.325\textwidth]{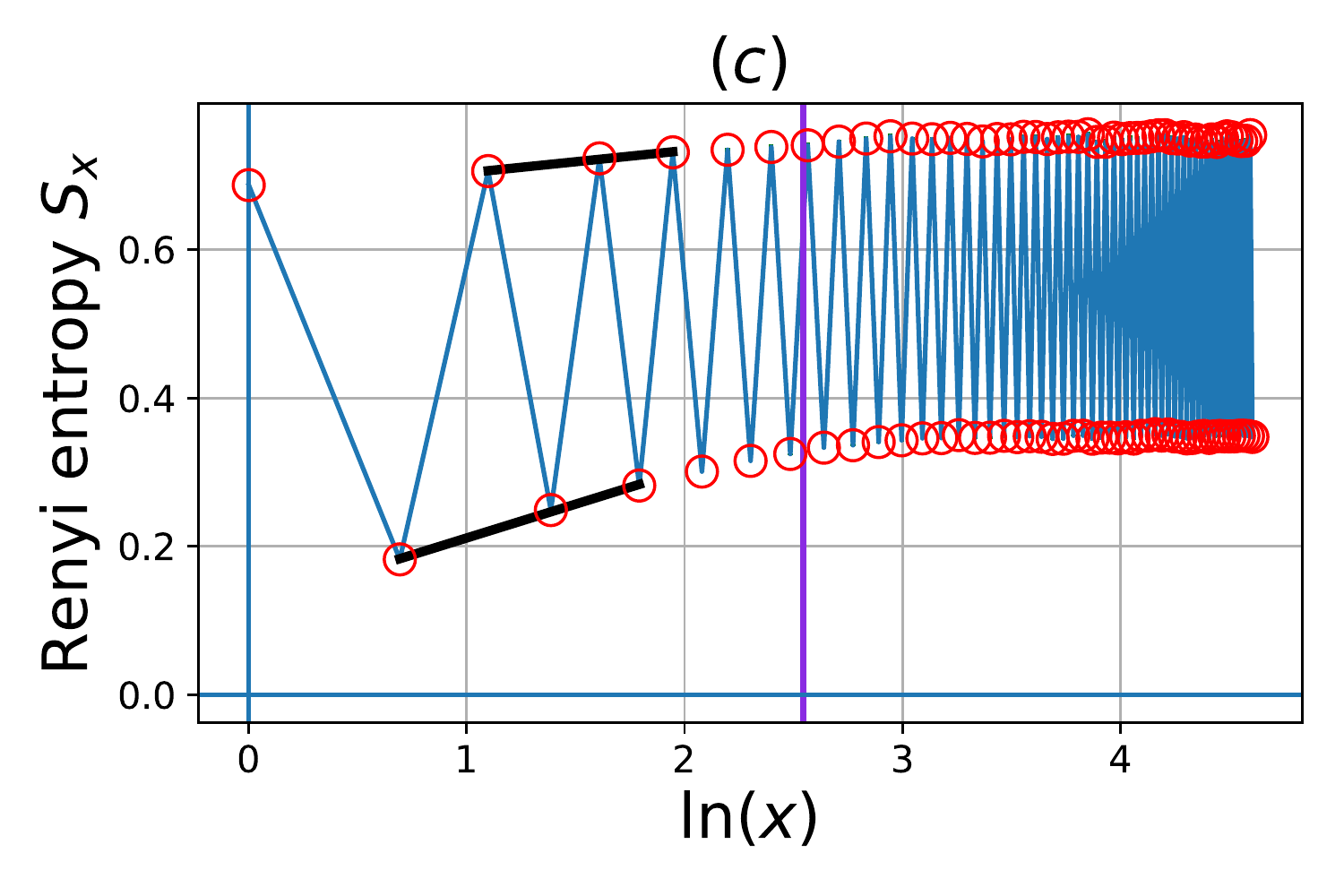}
\includegraphics[width=0.325\textwidth]{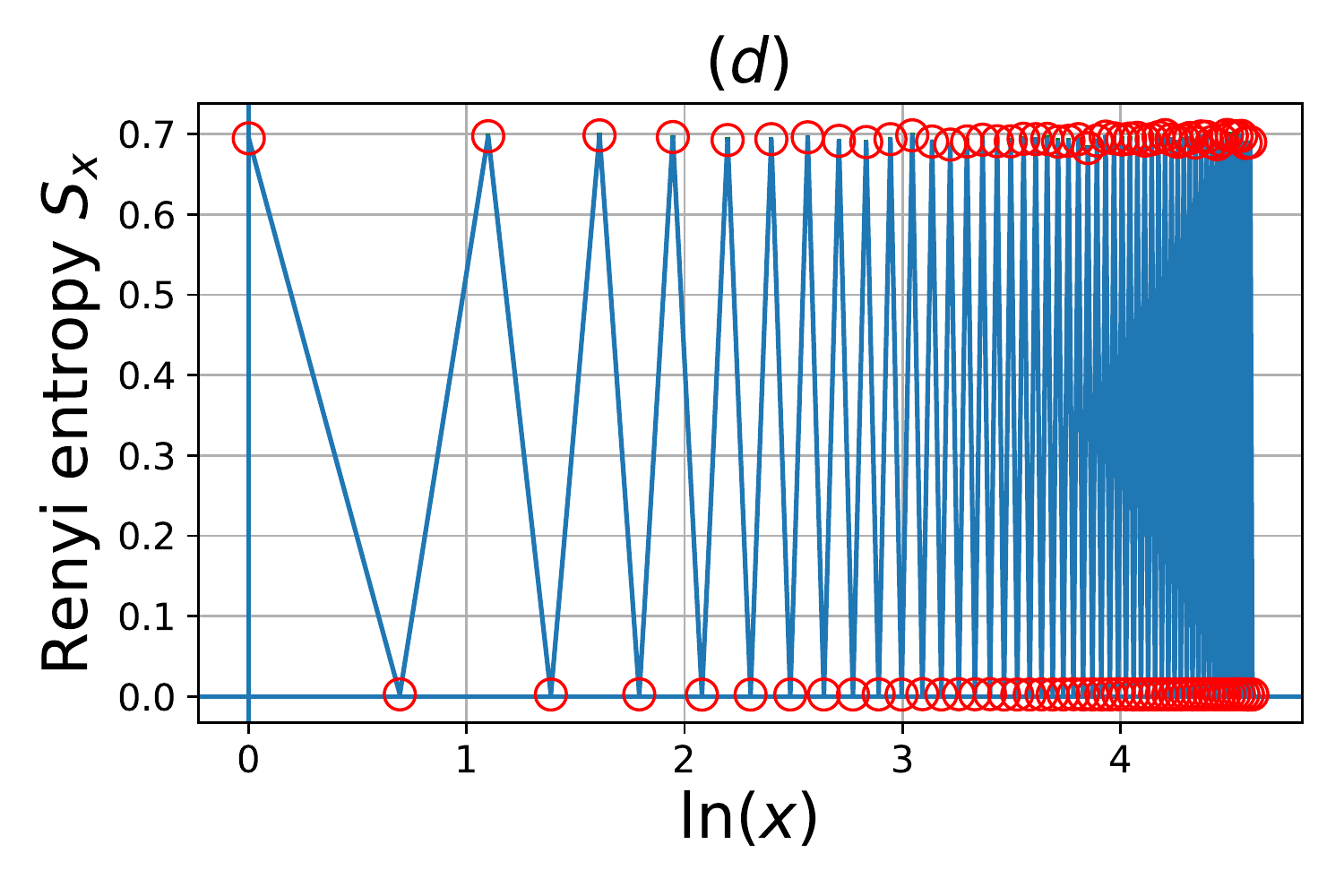}
\caption{Renyi entanglement entropy $S_x$ of order two for the uniform 1D chain versus $\ln(x)$, when the subsystem consists of the spins $0$, $1$, $\ldots$, $x-1$. There are $N = 200$ spins in the chain, and in (a) $\lambda = 0.25$, in (b) $\lambda = 1$, in (c) $\lambda = 5$, and in (d) $\lambda = 100$. The vertical lines in (b) and (c) are at $x=N/(\pi\lambda)$. (In (a) this line is to the right of the plotted region, and in (d) it is to the left of the plotted region.) In (a-c), we show two linear fits to the upper and the lower set of points, respectively. In (a, b, c), the slope of the upper line is (0.059, 0.047, 0.030), and the slope of the lower line is (0.126, 0.122, 0.091).} \label{g-10-to-11}
\end{figure*}

In the computations below, we take part $A$ of the system to be the first $x$ spins in the chain and part $B$ to be the remaining spins. Since the chain is symmetric under inversion of the direction of the spin chain, we have that the Renyi entropy of the first $x$ spins is the same as the Renyi entropy of the first $N-x$ spins. This statement is explained pictorially in Fig.\ \ref{RE_demo}.  We therefore only compute the Renyi entropy for $x\leq N/2$. It is more time consuming to compute the Renyi entropy than the correlations, since we use Monte Carlo simulations. We shall therefore restrict ourselves to $N=200$ throughout. The results are shown in Fig.\ \ref{g-10-to-11}.

For $\lambda=100$, we observe that the Renyi entropy is close to zero whenever $x$ is even and close to $0.7$ whenever $x$ is odd. This is a consequence of the results in Sec.\ \ref{SEC:singlet}. When $x$ is even, we do not cut any of the singlets apart, and there are almost no correlations between the two parts. When $x$ is odd, we break one singlet into two when cutting the chain, and the entropy is close to $\ln(2)\approx 0.693$.

For $\lambda=0.25$, the correlations follow a power law decay, and we hence expect that the Renyi entropy is linear in $\ln(x)$ for $1\ll x\ll N$, possibly plus some oscillations. From \eqref{Scritical}, we get the leading order behavior $S_x \approx c\ln(x)/8+\mathrm{constant}$. For the HS model, the central charge is $c=1$, and it is hence relevant to compare the entropy plot to a straight line with slope $1/8$. Figure \ref{g-10-to-11}(a) shows that the entropy oscillates with period 2. If we look only at the points with $x$ even in the region $1\ll x\ll N$, the points approximately fall on a straight line with slope $0.126$. This fits with the expected value $1/8$ within the uncertainty of choosing the fitting region. If we look at the points with odd $x$, however, the slope of the line is around $0.059$, which does not fit with $1/8$. It may be that this discrepancy is related to the asymmetry observed in Fig.\ \ref{g-5-to-6}. For $\lambda=1$, the slopes of the two fitted lines have changed to $0.122$ and $0.047$. It is interesting that the slope for $x$ even is again close to $1/8$, while the slope for $x$ odd seems to change with $\lambda$. For $\lambda=5$, both slopes are reduced, but the results are likely inaccurate, since the number of points in the region with linear increase is small.

For intermediate values of $\lambda$, we observe a transition from a linear increase with $\ln(x)$ for small $x$ to an area law behavior for large $x$. The figure shows that the transition occurs approximately at $x=N/(\pi\lambda)$. This fits with the behavior of the correlations, where we saw a transition from power law decay to exponential decay.

\subsection{Strengths of the spin-spin interactions} \label{SEC:interaction1D}

To investigate the Hamiltonian that gives rise to the physics discussed above, we now take a closer look at the spin-spin interaction strengths \eqref{b_ij} for the choice $w_{ij}=(z_i+z_j)/(z_i-z_j)$. We first investigate some limiting cases analytically, and after that present numerical results for different values of $N$ and $\lambda$.

\subsubsection{Behavior for small and large $\lambda$ with $N$ fixed}

We first consider the limit, where $2\pi\lambda\ll 1$. In this case
\begin{equation}\label{apw}
w_{ij}=\frac{e^{2\pi i\lambda/N}+e^{2\pi j\lambda/N}} {e^{2\pi i\lambda/N}-e^{2\pi j\lambda/N}}\approx \frac{N}{\pi (i-j)\lambda}
\end{equation}
and hence
\begin{equation}\label{b_ij_limit_general}
b_{i,i+d} \approx \frac{N^2}{6\pi^2\lambda^2d^2}
\left(1+\sum_{k(\neq i \neq i+d)}
\frac{d^2}{(k-i)(k-i-d)}\right).
\end{equation}
The sum can be simplified by utilizing
\begin{equation}\label{b_ij_2}
\frac{d}{(k-i)(k-i-d)} = \frac{1}{k-i-d} - \frac{1}{k-i}.
\end{equation}
For $d>0$ this leads to
\begin{multline}\label{b_ij_limit_dp}
b_{i,i+d} \approx \\
\frac{N^2}{6\pi^2\lambda^2d^2}
\left[3-\sum_{k=1}^d \left(\frac{d}{k+i-i_0}
-\frac{d}{k-N+i-i_0}\right)\right],
\end{multline}
where $i_0$ is the lowest possible value of $i$ (i.e., $i=i_0$ for the left most spin in the chain). The result for $d<0$ is obtained by taking $d\to -d$ and $i-i_0\to N-1-(i-i_0)$ in \eqref{b_ij_limit_dp}. If we consider a spin in the bulk of the chain, the expression for $b_{i,i+d}$ simplifies further to
\begin{equation}\label{b_ij_limit}
b_{i,i+d} \approx
\frac{N^2}{2\pi^2\lambda^2d^2}.
\end{equation}
In this limit, we hence observe that the interaction strength is inversely proportional to the square of the distance between the spins as in the original HS model. In the HS model, the spins are sitting on a circle, but as long as $|d|\ll N$, the chord distance is approximately the same as $|d|$, as already noted in \eqref{bapp}. For small $\lambda$ and large $N$, we hence expect that the uniform 1D chain model behaves similarly to the 1D HS model, except for possible edge effects. This is consistent with the observations made in the last two sections.

The result derived above for $2\pi\lambda\ll 1$ is also a good approximation for spins in the bulk under the less strict condition $2\pi\lambda|i-j|\ll N$. Although in this case there are some values of $k$ for which \eqref{apw} does not provide a good approximation for $w_{ki}$ and $w_{kj}$, those terms are much smaller than those for which \eqref{apw} is a good approximation. The error made by nevertheless using \eqref{apw} for all $k$ is hence small.

Next we consider the limit $2\pi\lambda\gg N$. We have
\begin{multline}\label{b_ij_3}
b_{ij} =\frac{1}{6}\frac{(e^{\frac{2\pi\lambda i}{N}} + e^{\frac{2\pi\lambda j}{N}})^2}{(e^{\frac{2\pi\lambda i}{N}} - e^{\frac{2\pi\lambda j}{N}})^2}\\
+\frac{1}{6}\sum_{k(\neq i\neq j)}\frac{( e^{\frac{2\pi\lambda k}{N}}+e^{\frac{2\pi\lambda i}{N}})}{( e^{\frac{2\pi\lambda k}{N}}-e^{\frac{2\pi\lambda i}{N}})} \frac{( e^{\frac{2\pi\lambda k}{N}}+e^{\frac{2\pi\lambda j}{N}})}{( e^{\frac{2\pi\lambda k}{N}}-e^{\frac{2\pi\lambda j}{N}})}.
\end{multline}
Now, for $2\pi\lambda\gg N$, we have
\begin{equation}
\frac{e^{\frac{2\pi\lambda k}{N}}+e^{\frac{2\pi\lambda j}{N}}}
{e^{\frac{2\pi\lambda k}{N}}-e^{\frac{2\pi\lambda j}{N}}} \approx \mathrm{sign}(k-j),
\end{equation}
and hence
\begin{equation}\label{b_ij_4}
b_{ij} \approx \frac{1}{6}(N-2|j-i|+1).
\end{equation}
In this case, the interaction strength is decaying linearly, and the range of the interaction is determined by the system size. We would hence expect a behavior of the system that is different from the HS model. This is consistent with the observation that the ground state is a product of singlets in that limit.

\subsubsection{Numerical results}

\begin{figure*} \includegraphics[width=0.325\textwidth]{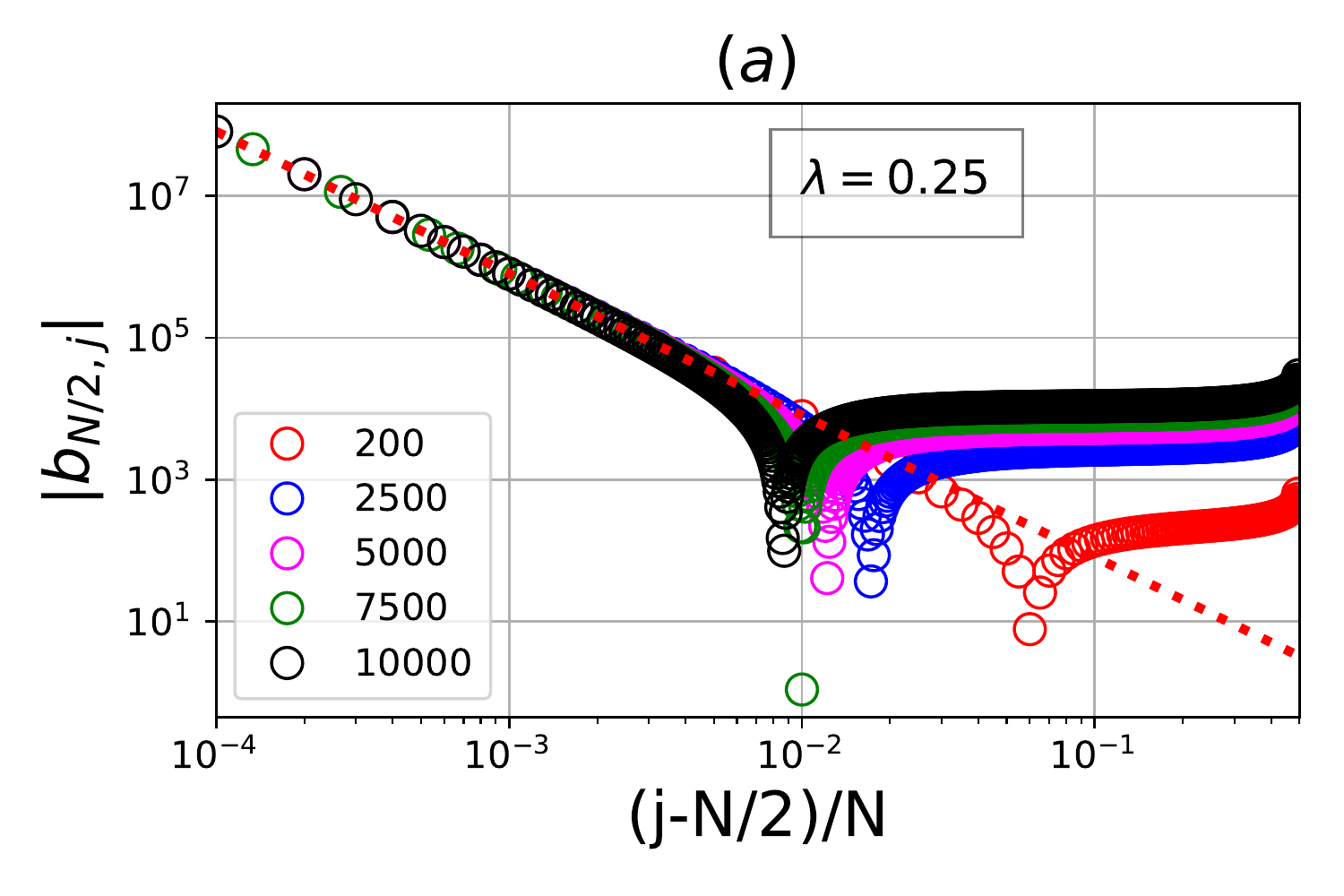}\hfill
\includegraphics[width=0.325\textwidth]{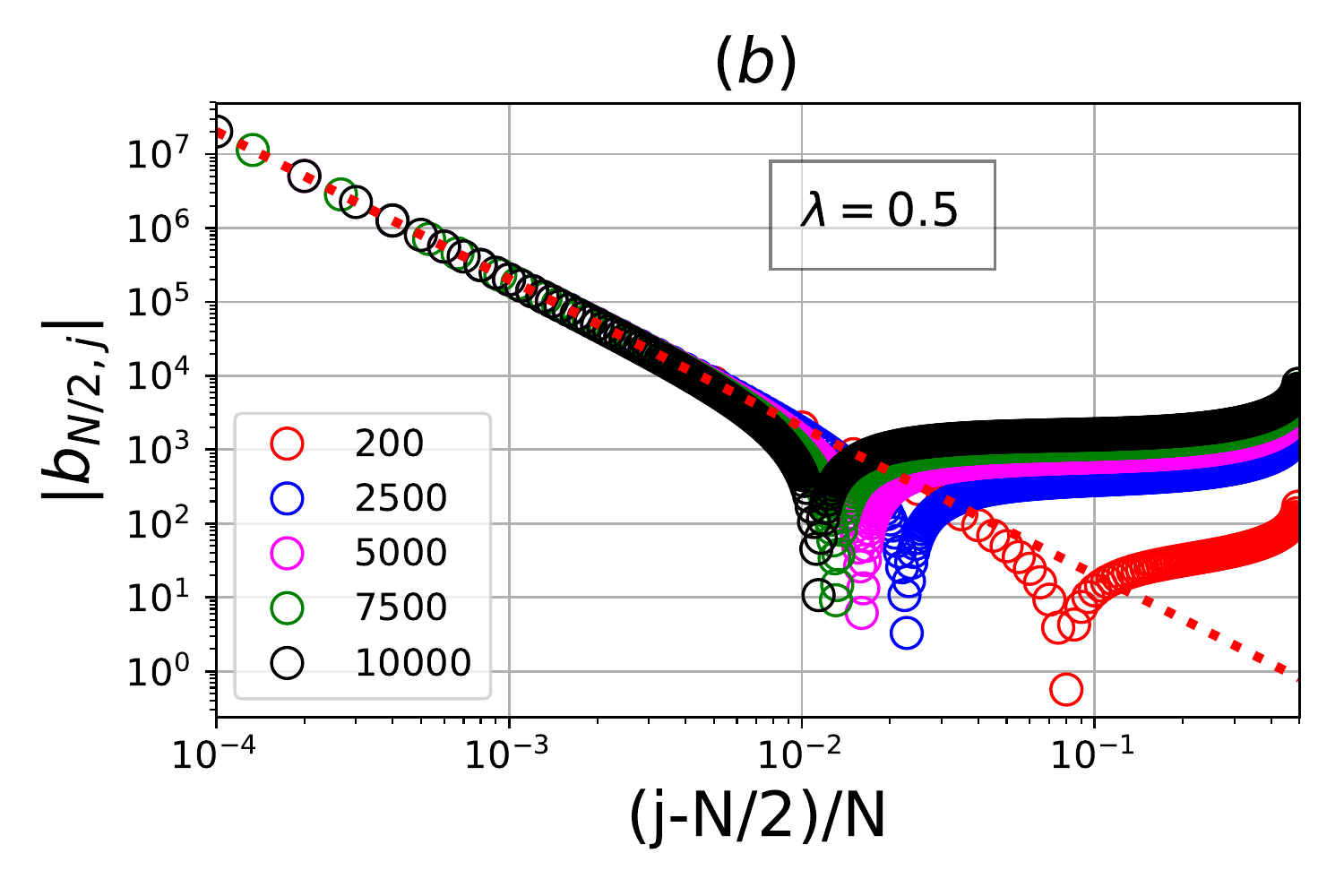}\hfill
\includegraphics[width=0.325\textwidth]{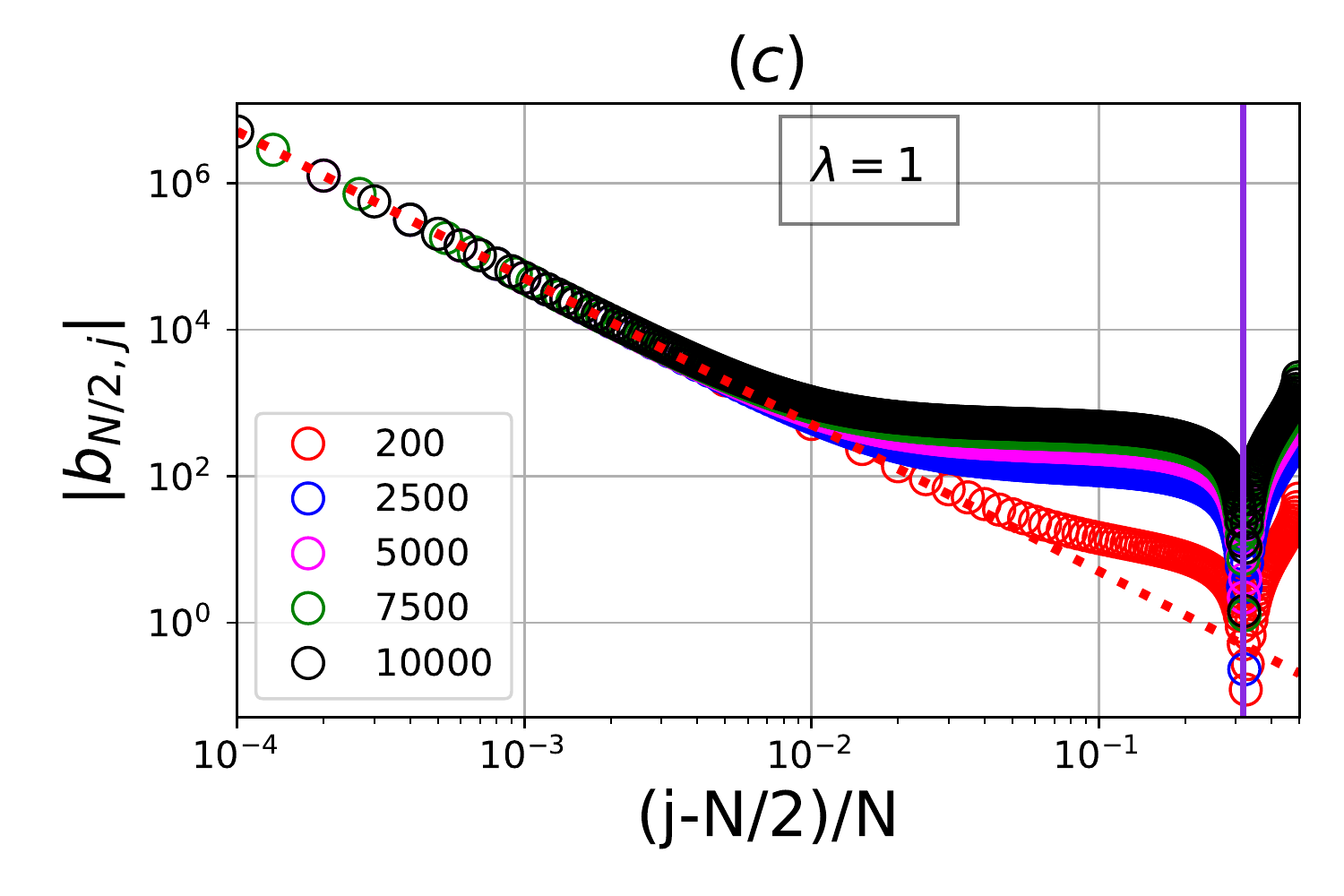}\\
\includegraphics[width=0.325\textwidth]{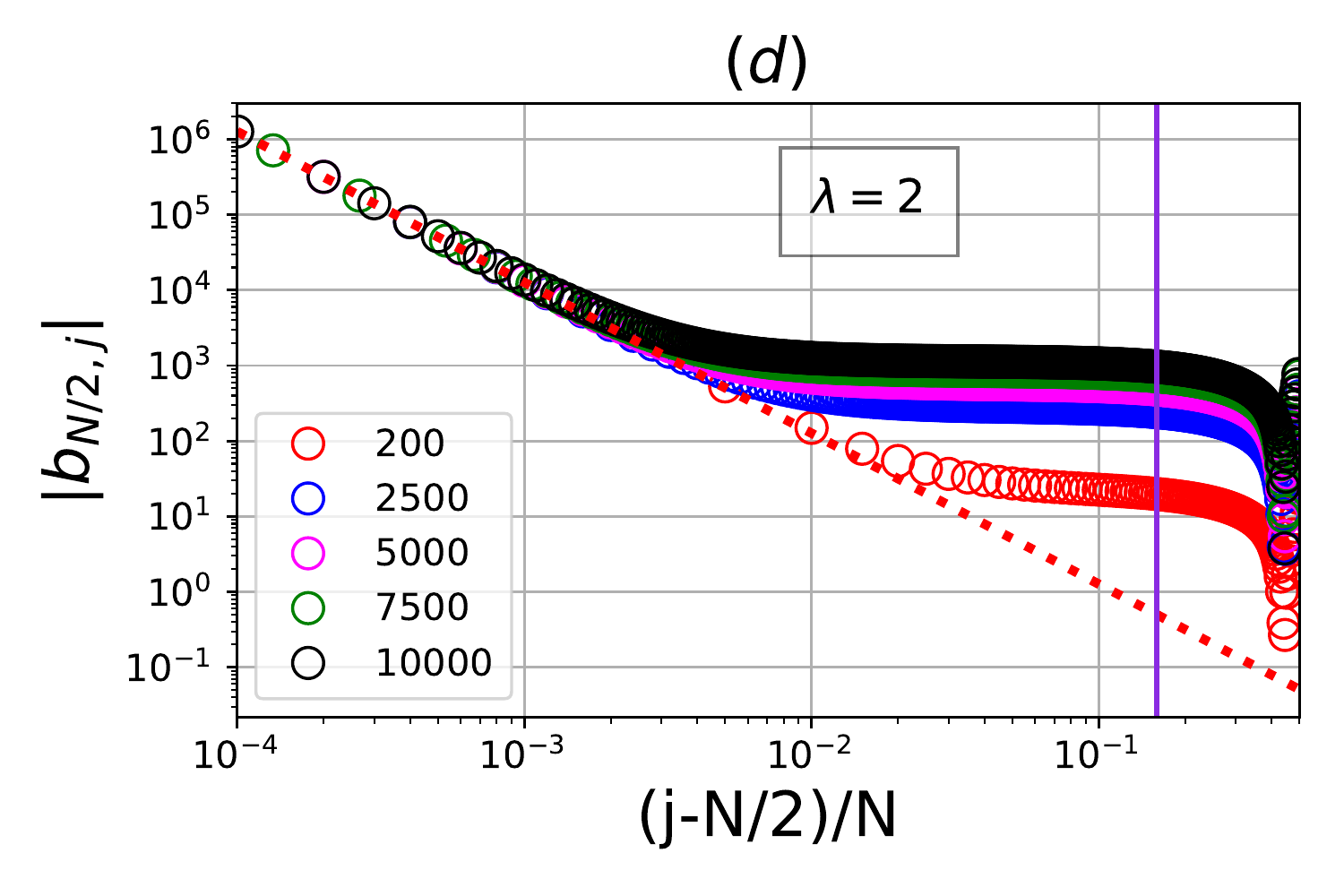}\hfill
\includegraphics[width=0.325\textwidth]{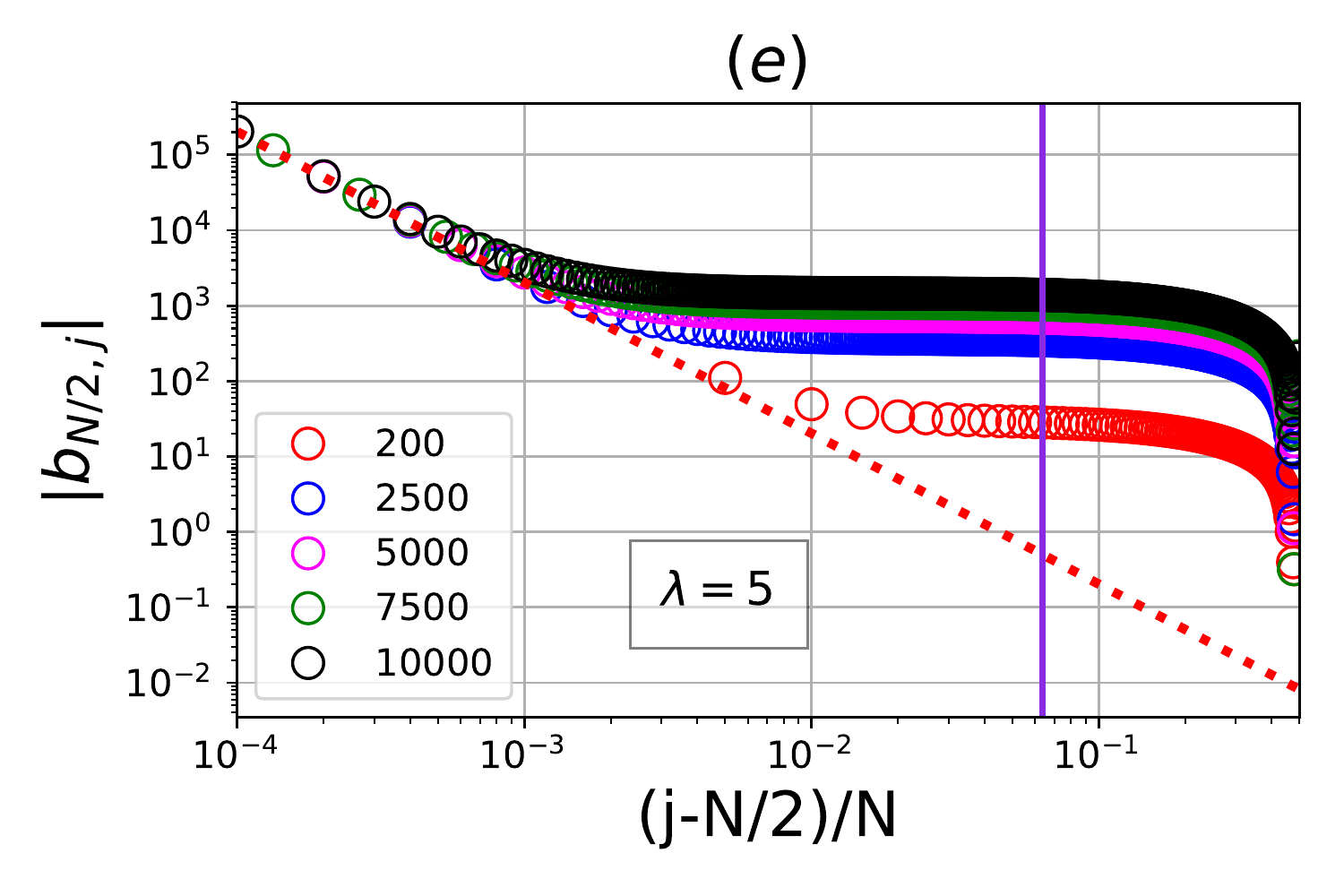}\hfill
\includegraphics[width=0.325\textwidth]{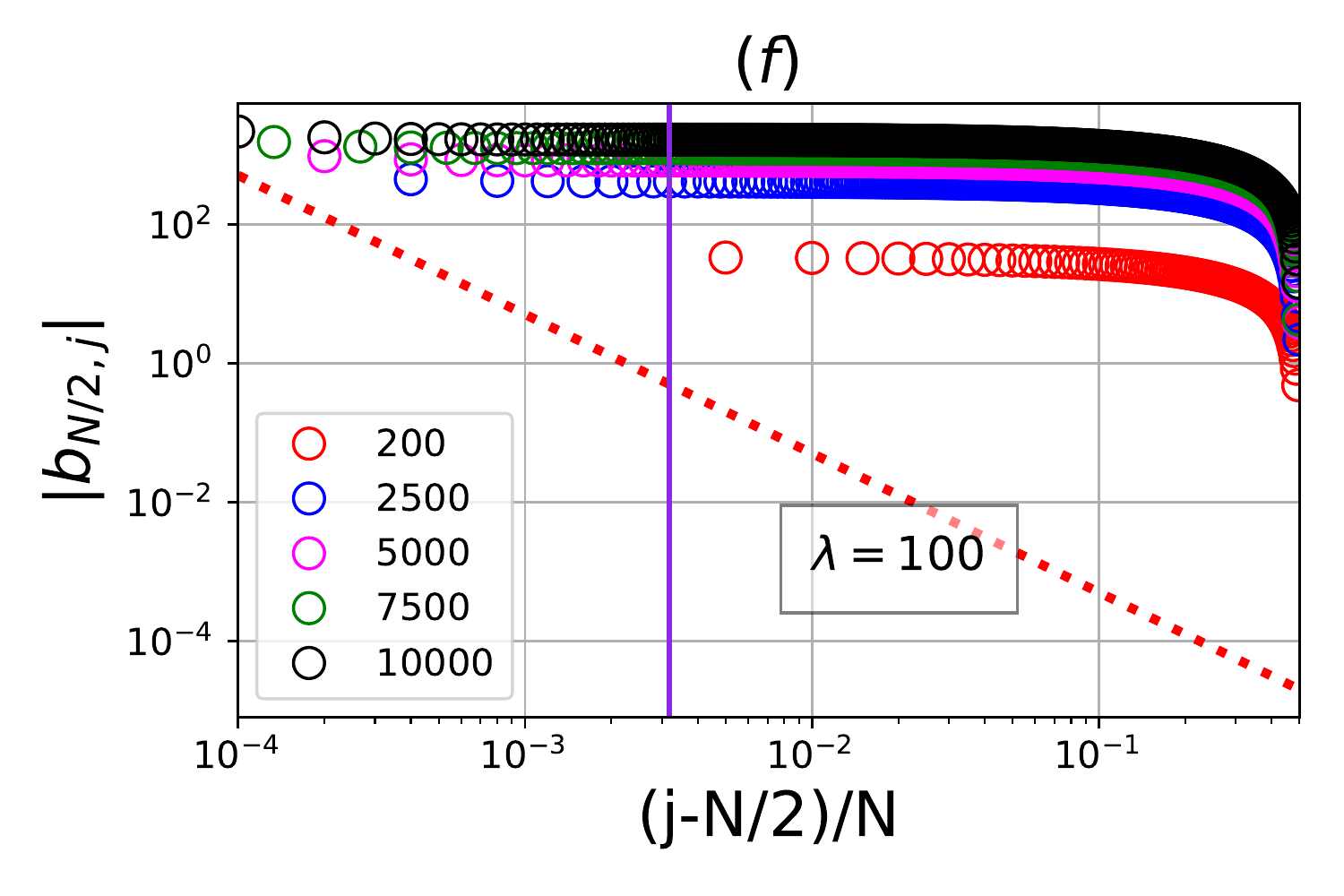}
\caption{Absolute value of the spin-spin interaction strength $b_{kj}$ (Eq.\ \eqref{b_ij}) for the uniform 1D chain as a function of $|j-k|/N$ for $k=N/2$ (bulk spin) and $j\in\{N/2+1,N/2+2,\ldots,N-1\}$ (for clarity we plot only some of these $j$ values). The different plots are for different values of $\lambda$, and there are $N=200$ (red), $N=2500$ (blue), $N=5000$ (magenta), $N=7500$ (green), or $N=10000$ (black) spins in the chain. The red dotted line in each plot is the limit \eqref{b_ij_limit}, and the vertical lines in the plots (c-f) are positioned at $(j-N/2)/N=1/(\pi\lambda)$.} \label{middle_spin_RH_scaled}
\end{figure*}

We plot results for $|b_{N/2,j}|$ for different values of $\lambda$ and $N$ in Fig.\ \ref{middle_spin_RH_scaled}. The limit \eqref{b_ij_limit} is shown as the red dotted line in the plots. This behavior is followed as long as $|j-N/2|/N$ is small enough, and this suggests that there is a connection between this behavior of the interaction strengths and the power law decay of correlations in the ground state. The limiting behavior \eqref{b_ij_4} is approximately followed in panel (f).

An important conclusion from the plots is that the Hamiltonian is, generally, nonlocal. We also see that $|b_{N/2,j}|$ does not follow a simple decay law over the entire range of $|j-N/2|/N$ values, but changes behavior qualitatively depending on the distance between the spins. Motivated by the observations for the spin-spin correlations, one may speculate if there is a change of behavior at $|j-N/2|/N=1/(\pi\lambda)$. We do, however, not observe sharp transitions at these points in the plots. This may happen since the correlations between spin number $N/2$ and spin number $j$ are not determined by $b_{N/2,j}$ alone, but depend on all the $b_{jk}$. The fact that $|b_{N/2,j}|$ changes behavior depending on distance in this model suggests that such changes may be a general mechanism to obtain models, where the correlations follow different decay laws depending on the distance between the spins.

Finally, we note that there is a whole family of two-body Hamiltonians having the analytical state as ground state. There are hence many different, possible behaviors of $b_{jk}$, and the results presented here show only one example.

\subsection{Spin chains with an odd number of spins}

We have only considered spin chains with an even number of spins so far, since the wavefunction \eqref{wf} is zero unless the total number of spins is even. One may speculate, however, if it is possible to decouple one of the spins from all the others by moving it infinitely far away and in this way obtain a model for a spin chain with an odd number of spins. We show here that this is possible for general $\lambda$, but the resulting model does not have the natural property to be symmetric under inversion of the direction of the spin chain.

We move the $N$th spin infinitely far away from the others by taking $z_N\to\infty$ along the positive real axis in the complex plane. With the definition $w_{ij}=(z_i+z_j)/(z_i-z_j)$, we have $w_{iN}\to -1$, and with the definition $w_{ij}=1/(z_i-z_j)$, we have $w_{iN}\to 0$. It follows from \eqref{b_ij} that the spin interaction $b_{iN}$ between the $i$th and the $N$th spin is zero for all $i$ for the choice $w_{ij}=1/(z_i-z_j)$, but not for the choice $w_{ij}=(z_i+z_j)/(z_i-z_j)$. The $N$th spin hence decouples from the others in the former case, but not in the latter. It is, however, the latter choice that gives a Hamiltonian that is symmetric under inversion of the direction of the chain.

For small $\lambda$, it is possible to have a chain with an odd number of spins and a Hamiltonian that is symmetric. This follows from \eqref{b_ij} and $w_{ij}=1/(z_i-z_j)\approx N/[2\pi\lambda(i-j)]$. Another way to obtain chains with an odd number of spins for small $\lambda$ is to consider a ladder model with an odd number of spins on each leg as already demonstrated in Sec.\ \ref{SEC:decoupling}.

\section{Uniform ladder model}\label{SEC:ladder}

\begin{figure}
\includegraphics[width=0.8\columnwidth]{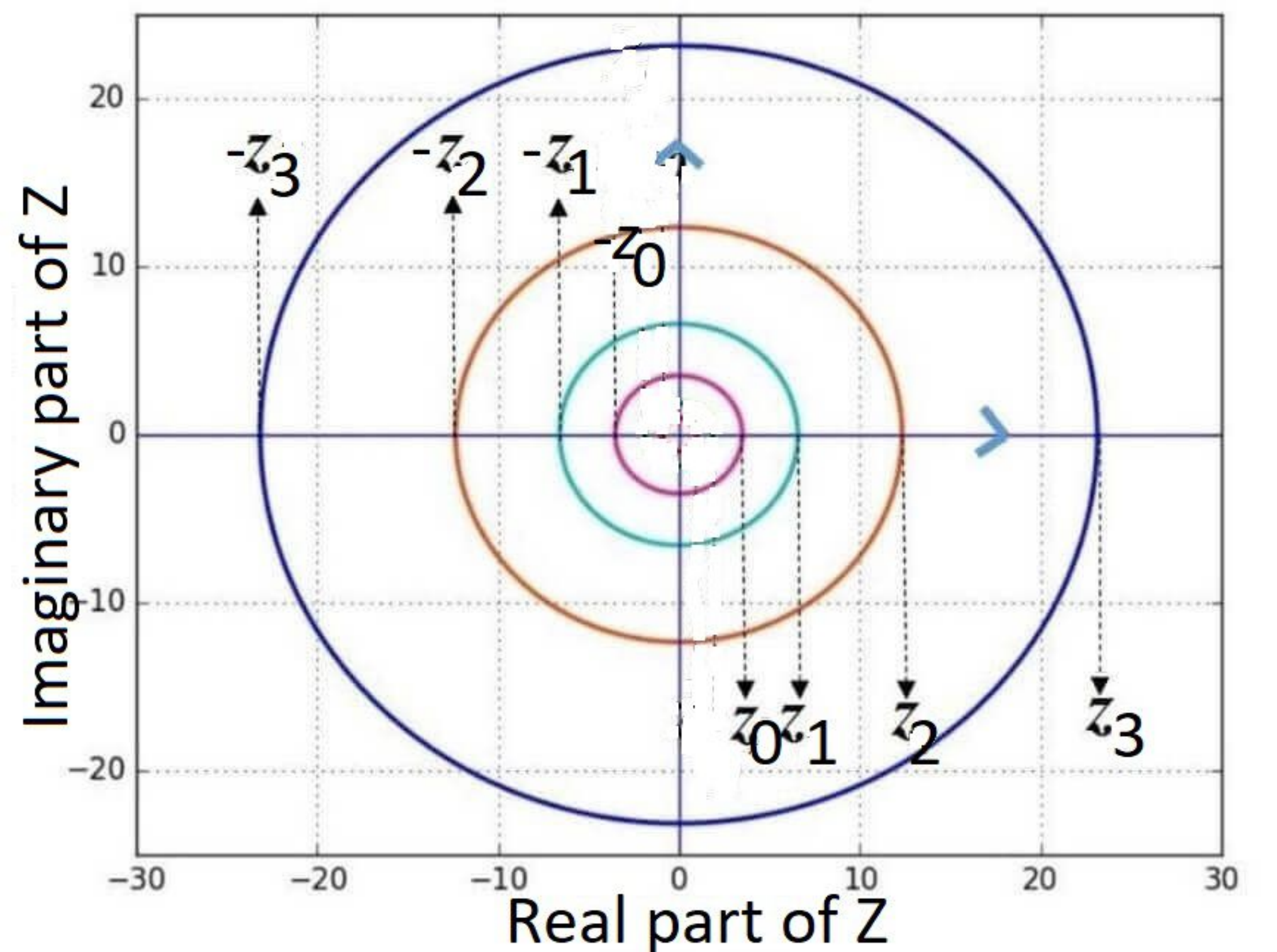}\vfill
\includegraphics[width=0.8\columnwidth]{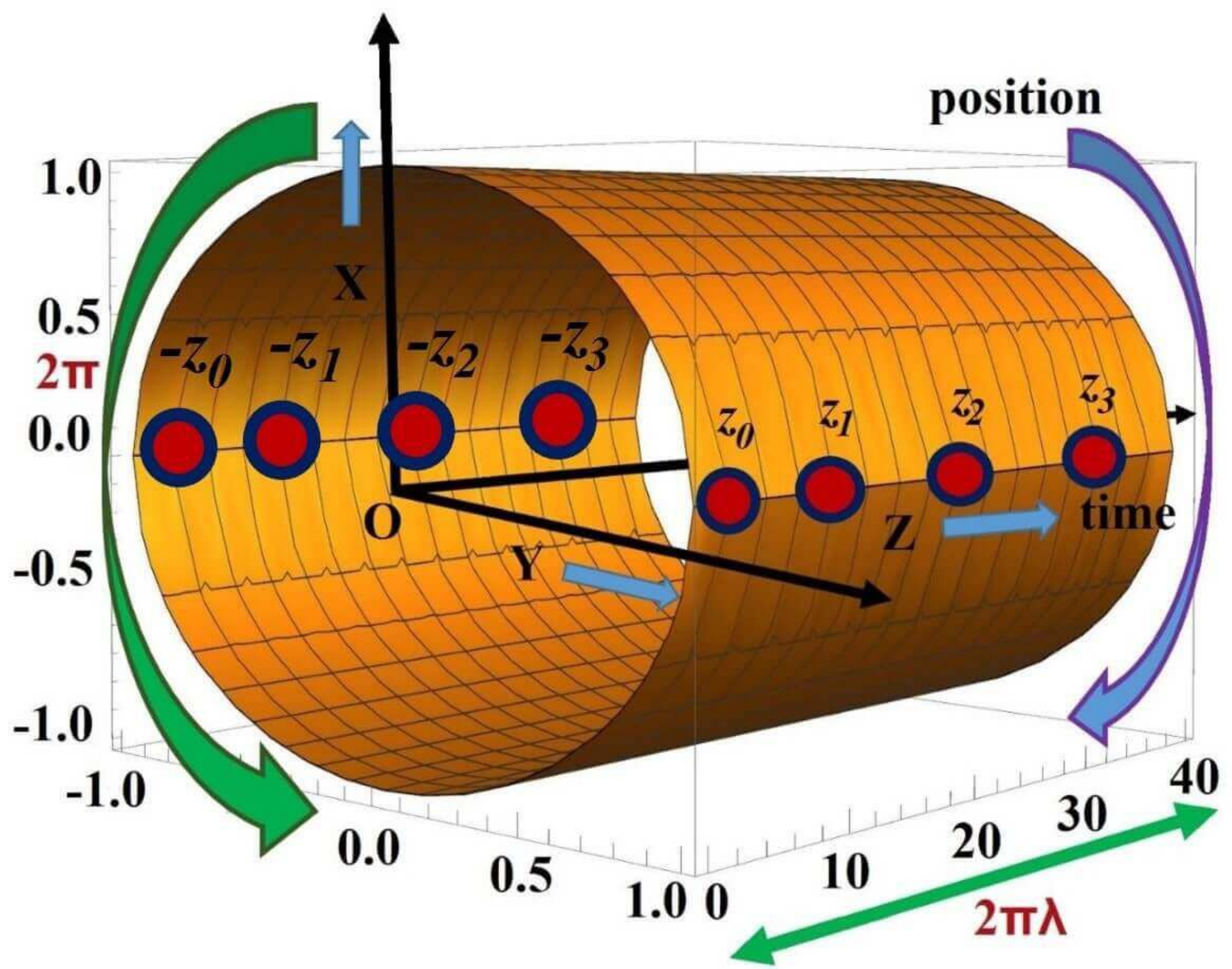}\vfill
\caption{Mapping of the spin positions from the complex plane (upper plot) to the cylinder surface (lower plot) for the uniform ladder. The radii of the consecutive circles in the plane are $\exp(2\pi \lambda j/N)$.}\label{fig-10}
\end{figure}

In this section, we investigate the uniform ladder model obtained by choosing $z_{j\pm}=\pm\exp({2\pi \lambda j/N})$. Here, $N$ is the total number of spins, which must be even, and $j\in\{0,1,\ldots,N/2-1\}$. Note that $j+$ ($j-$) refers to spin number $j$ on the front (back) of the cylinder. The parameter $\lambda/2$ determines the ratio between the length of the ladder, which is $\pi\lambda$, and the circumference of the cylinder, which is $2\pi$. The mapping from the complex plane to the cylinder is shown in Fig.\ \ref{fig-10}. In the complex plane, the spins are along both the positive and the negative part of the real axis, and on the cylinder they are placed on opposite sides.

\subsection{Strengths of the spin-spin interactions}

\begin{figure*}
\includegraphics[width=0.33\textwidth]{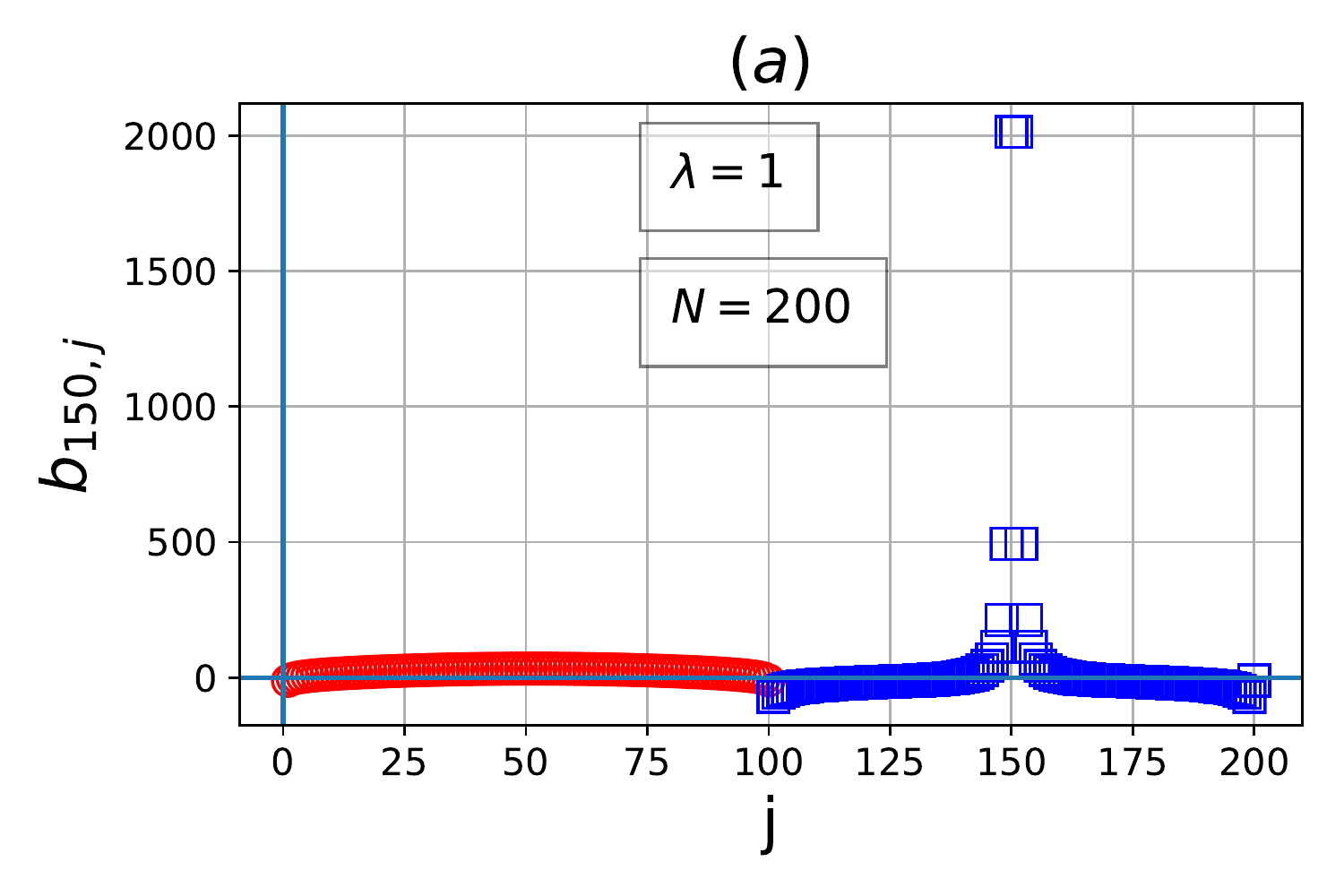}\hfill
\includegraphics[width=0.33\textwidth]{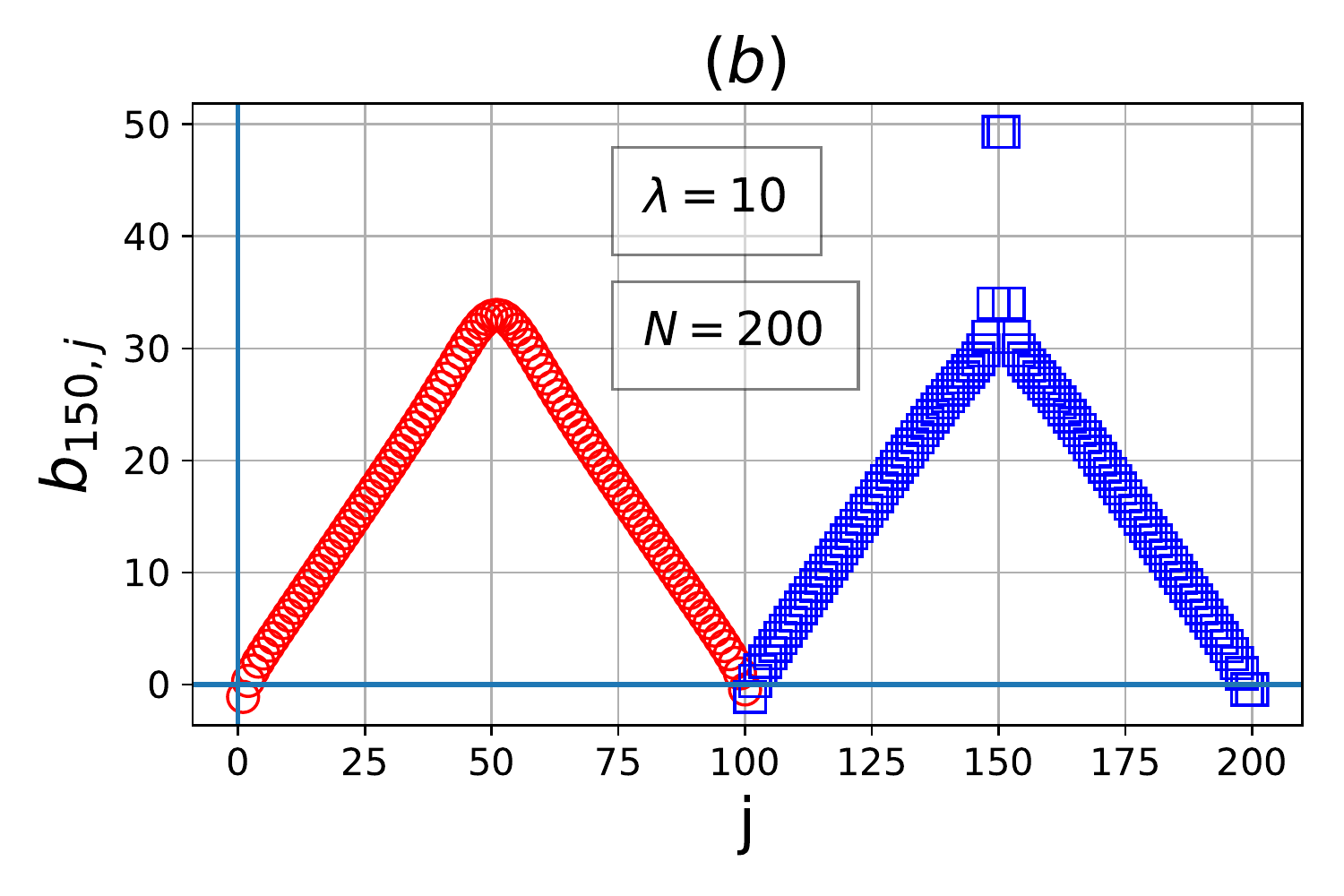}\hfill
\includegraphics[width=0.33\textwidth]{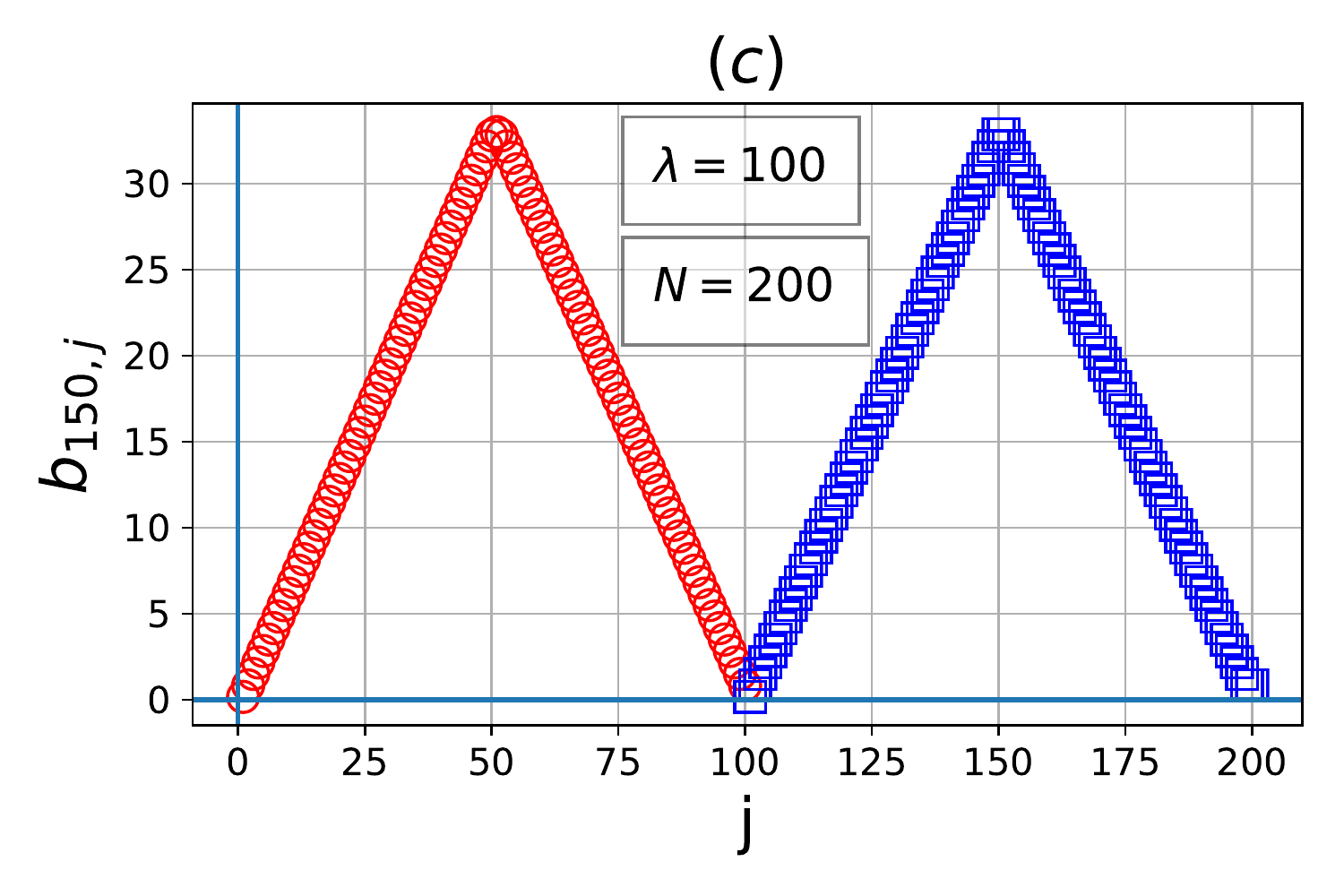}
\caption{Spin-spin interaction \eqref{b_ij} between spin number $49-$ and all other spins for a uniform ladder with $N=200$ spins. In this plot, we use a numbering such that $1$ to $100$ are the spins $0+$ to $99+$ on the front of the cylinder and $101$ to $200$ are the spins $0-$ to $99-$ on the back of the cylinder.}\label{g-13-to-14}
\end{figure*}

From Secs.\ \ref{SEC:decoupling} and \ref{SEC:singlet}, we know that for $\lambda$ very small, the ladder decouples into two chains, and for $\lambda$ very large, each spin on one of the legs forms a singlet with the neighboring spin on the other leg. We hence expect that the legs of the ladder are weakly coupled for small $\lambda$ and strongly coupled for large $\lambda$. To see what the coupling looks like, we plot the spin-spin interaction strength \eqref{b_ij} for different values of $\lambda$ in Fig.\ \ref{g-13-to-14}. For $\lambda=1$, we indeed observe that interactions between spins on different legs are much weaker than the strongest interactions between spins on the same leg. For $\lambda=100$, the interactions with the neighboring spin on the opposite leg are the strongest. For the intermediate case $\lambda=10$, the interactions are strongest for neighboring spins on the same leg, but there are also considerable interactions between spins on different legs. Another important conclusion from the plot is that the spin-spin interactions between spins on the same leg qualitatively display the same behavior as for the chain. We can hence, at least for the middle spin, roughly think of the ladder as two copies of the chain model plus interactions between the two legs. It is also interesting to note that for the larger values of $\lambda$, the strength of the spin-spin interaction is approximately the same for spins on the same leg as for spins on opposite legs, except when the distance between the spins is small. Finally, the plots show that the interactions are highly nonlocal for $\lambda$ large.

\subsection{Weak coupling}\label{SEC:weak}

\begin{figure}
\includegraphics[width=0.8\columnwidth]{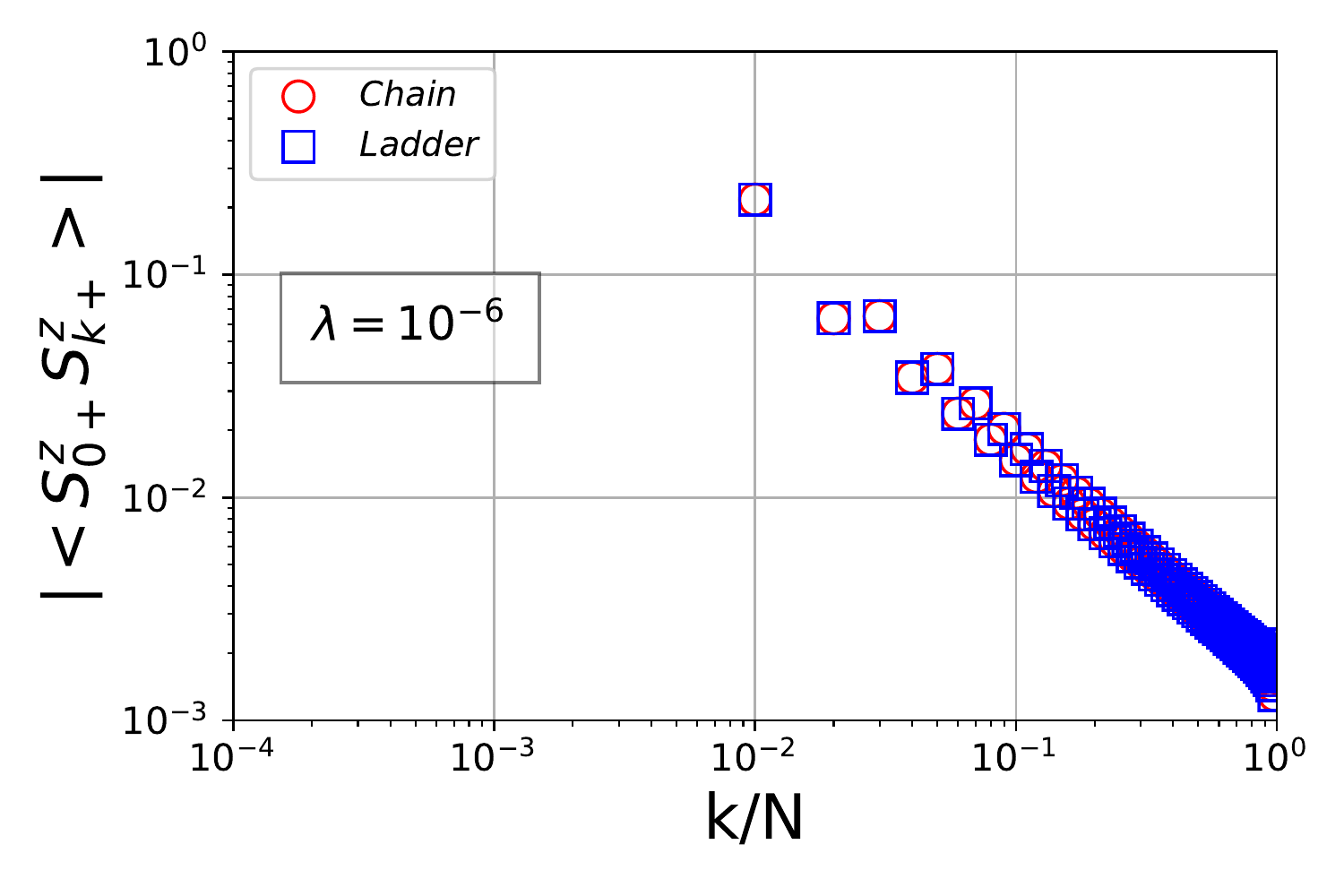}
\caption{Comparison of the spin-spin correlations for the ladder and the chain. The blue squares show the absolute value of the spin-spin correlation $\langle S^z_{0+}S^z_{k+}\rangle$ between spins on the front of the cylinder for the uniform ladder with $N = 200$ and $\lambda=10^{-6}$ as a function of $k\in\{1,2,\ldots,99\}$. The red circles show the same correlations for the chain model obtained by removing all the spins on the back of the cylinder.} \label{fig-cor-weak-coup}
\end{figure}

\begin{figure}
\includegraphics[width=0.8\columnwidth]{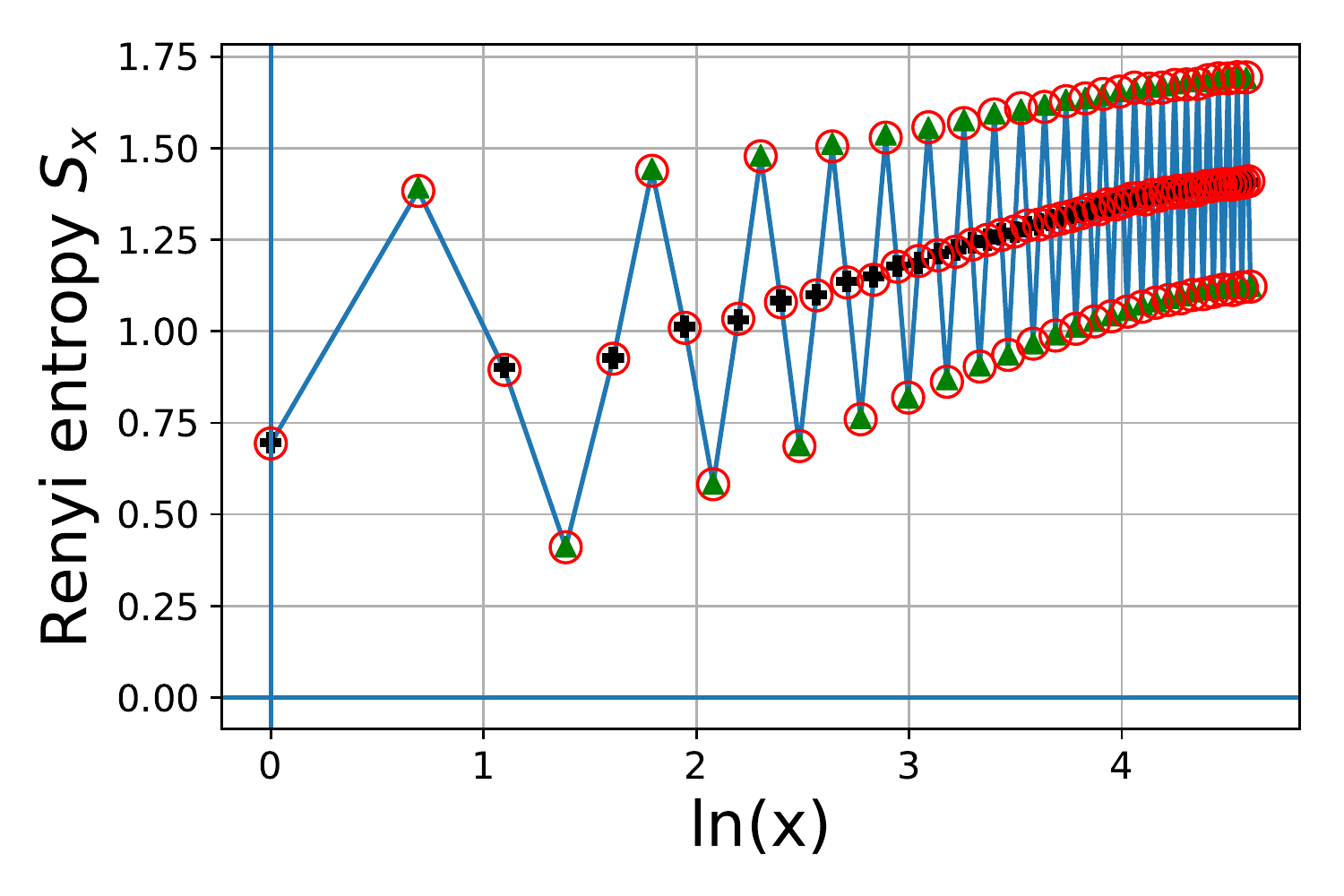}
\caption{Renyi entropy $S_x$ of order two for the uniform ladder with $N=200$ and $\lambda = 10^{-6}$. For $x$ even (odd), part $A$ of the system consists of the first $x/2$ (the first $(x+1)/2$) spins on the front leg and the first $x/2$ (the first $(x-1)/2$) spins on the back leg of the ladder. The green triangles and the black pluses show the sum of the entropies for two independent spin chains, when the spin chains are cut at the same positions as the legs of the ladder.} \label{fig-ent-weak-coup}
\end{figure}

We first consider the case of small $\lambda$, where the interactions between the two legs of the ladder are weak. We found in Sec.\ \ref{SEC:decoupling} that the ladder decouples into two independent spin chain models in the limit of small $\lambda$. Here, we take the small, but finite, value $\lambda=10^{-6}$ and plot the spin-spin correlations and the Renyi entropy in Figs.\ \ref{fig-cor-weak-coup} and \ref{fig-ent-weak-coup}, respectively. The plots show results both for the ladder and for two independent spin chains, and we indeed see that these two cases give practically the same values.

\subsection{Spin-spin correlations}

\begin{figure*}
\includegraphics[width=0.325\textwidth]{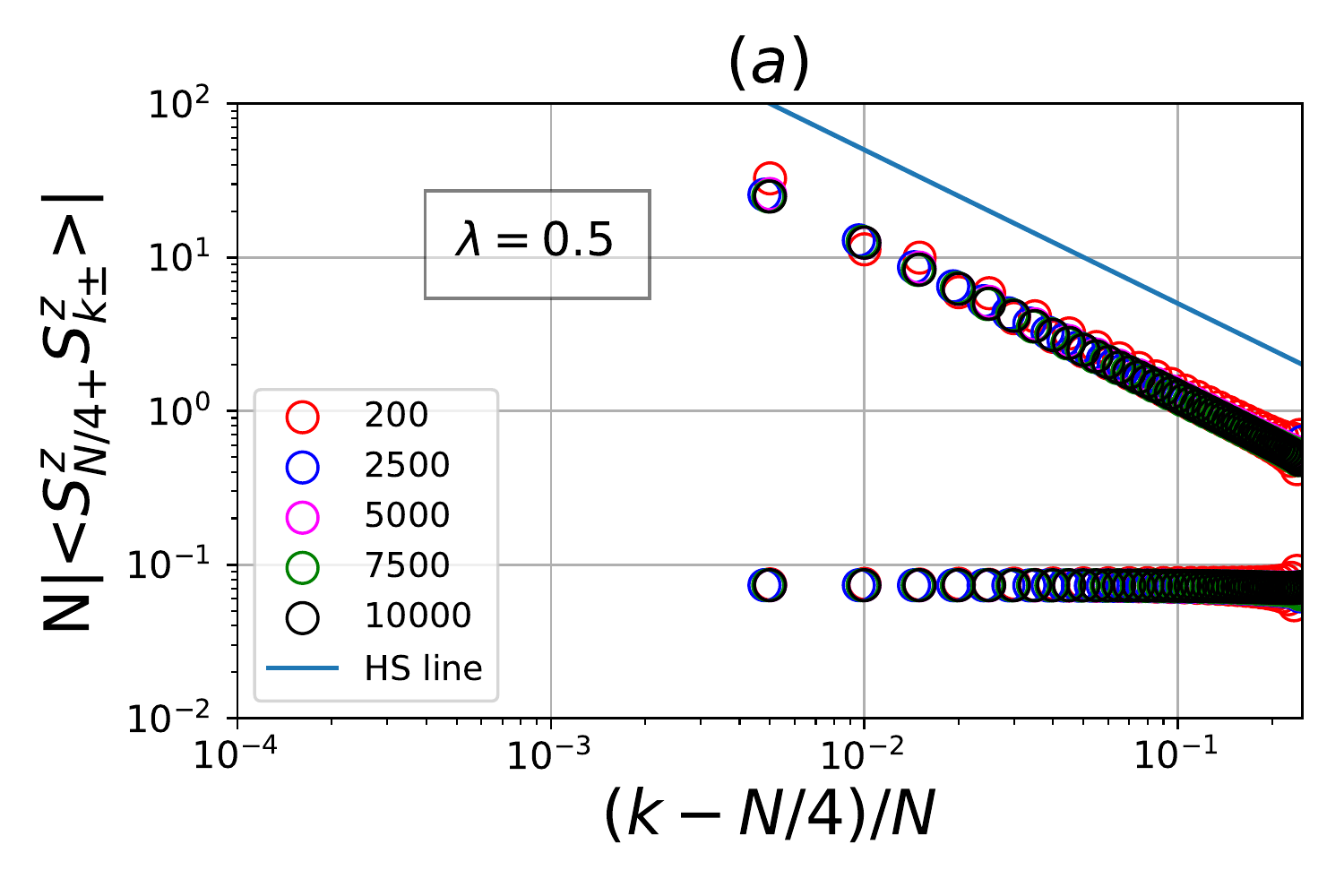}\hfill
\includegraphics[width=0.325\textwidth]{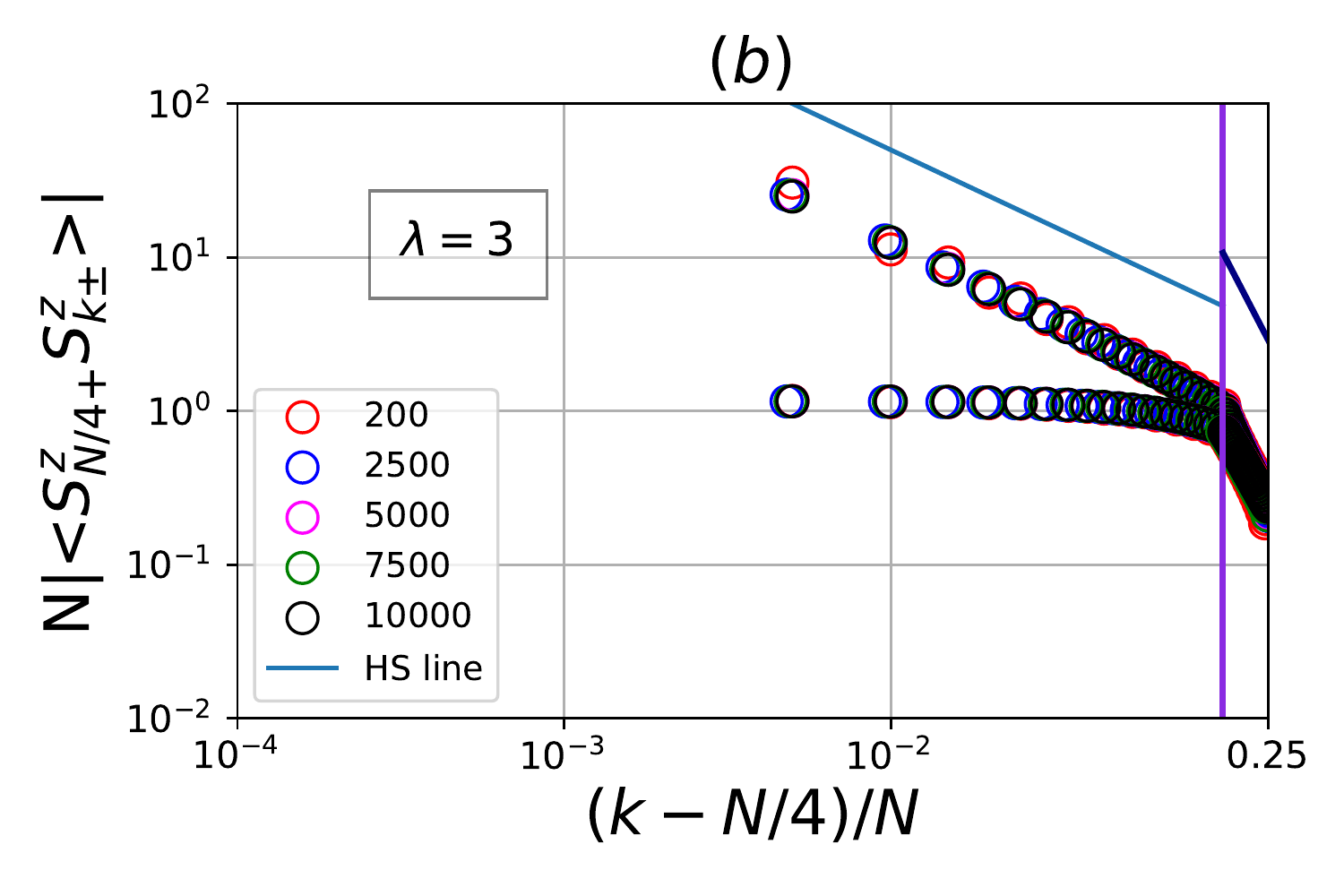}\hfill
\includegraphics[width=0.325\textwidth]{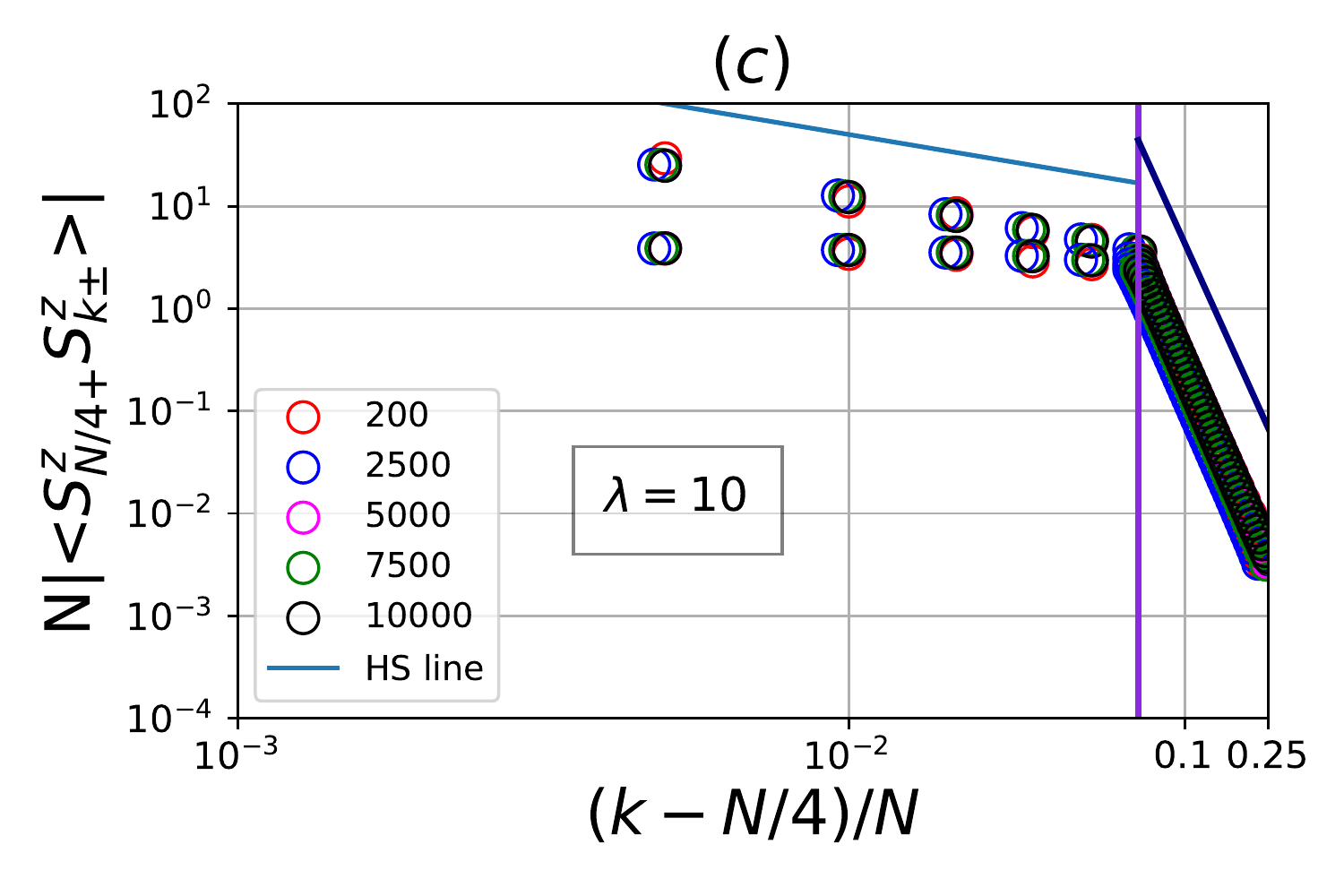}
\caption{Absolute value of the spin-spin correlation $\langle S^z_{j+} S^z_{k\pm} \rangle$ for the uniform ladder as a function of $(k-j)/N$ for $j=N/4$ (bulk spin) and $k\in\{N/4+1,N/4+2,\ldots,N/2-1\}$ (for clarity we plot only some of these $k$ values). In each plot, the upper (lower) data points show the correlations of the bulk spin with other spins on the same (opposite) leg. The different plots are for different values of $\lambda$, and there are $N=200$ (red), $N=2500$ (blue), $N=5000$ (magenta), $N=7500$ (green), or $N=10000$ (black) spins in the chain. The vertical lines in (b-c) are positioned at $(k-N/4)/N=1/(\pi\lambda)$. Note that the $x$-axis is in log scale to the left of these lines and in linear scale to the right of these lines. For $\lambda=0.5$, the correlations between spins on the same leg are seen to follow a power law decay, while the correlations between spins on opposite legs are much smaller and almost independent of distance. For larger values of $\lambda$, we still see a power law decay for short distances, but at longer distances the correlations decay exponentially, both for correlations between spins on the same leg and for correlations between spins on opposite legs. The transition from power law to exponential decay is seen to occur approximately at the vertical lines. In the standard 1D HS model the correlations decay as the inverse of the distance, and in (a-c) we plot a straight line of slope $-1$ for comparison in the region where the $x$-axis is in log scale. The straight lines in the region to the right of the vertical line in the plots (b-c) are proportional to $\exp(-\pi\lambda (k-N/4)/N)$.} \label{g-15-to-16}
\end{figure*}

Results for the spin-spin correlations for a bulk spin and different values of $\lambda$ and $N$ are provided in Fig.\ \ref{g-15-to-16}. We find that the sign of the correlations is generally positive (negative) if the two spins are separated by an even (odd) number of nearest neighbor links. We hence plot only the absolute value of the correlations. For spins on the same leg, it is seen that the correlations follow the same pattern as for the 1D chain. In the region well to the left of the line $(k-N/4)/N=1/(\pi\lambda)$, the correlations decay as
\begin{equation}
|\langle S^z_{j+}S^z_{k+}\rangle| \propto |j-k|^{-1},
\end{equation}
and in the region well to the right of the line $(k-N/4)/N=1/(\pi\lambda)$, they decay as
\begin{equation}\label{corlonglad}
|\langle S^z_{j+}S^z_{k+}\rangle| \propto \frac{1}{N}
\exp\left(-\frac{\pi\lambda|k-N/4|}{N}\right).
\end{equation}
For spins on different legs, we see that the correlations are almost independent of distance in the region well to the left of the line $(k-N/4)/N=1/(\pi\lambda)$, and well to the right of the line they follow \eqref{corlonglad} with practically the same proportionality constant as for spins on the same leg. The conclusion is hence that also for the ladder model, we can have a situation, where the nature of the decay changes depending on the distance between the spins.

\subsection{Renyi Entropy of order two}

\begin{figure*}
\includegraphics[width=0.33\textwidth]{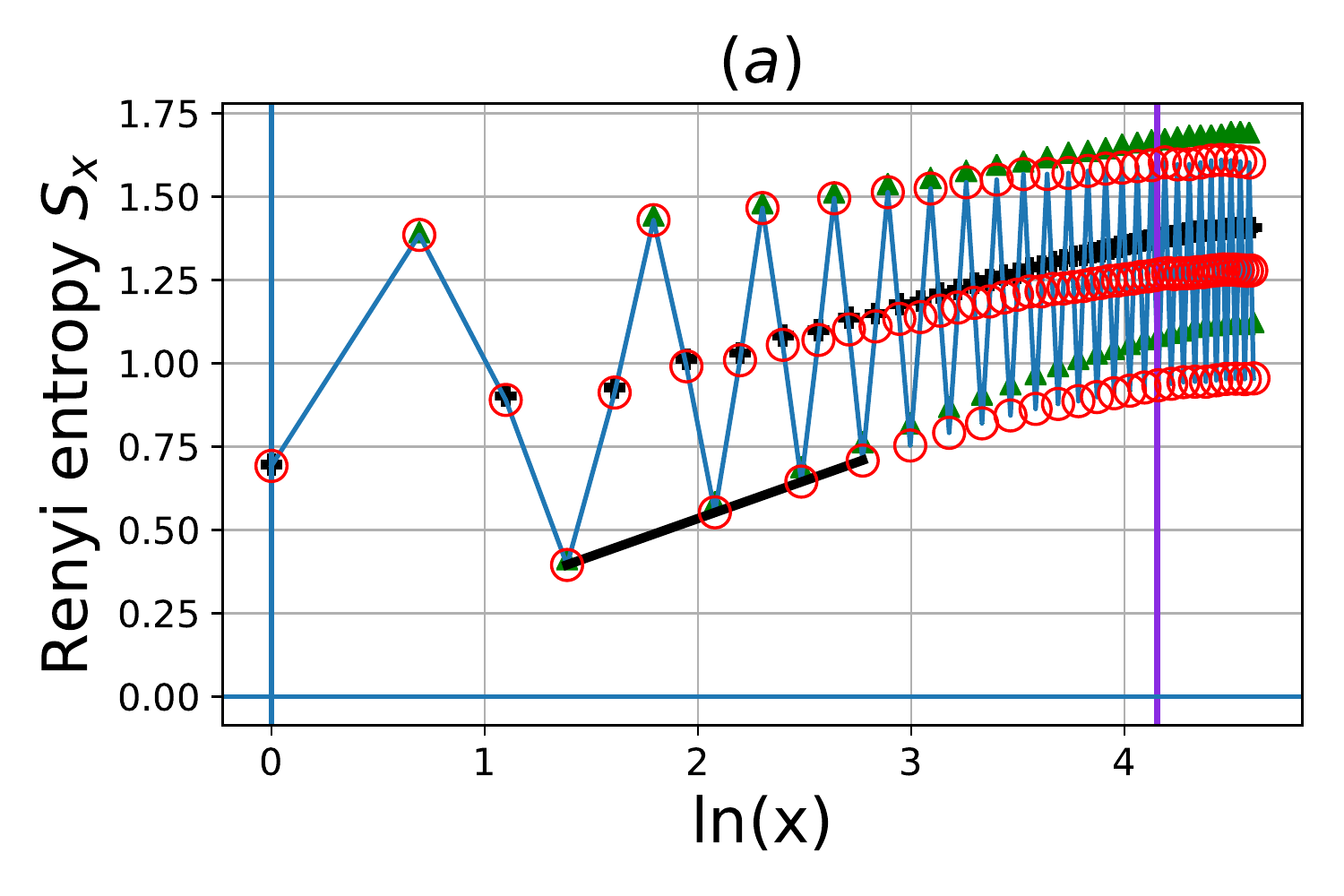}\hfill
\includegraphics[width=0.33\textwidth]{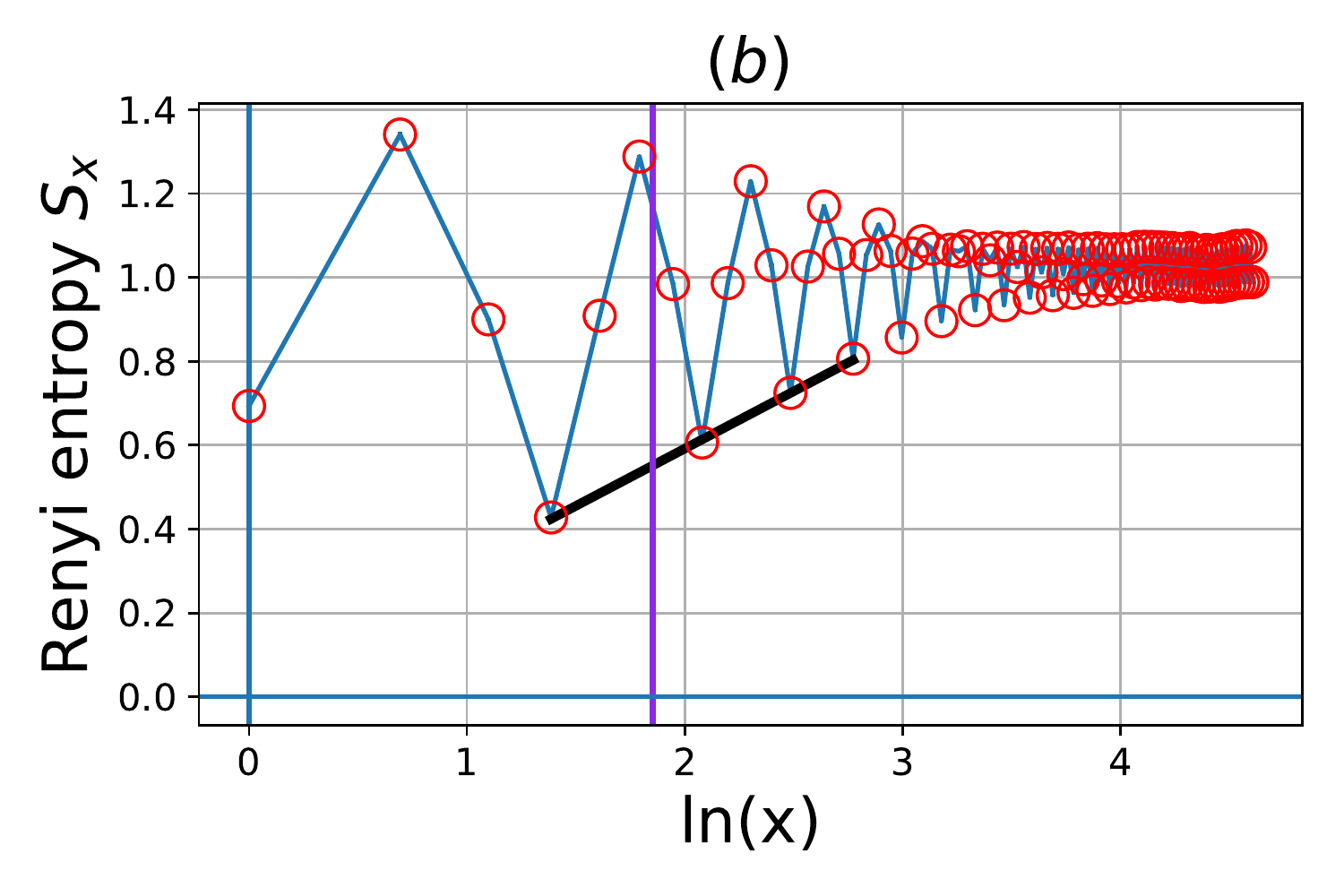}\hfill
\includegraphics[width=0.33\textwidth]{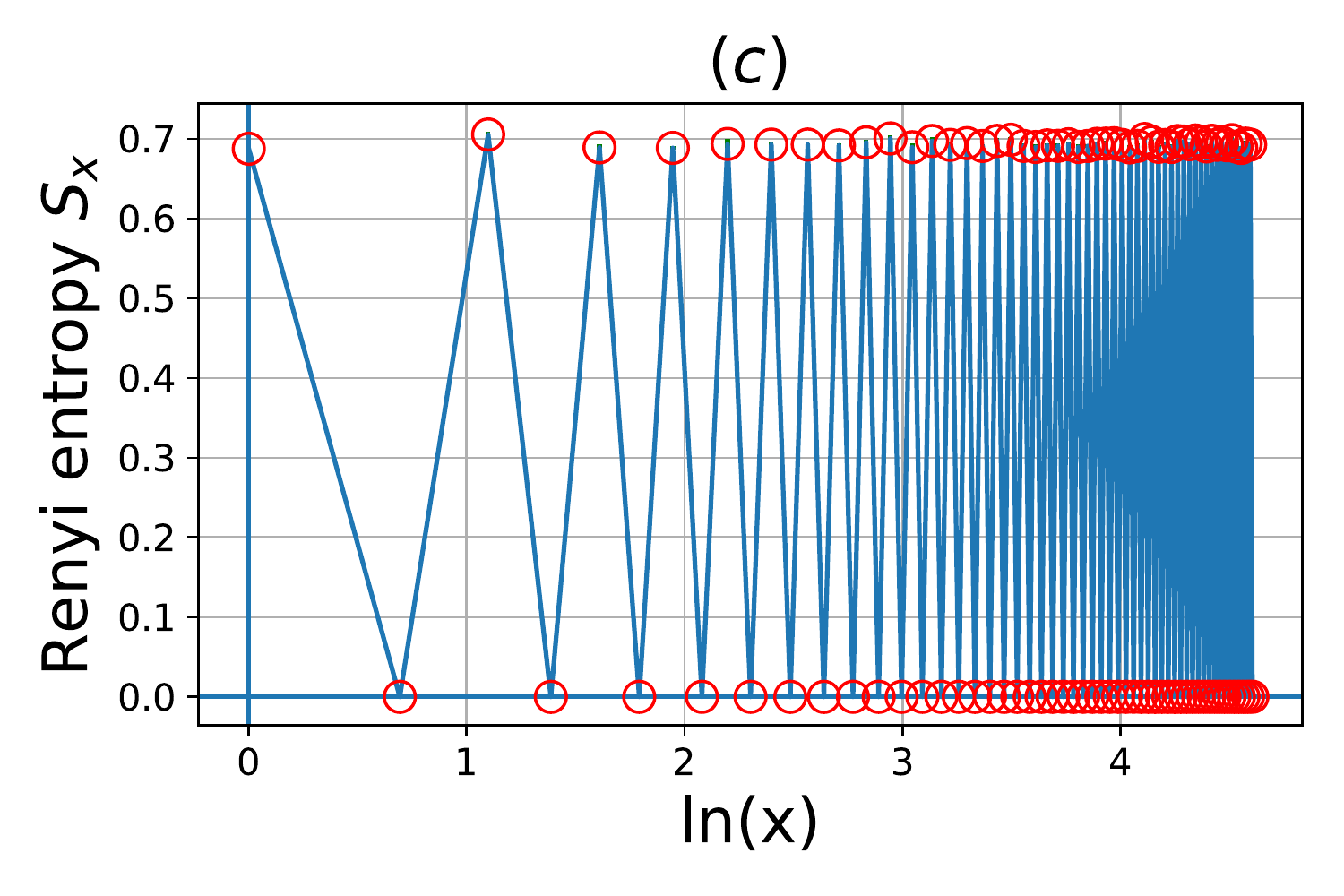}
\caption{Renyi entropy $S_x$ of order two for the uniform ladder with $N=200$ and (a) $\lambda = 1$, (b) $\lambda = 10$, and (c) $\lambda = 200$. For $x$ even (odd), part $A$ of the system consists of the first $x/2$ (the first $(x+1)/2$) spins on the front and the first $x/2$ (the first $(x-1)/2$) spins on the back of the cylinder. In (a), we also plot (green triangles and black pluses) the sum of the entropies for two independent spin chains with the same $\lambda$ and cut at the same positions as the legs of the ladder. The discrepancies show that the interchain interactions in the ladder model are important for $\lambda=1$. The straight line fits in (a) and (b) have slope  $0.23$ and $0.27$, respectively. The vertical lines in (a) and (b) are at $x=N/(\pi\lambda)$.} \label{g-22-to-25}
\end{figure*}

Results for the Renyi entropy of order two are shown in Fig.\ \ref{g-22-to-25} for $N=200$ and different values of $\lambda$. For $\lambda=200$, the entropy is close to $\ln(2)\approx0.693$, whenever the partition cuts a singlet apart, and it is close to zero, whenever none of the singlets are cut apart. For $\lambda=1$, we see that the entropy grows linearly with $\ln(x)$ in the region $1\ll x\ll N$, except for oscillations. Considering only the points for which $x$ is a multiple of four (this corresponds to both legs being cut after an even number of spins), the fitted slope is $0.23$. In the limit of $\lambda$ going to zero, the two legs of the ladder decouple into two independent spin chains, and the entanglement entropy for the ladder is twice the entanglement entropy for a single chain. The relevant slope to compare to is hence $c/4=0.25$ rather than $c/8$. It is also seen that the vertical line $x=N/(\pi\lambda)$ is approximately at the transition between linear growth with $\ln(x)$ for small $x$ and area law behavior for large $x$ (after averaging out the oscillations). This is consistent with the results for the correlations in the previous section.

\section{Conclusion}\label{SEC:conclusion}

We have constructed and studied a family of two-body spin models on a cylinder that are related to the HS model. The usual HS model corresponds to placing the spins uniformly on a circle around the cylinder. Here, we have instead placed the spins along one or two lines on the cylinder that are parallel to the cylinder axis. This gives rise to chain and ladder models, respectively. The construction allows us to scale the distance between the spins and hence the length of the chain or ladder independently from the circumference of the cylinder, and we have studied the significance of this extra parameter $\lambda$ on the physics.

When the length of the chain or ladder is small compared to the circumference of the cylinder (small $\lambda$), the properties of the ground state are described by the SU(2)$_1$ Wess-Zumino-Witten universality class. The spin-spin correlations decay as a power law with exponent $-1$, and for subsystems consisting of an even number of spins in each of the legs or in the chain, the Renyi entropy of order two grows as the logarithm of the subsystem size with a proportionality constant consistent with a central charge of $c=1$. There are also some edge effects. A spin in the chain is more strongly correlated with the neighboring spin on the side, where there is an odd number of spins, than with the neighboring spin on the side, where there is an even number of spins. In addition, when the number of spins in the subregion is odd, the proportionality constant in the Renyi entropy is lower than predicted by a critical theory with central charge $c=1$, and the slope varies with $\lambda$. In the small $\lambda$ limit, the ladder model reduces to a product of two chain models, and the spin-spin interaction strengths for spins in the bulk are inversely proportional to the square of the distance between the spins as in the HS model. The conclusion is hence that for small $\lambda$, the physics of the investigated model is the same as for the HS model, except for edge effects.

When keeping the number of spins $N$ fixed and taking the limit, where the length of the chain or ladder is large compared to the circumference of the cylinder (large $\lambda$), the wavefunction of the ground state reduces to a product of singlets, and the singlets are formed between neighboring spins. In this limit, the correlations decay exponentially, and the Renyi entropy follows an area law. The model hence enables us to transform between one or two copies of an HS-like model and a product of singlets with a Hamiltonian that contains only two-body interactions. All the way along this path the ground state is known analytically and various properties can be computed for large system sizes using Monte Carlo simulations or analytical tools.

When changing $\lambda$ from small to large, we do not observe a sharp transition between the two behaviors described above. Instead the transition occurs gradually, in the sense that the chain shows different behaviors depending on the distances considered. For small distances and small subsystem sizes, the system behaves as in the critical phase. For large distances and large subsystem sizes, the correlations decay exponentially, and the entropy follows an area law. As $\lambda$ changes, the border between small and large distances moves. The ladder model shows a similar behavior.

The results presented in this paper are interesting, because they show that it is possible to have a system, where the correlations and the entropy behave in different ways depending on the distances considered. Although the precise pattern of interaction strengths present in the considered models is difficult to realize in experiments, the study suggests that having interaction strengths in the Hamiltonian that change behavior depending on the distance may be a mechanism to obtain a model, where the correlations and the entropy change behavior depending on distance.

The investigated models contain several parameters, since the spin positions can be chosen freely on two lines, and for each choice there is a family of two-body Hamiltonians having the same ground state. Several further investigations could hence be done within the same framework. Apart from being a nontrivial generalization of the HS model with only two-body interactions, the models presented in this work provide an interesting playground for testing numerical approximation schemes. The models display a variety of physical properties, and they have the unusual feature of combining possibly long-range two-body interactions with an analytically known ground state for which various properties can easily be computed.

\begin{acknowledgments}
We would like to thank Germ\'an Sierra and Hong-Hao Tu for helpful discussions and Germ\'an Sierra for pointing our attention to the work by Inozemtsev.
\end{acknowledgments}

\appendix

\section{Wavefunction for small $\Lambda$}\label{appB}

In this section, we derive an expression for the wavefunction when \eqref{limit} applies. As mentioned in the main text, we number the spins with $\sigma_j>0$ from $1$ to $N_+$ and the spins with $\sigma_j<0$ from $N_++1$ to $N=N_++N_-$. First note that
\begin{multline}
z_j-z_k = \sigma_j e^{\Lambda f(j)} - \sigma_k e^{\Lambda f(k)}
=e^{\Lambda[f(j)+f(k)]/2}\\
\times\{\sigma_j e^{\Lambda [f(j)-f(k)]/2} - \sigma_k e^{-\Lambda [f(j)-f(k)]/2}\}\\
\approx\left\{\begin{array}{ll}
\Lambda[f(j)-f(k)]\sigma_j e^{\Lambda[f(j)+f(k)]/2} &\textrm{for } \sigma_j=\sigma_k\\
2\sigma_j e^{\Lambda[f(j)+f(k)]/2} &\textrm{for } \sigma_j=-\sigma_k
\end{array}\right..
\end{multline}
The factor
\begin{multline}
\prod_{j<k}\{2e^{\Lambda[f(j)+f(k)]/2}\}^{(s_js_k-1)/2}\\
=e^{\sum_{j<k}\{\Lambda[f(j)+f(k)]/2+\ln(2)\}(s_js_k-1)/2}\\
=e^{\sum_{j,k}\{\Lambda[f(j)+f(k)]/2+\ln(2)\}(s_js_k-1)/4}\\
=e^{\sum_{j}[-N\Lambda f(j)-N\ln(2)]/4}
\end{multline}
does not depend on $s_j$, so it will be absorbed in the normalization of the wavefunction and can be ignored. We are hence left with
\begin{multline}\label{wfprod}
\psi_{s_1,\ldots,s_N}(z_1,\ldots,z_N)\approx \textrm{constant}
\times\delta_\mathbf{s} \prod_{p=1}^N\chi_{p,s_p}\\
\times\prod_{j<k} \sigma_j^{\frac{1}{2} (s_js_k-1)}
\prod_{\{j<k|\sigma_j=\sigma_k\}} \{\Lambda[f(j)-f(k)]/2\}^{\frac{1}{2} (s_js_k-1)}.
\end{multline}
Let us introduce the notation
\begin{equation}
s_+\equiv \sum_{\{j|\sigma_j=+1\}} s_j \qquad \textrm{and} \qquad
s_-\equiv \sum_{\{j|\sigma_j=-1\}} s_j.
\end{equation}
We then have
\begin{multline}\label{sigmaprod}
\prod_{j<k}\sigma_j^{\frac{1}{2}(s_js_k-1)}
=(-1)^{\sum_{\{j<k|\sigma_j=-1\}}(s_js_k-1)/2}\\
=(-1)^{\sum_{\{j<k|\sigma_j=\sigma_k=-1\}}(s_js_k-1)/2}\\
=(-1)^{\sum_{\{j,k|\sigma_j=\sigma_k=-1\}}(s_js_k-1)/4}
=(-1)^{(s_-^2-N_-^2)/4},
\end{multline}
where we have used that the spins with $\sigma_j<0$ have higher indices than those with $\sigma_j>0$. We also have
\begin{multline}\label{chiprod}
\prod_{p=1}^N\chi_{p,s_p}
=\prod_{p=1}^{N_+}\chi_{p,s_p}\prod_{p=1}^{N_-}e^{i\pi(N_++p-1)(1+s_{N_++p})/2}\\
=\prod_{p=1}^{N_-} (-1)^{N_+(1+s_{N_++p})/2} \prod_{p=1}^{N_+}\chi_{p,s_p}\prod_{p=1}^{N_-}\chi_{p,s_{N_++p}}\\
=(-1)^{N_+(N_-+s_-)/2} \prod_{p=1}^{N_+}\chi_{p,s_p}\prod_{p=1}^{N_-}\chi_{p,s_{N_++p}}.
\end{multline}

Now note that
\begin{align}
\prod_{\{j<k|\sigma_j=\sigma_k\}} \Lambda^{\frac{1}{2} (s_js_k-1)}
&=\prod_{\{j,k|\sigma_j=\sigma_k\}} \Lambda^{\frac{1}{4} (s_js_k-1)}\nonumber\\
&=\Lambda^{\frac{1}{4}(s_+^2-N_+^2+s_-^2-N_-^2)}.
\end{align}
Therefore, in the limit \eqref{limit}, only configurations that minimize $s_+^2+s_-^2$ will remain. If $N_+$ and $N_-$ are both even, we have that $s_+^2+s_-^2$ is minimized for $s_+=s_-=0$. In other words, $\delta_{\mathbf{s}}$ is replaced by $\delta_{s_+}\delta_{s_-}$. If $N_+$ and $N_-$ are both odd, it is not possible to have $s_+=0$ or $s_-=0$, and we minimize $s_+^2+s_-^2$ for the choice $s_+=-s_-=+1$ and for the choice $s_+=-s_-=-1$. In that case, $\delta_{\mathbf{s}}$ is replaced by $\delta_{s_+=1}\delta_{s_-=-1}$ and $\delta_{s_+=-1}\delta_{s_-=1}$, respectively. The relative sign of the two terms in the wavefunction is $(-1)^{N_+(N_--1)/2-N_+(N_-+1)/2}=-1$. Inserting the above observations into \eqref{wfprod}, we obtain \eqref{wfeven} and \eqref{wfodd} in the main text.

\section{Wavefunction for large $\Lambda$}\label{appC}

To determine the form of the wavefunction \eqref{wf} with coordinates \eqref{zj} in the limit \eqref{condition}, we first write the wavefunction as follows
\begin{multline}\label{1D_singlet_1}
\delta_\mathbf{s}\prod_{p=1}^N\chi_{p,s_p} \prod_{j<k}(z_j-z_k)^{\frac{1}{2} (s_js_k-1)}=\\
\delta_\mathbf{s}\prod_{p=1}^N\chi_{p,s_p}
\prod_{j<k}\left[\cancel{\sigma_{2j}e^{\Lambda f(2j)}}-\sigma_{2k}e^{\Lambda f(2k)}\right]^{(s_{2j}s_{2k}-1)/2}\\
\times\prod_{j<k}\left[\cancel{\sigma_{2j-1}e^{\Lambda f(2j-1)}}-\sigma_{2k-1}e^{\Lambda f(2k-1)}\right]^{(s_{2j-1}s_{2k-1}-1)/2}\\
\times\prod_{j<k}\left[\cancel{\sigma_{2j}e^{\Lambda f(2j)}}-\sigma_{2k-1}e^{\Lambda f(2k-1)}\right]^{(s_{2j}s_{2k-1}-1)/2}\\
\times\prod_{j<k}\left[\cancel{\sigma_{2j-1}e^{\Lambda f(2j-1)}}-\sigma_{2k}e^{\Lambda f(2k)}\right]^{(s_{2j-1}s_{2k}-1)/2}\\
\times\prod_{k=1}^{N/2}\left[\sigma_{2k-1}e^{\Lambda f(2k-1)}-\sigma_{2k}e^{\Lambda f(2k)}\right]^{(s_{2k-1}s_{2k}-1)/2}.
\end{multline}
For the sake of generality, we shall here assume that the $\sigma_k$ are general phase factors not restricted to being plus or minus one. Utilizing \eqref{condition}, we can ignore the terms that are crossed out in the above expression. Let us define $\tilde{\sigma}(2k)$ and $\tilde{f}(2k)$ such that
\begin{equation}\label{ftilde}
-\tilde{\sigma}_{2k}e^{\Lambda \tilde{f}(2k)}=
-\sigma_{2k}e^{\Lambda f(2k)}+\sigma_{2k-1}e^{\Lambda f(2k-1)}.
\end{equation}
Since $f(2k)\geq f(2k-1)$, we must have $\tilde{f}(2k)\leq f(2k)+\ln(2)/\Lambda$.

With the definition \eqref{ftilde} we get that \eqref{1D_singlet_1} simplifies to
\begin{multline}
\delta_\mathbf{s}\prod_{p=1}^N\chi_{p,s_p} \prod_{j<k}(z_j-z_k)^{\frac{1}{2} (s_js_k-1)}\approx
\textrm{constant}\times \delta_\mathbf{s} e^{\Lambda F}\\
\times \prod_{p=1}^N\chi_{p,s_p} \prod_{j<k}(-\sigma_k)^{(s_js_k-1)/2}
\prod_{k=1}^{N/2}\left(\frac{\tilde{\sigma}_{2k}} {\sigma_{2k}}\right)^{(s_{2k-1}s_{2k}-1)/2},
\end{multline}
where
\begin{multline}
F\equiv\sum_{k=2}^{N/2} \left[f(2k-1)s_{2k-1}+f(2k)s_{2k}\right] \sum_{j=1}^{k-1}(s_{2j-1}+s_{2j})\\
+\sum_{k=1}^{N/2}\tilde{f}(2k)s_{2k-1}s_{2k}
\end{multline}
The configurations with the highest weight in the wavefunction are hence those that maximize $F$ under the constraint $\sum_js_j=0$. In appendix \ref{appD}, we show that, under the constraint $\sum_js_j=0$, $F$ is maximal for all configurations fulfilling $s_{2j-1}+s_{2j}=0$ for all $j\in\{1,2,\ldots,N/2\}$. We also show that all other configurations with $\sum_js_j=0$ have negligible weight, when the approximation \eqref{condition} applies.

To show that the wavefunction is a product of singlets, we additionally need to show that the wavefunction has the right phase factors. The phase of the wavefunction for a given configuration is
\begin{equation}\label{phase}
\prod_{p=1}^N\chi_{p,s_p}\prod_{j<k}(-\sigma_k)^{(s_js_k-1)/2}
\prod_{k=1}^{N/2}\left(\frac{\tilde{\sigma}_{2k}} {\sigma_{2k}}\right)^{(s_{2k-1}s_{2k}-1)/2}.
\end{equation}
Utilizing that $s_{2j-1}+s_{2j}=0$ for all contributing configurations, we get
\begin{equation}
\sum_{j=1}^{k-1}s_js_k=\left\{\begin{array}{cl}
-1 & \textrm{for } k \textrm{ even}\\
0 & \textrm{for } k \textrm{ odd}
\end{array}\right.,
\end{equation}
and it follows that the latter two products in \eqref{phase} do not depend on the configuration. From the definition of $\chi_{p,s_p}$, one can check that
$\chi_{2j-1,+1}\chi_{2j,-1}=-\chi_{2j-1,-1}\chi_{2j,+1}$. The phase factors of the terms in the wavefunction are hence precisely those for a product of singlets.

\section{Derivation of an inequality}\label{appD}

To show that precisely the configurations with $s_{2j-1}+s_{2j}=0$ for all $j\in\{1,2,\ldots,N/2\}$ maximize $F$ under the constraint $\sum_js_j=0$, we first note that
\begin{widetext}
\begin{align}\label{rewrite}
F&=\sum_{k=2}^{N/2} \left[f(2k-1)s_{2k-1}+f(2k)s_{2k}\right] \sum_{j=1}^{k-1}(s_{2j-1}+s_{2j})+\sum_{k=1}^{N/2}\tilde{f}(2k)s_{2k-1}s_{2k}\nonumber\\
&=\frac{1}{2}\sum_{k=2}^{N/2} \left[f(2k)+f(2k-1)\right] (s_{2k}+s_{2k-1}) \sum_{j=1}^{k-1}(s_{2j-1}+s_{2j})
+\sum_{k=1}^{N/2}\tilde{f}(2k)(s_{2k-1}s_{2k}+1)\nonumber\\
&+\frac{1}{2}\sum_{k=2}^{N/2} \left[f(2k)-f(2k-1)\right]
(s_{2k}-s_{2k-1}) \sum_{j=1}^{k-1}(s_{2j-1}+s_{2j})
-\sum_{k=1}^{N/2}\tilde{f}(2k).
\end{align}
\end{widetext}
With this expression, it is natural to group the spins together in pairs. We define
\begin{equation}
t_j=\frac{1}{2}(s_{2j-1}+s_{2j}), \quad j\in\{1,2,\ldots,N/2\}.
\end{equation}
Note that $t_j$ can take the values $-1$, $0$, or $+1$. We shall refer to these as negative defect, no defect, and positive defect, respectively. Note also that the condition $\sum_{j=1}^N s_j=0$ translates into $\sum_{j=1}^{N/2}t_j=0$, so for a given choice of configuration, the number of positive defects must equal the number of negative defects. Let us consider a configuration with defects at $a_1<a_2<\ldots<a_D$, where $a_j\in\{1,2,\ldots,N/2\}$. The factor $(s_{2k}-s_{2k-1})$ is zero if there is a defect at position $k$, and can be either plus or minus two if there is no defect. The choice of sign does not affect other parts of the right hand side of \eqref{rewrite}, and we get the largest value of the right hand side if the sign of $(s_{2k}-s_{2k-1})$ always cancels the sign of $\sum_{j=1}^{k-1}(s_{2j-1}+s_{2j})$. We can therefore rewrite \eqref{rewrite} into
\begin{multline}\label{rewrite2}
\frac{1}{2}F+\frac{1}{2}\sum_{k=1}^{N/2}\tilde{f}(2k)
\leq \\
\sum_{k=2}^{N/2} \left[f(2k)+f(2k-1)\right] t_{k} \sum_{j=1}^{k-1}t_{j}
+\sum_{k=1}^{N/2}\tilde{f}(2k)|t_k|\\
+\sum_{k=2}^{N/2} \left[f(2k)-f(2k-1)\right](1-|t_k|)
\left|\sum_{j=1}^{k-1}t_j\right|.
\end{multline}
Note that the right hand side of this expression is zero if no defects are present.

\begin{figure*}
\includegraphics[width=0.8\textwidth]{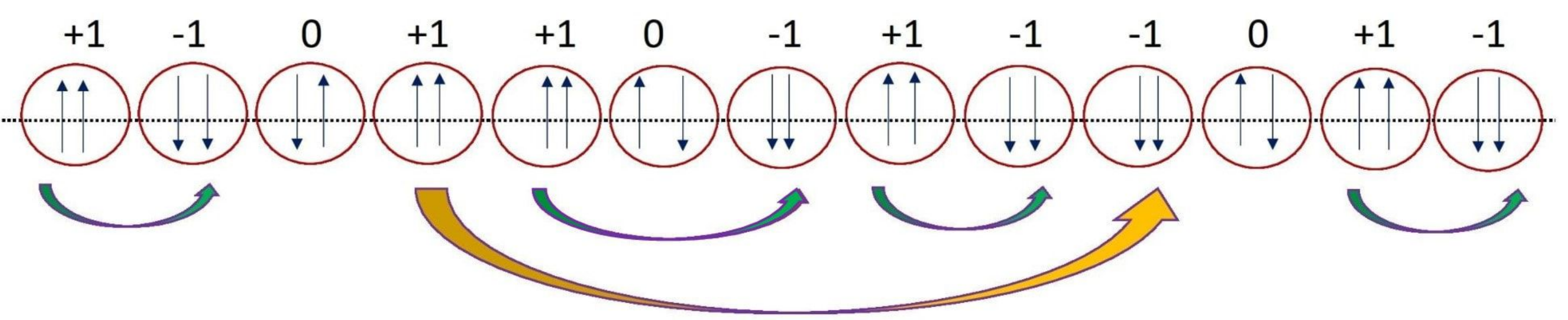}
\caption{Example of the pairing of spins and defects used in the derivation. The figure shows the spins $1,2,\ldots,N$ in the wavefunction \eqref{wf}. Each spin is shown as a black arrow. An arrow pointing up (down) represents the spin state $s_j=+1$ ($s_j=-1$). The red circles illustrate the pairing of neighboring spins (spin 1 with spin 2, spin 3 with spin 4, $\ldots$ , spin $N-1$ with spin $N$). If the neighboring spins are both up, this leads to a positive defect (labelled $+1$), if the neighboring spins are both down, this leads to a negative defect (labelled $-1$), and if one spin is up and one is down, it leads to a neutral site with no defect (labelled $0$). Defects with opposite signs are then paired. We first pair the defects, for which the two defects are neighbors or only have neutral sites between them. These pairs are marked with the four green arrows. We then pair defects, for which the two defects are neighbors if we ignore neutral sites and defects that we have already paired. There is one such pair in the figure, and it is marked with a yellow arrow. We continue this procedure until all defects have been paired.}\label{defect-min}
\end{figure*}

We now pair up the positive and negative defects in an iterative process as follows. In each iteration step, we pair $a_q$ with $a_p$ according to the following rules:
\begin{enumerate}
\item{$q$ is the lowest possible number such that $a_q$ is a defect that has not yet been paired.}
\item{$p$ is the lowest possible number such that
\begin{itemize}
\item{$a_p$ is a defect that has not yet been paired}
\item{$t_{a_p}=-t_{a_q}$}
\item{$\sum_{x=q+1}^{p-1}t_{a_x}=0$}
\end{itemize}}
\end{enumerate}
We repeat this process until all defects have been paired. An example is shown in Fig.\ \ref{defect-min}

Unless there are no defects in the system, there will always be at least one value of $p$ for which $a_p$ and $a_{p+1}$ have been paired. Let us first consider such a pair. We introduce the notation
\begin{equation}
v_p=\sum_{j=1}^{a_p-1}t_j=\sum_{j=1}^{p-1}t_{a_j}.
\end{equation}
Note that $v_{p+1}=v_p+t_{a_p}$. With the choice of pairing we have made, $v_{p+1}$ is nonzero and $v_{p+1}$ and $t_{a_{p+1}}$ have opposite signs. Since $v_{p+1}$ and $v_p$ are integers fulfilling $|v_{p+1}-v_p|=1$, $v_p$ is either zero or has the same sign as $v_{p+1}$. Furthermore, since $t_{a_p}=-t_{a_{p+1}}$, it also follows that $t_{a_p}$ and $v_p$ have the same sign if $v_p$ is nonzero. Therefore $|v_{p+1}|=|v_p+t_{a_p}|=|v_p|+1$. It follows that
\begin{align}
t_{a_p}v_p&=|t_{a_p}||v_p|=|v_p|,\\
t_{a_{p+1}}v_{p+1}&=-|t_{a_{p+1}}||v_{p+1}|=-|v_p|-1.
\end{align}
Utilizing this result, we find that the terms on the right hand side of \eqref{rewrite2}, which have $k\in\{a_p,a_p+1,\ldots,a_{p+1}\}$, add up to
\begin{widetext}
\begin{multline}\label{rewrite3}
\sum_{k=a_p}^{a_{p+1}} \left[f(2k)+f(2k-1)\right] t_{k} \sum_{j=1}^{k-1}t_{j}+\sum_{k=a_p}^{a_{p+1}}\tilde{f}(2k)|t_k|
+\sum_{k=a_p}^{a_{p+1}} \left[f(2k)-f(2k-1)\right](1-|t_k|)
\left|\sum_{j=1}^{k-1}t_j\right|\\
=\left[f(2a_p)+f(2a_p-1)\right] |v_p|
-\left[f(2a_{p+1})+f(2a_{p+1}-1)\right] (|v_p|+1)
+\tilde{f}(2a_p)+\tilde{f}(2a_{p+1})
+\sum_{k=a_p+1}^{a_{p+1}-1} \left[f(2k)-f(2k-1)\right]
(|v_p|+1)\\
=\tilde{f}(2a_p)-f(2a_p)
+\tilde{f}(2a_{p+1})-f(2a_{p+1})
-[f(2a_{p+1})-f(2a_p-1)]|v_p|
-\sum_{k=a_p}^{a_{p+1}-1} \left[f(2k+1)-f(2k)\right]
(|v_p|+1)
\end{multline}
\end{widetext}
So far we have not made assumptions about $\tilde{f}$. For the case of interest here, however, we know that $\tilde{f}(2a_p)-f(2a_p)$ and $\tilde{f}(2a_{p+1})-f(2a_{p+1})$ are both at most $\ln(2)/\Lambda$. On the other hand, we know due to \eqref{condition} that the last term on the right hand side of \eqref{rewrite3} is much more negative than $-2\ln(2)/\Lambda$ even for $|v_p|=0$. Hence the right hand side of \eqref{rewrite3} is negative.

We then repeat the same computation for all other pairs of defects $a_q$ and $a_p$ for which $|p-q|=1$. When we have done that, we remove all defects from the set $\{a_1,a_2,\ldots,a_D\}$ that we have already taken into account and repeat the same computations for the pairs of defects, for which the defects in the pair are neighbors in the new set (an example is the pair marked with the yellow arrow in Fig.\ \ref{defect-min}). The computation is again the same as above, except that some values of $k$ between $a_q$ and $a_p$ are omitted from the sums, since we have already taken them into account. This does, however, not change the conclusion that the result is negative. We repeat this procedure until all defect pairs have been taken into account. Since the contribution from each pair is negative, we conclude that
\begin{equation}\label{Fresult}
F\leq -\sum_{k=1}^{N/2}\tilde{f}(2k),
\end{equation}
and equality is only obtained if there are no defects, i.e.\ if $s_{2j-1}+s_{2j}=0$ for all $j\in\{1,2,\ldots,N/2\}$. It also follows from the above derivation and \eqref{condition} that the next highest value of $e^{\Lambda F}$ is much lower than the highest value of $e^{\Lambda F}$, so only configurations fulfilling $s_{2j-1}+s_{2j}=0$ contribute significantly to the wavefunction.

Let us finally comment that the result \eqref{Fresult} is valid independent of \eqref{condition} if $f(N)\geq f(N-1)> f(N-2)\geq f(N-3)>\ldots\geq f(1)$ and $\tilde{f}(2k)\leq f(2k)$ for all $k$. The particular case $\tilde{f}(2k)=f(2k)$ leads to
\begin{equation}
\sum_{j<k}f(k)s_js_k\leq -\sum_{k=1}^{N/2}f(2k) \quad \textrm{for} \quad \sum_js_j=0,
\end{equation}
with equality obtained only for spin configurations fulfilling $s_{2j-1}+s_{2j}=0$.

\end{document}